\documentclass[11pt,a4paper]{article}

\usepackage{xcolor}
\usepackage{amsmath}
\usepackage{amsthm}
\usepackage{amssymb}
\usepackage{latexsym}
\usepackage{graphicx}

\def\bC{\mathbf{C}}
\def\bX{\mathbf{X}}
\def\bzero{\mathbf{0}}
\def\bI{\mathbf{I}}
\def\bK{\mathcal{K}}
\def\bN{\mathcal{N}}
\def\bZ{\mathbf{Z}}
\def\bW{\mathbf{W}}
\def\bV{\mathbf{V}}
\def\bx{\mathbf{x}}
\def\bu{\mathbf{u}}
\def\bw{\mathbf{w}}
\def\bv{\mathbf{v}}
\def\COV{C\hspace{-0.05cm}ov}

\theoremstyle{plain}
\newtheorem{thm}{Theorem}
\newtheorem{lemma}{Lemma}

\newenvironment{apabib}{
  \noindent
{\large\bf References} \vspace{-3mm}
\begin{list}{}{
  \topsep          0mm
  \leftmargin     10mm
  \parsep          3mm
  \listparindent -10mm}
  \item[]\strut\par}{
\end{list}}

\begin{document}
\fontsize{12}{18pt}\selectfont


\thispagestyle{empty} \vspace*{10mm}
\begin{center}

{\Large\bf  Nonparametric Regression with Multiple Thresholds: 
Estimation and Inference}

\vspace*{10mm}
{\large\bf Yan-Yu Chiou$^{a}$, Mei-Yuan Chen$^{b, *}$, Jau-er Chen$^{c, *}$}

\vspace*{3mm}
\hspace*{1.24cm}{\small $^{a}$Institute of Economics, Academia Sinica, Taiwan. \hfill}

\hspace*{1.24cm}{\small $^{b}$Department of Finance, National Chung Hsing University, Taiwan. \hfill}

\hspace*{1.24cm}{\small $^c$Department of Economics, National Taiwan University, Taiwan. \hfill}

\vspace*{10mm}
2nd-round R\&R at the {\textit{Journal of Econometrics}}

\end{center}

\vfill
\hrule
\bigskip

\noindent
{\scriptsize 
We are grateful to the two anonymous referees for 
their constructive comments that have  greatly improved this paper.
We thank Ming-Yen Cheng for valuable discussions, and thank Zongwu Cai and the participants 
at the International Symposium on Recent Developments 
in Econometric Theory with Applications 
in Honor of Professor Takeshi Amemiya for their helpful comments.
The usual disclaimer applies.
 *Corresponding authors: 
National Chung Hsing University, Department of Finance,
250 Kuo Kuang Road, Taichung 402,  Taiwan.
Tel.: +886-4-22853323.
E-mail address: $\mbox{mei}_{-}\mbox{yuan}$@dragon.nchu.edu.tw (Mei-Yuan Chen);
National Taiwan University, Department of Economics, No.\ 1, Sec.\ 4, Roosevelt Road, Taipei 10617, Taiwan.
Tel.: +886-2-3366-8326. E-mail address: jauer@ntu.edu.tw (Jau-er Chen).
}

\pagebreak
\begin{center}
{\large\bf ABSTRACT}
\end{center}

This paper examines  nonparametric regression with an exogenous threshold variable, 
allowing for an unknown number of thresholds. Given the number of thresholds and corresponding threshold values, we first establish the asymptotic properties of the local constant estimator for a nonparametric regression with multiple thresholds. However, the number of thresholds and  corresponding threshold values are typically unknown in practice. We then use our testing procedure to determine the unknown number of thresholds and derive the limiting distribution of the proposed test. The Monte Carlo simulation results indicate the adequacy of the modified test and accuracy of the sequential estimation of the threshold values. We apply our testing procedure to an empirical study of the 401(k) retirement savings plan with income thresholds.

\vspace{10mm}



\pagebreak
\pagenumbering{arabic}
\setcounter{page}{1}

\section{Introduction}

Piecewise linearity has been widely used to model shifts in economic relationships under a regression framework. Most regressions with piecewise linearity can be represented as
linear regressions with thresholds. For example, linear regressions with structural changes can be written as linear threshold regressions with the time index as the threshold variable.
Among previous studies, 
Bai and Perron~(1998, 2003),  Qu and Perron~(2007), and Yamamoto and Perron~(2013)  estimate and test linear regressions with structural changes and
Chen~(2008), Qu~(2008), and Oka and Qu~(2011) estimate and test linear quantile regressions with structural changes. The threshold model splits the sample into classes based on the value of an observed variable (i.e., whether it exceeds a certain threshold). 
In empirical work, determining the threshold of economic variables such as taxes rates as well as the optimal public debt ratio is relevant for policy makers.
When the threshold is unknown as is typical in practice, it needs to be estimated, and this consequently increases the complexity of the econometric problem. Nonetheless, theories of estimation and inference are well developed for linear models with exogenous regressors, 
including the works by Chan~(1993), Hansen~(1996, 1999, 2000), and Caner~(2002).

The scope of threshold models has broadened considerably in  recent years.
In particular, discussions of piecewise linearity have been extended to nonparametric regressions. Su and Xiao~(2008), for instance, test for structural changes in time-series
nonparametric regression models, while Chen and Hong~(2012) investigate how to test for smooth structural changes in time-series models by using nonparametric regressions. 
In addition, Chen and Hong~(2013) extend their earlier study to test for smooth structural changes in panel data models. In economics, the regression discontinuity (RD) design has  gradually emerged as a common tool in applied research. The validity of RD estimates depends crucially both on the threshold variable (also termed the running variable in the RD literature) and on an adequate description of the conditional mean function of the outcome variable. Since what looks like a jump at the threshold might simply be  unaccounted for nonlinearity, the nonparametric approach plays an important role in the RD estimations (cf.\ Angrist and Pischke, 2009). For example, by allowing for an unknown threshold value in the RD framework, Henderson, Parmeter, and Su (2014) provide estimation and inference procedures for the threshold value in a nonparametric regression with one threshold. Although related to Henderson et al.\ (2014), which is a pioneering study examining the nonparametric regression with one threshold, our study analyzes nonparametric regression with multiple thresholds. Further, in contrast to Henderson et al.\ (2014), the threshold variable is excluded from the explanatory variables in our framework.
In empirical applications, multiple thresholds might be present; however, the number of thresholds and the corresponding threshold values are typically unknown in practice. Therefore, identifying the unknown number of thresholds and estimating the threshold values are critical issues in a nonparametric regression with multiple thresholds, especially when conducting empirical studies. We thus propose a testing procedure to determine the unknown number of thresholds and derive the limiting distribution of the proposed test. To the best of our knowledge, the present study is the first to  comprehensively investigate the aforementioned issues. This study develops a test procedure for testing the existence of thresholds, determining the number of thresholds, and estimating the values of thresholds in nonparametric regression. Specifically, this procedure is a modified significance test based on the work of  A\"{\i}t-Sahalia et al.~(2001). In addition, 
we establish the consistency and asymptotic normality of the threshold value estimators by using the sequential method. Hence, this study complements the existing literature on estimating and testing multiple thresholds in nonparametric regression models. Further, we apply our testing procedure to an empirical study of the 401(k) retirement savings plan
with income thresholds and identify four threshold values. Those crucial income threshold values are all above the median income value.

The rest of the paper is organized as follows. The model specification and estimation for a nonparametric regression with thresholds are introduced in Section 2.  This section  also summarizes the necessary assumptions for deriving our theoretical results of the test statistics and estimators under the known thresholds. Section 3 provides the test determining the unknown number of thresholds. Section 4 presents the statistical properties of the multiple threshold estimator. Section 5 investigates the performance of these tests by using Monte Carlo studies, while Section 6 presents an empirical application. Section 7 concludes. All the technical proofs are collected in the Appendix.

\section{Model, Assumptions, and Asymptotics}

We first fix the notations and consider the following threshold model, which is
a nonparametric regression with $s$ thresholds and known threshold values: 
\begin{eqnarray*}
E(Y |\bX, Q ) = \sum^{s+1}_{j=1} m_{\gamma_{j}}(\bX) I_{\gamma_{j}}(Q),
\end{eqnarray*}
where 
$Y$ is the outcome variable, $\bX$ is a vector of the covariates, 
$Q$ is the threshold variable, 
which is used to split the sample into distinct $s$ thresholds,
$\gamma_1, \gamma_2, \ldots, \gamma_{s+1}$ are the corresponding threshold values, and
$I_{\gamma_{j}}(Q_{i})$ denotes an indicator function defined as
\begin{eqnarray*}
I_{\gamma_{j}}(Q) &=& \left\{ \begin{array}{ll}
                                   1 & Q \in [ \gamma_{j-1} , \gamma_{j} ) , \\
                                   0 & \mbox{otherwise},
                                   \end{array} \right.
\end{eqnarray*}
with $\gamma_0 = -\infty$ and $\gamma_{s+1} = \infty$.
Accordingly the conditional mean of the $j$th regime at a grid point $\bx = [x_1, \ldots , x_p]'$ 
can be represented as
\begin{eqnarray*}
m_{\gamma_{j}}(\bx) &=& \mbox{E}(Y | \bX = \bx , I_{\gamma_{j}}(Q) = 1 ) \\
                &=& \int y \frac{f_{\gamma_{j}}(y, \bx)}{f_{\gamma_{j}}(\bx)} d\bx 
\end{eqnarray*}
where $f_{\gamma_{j}}(y, \bx) = \int I_{\gamma_{j}}(q) f(y, \bx, q) dq$ and
$f_{\gamma_{j}}(\bx) = \int I_{\gamma_{j}}(q) f(\bx,q) dq$ denote the joint density function of $Y$ and $\bX$
and the marginal density of $\bX$ in the $j$th regime, respectively.

Given a sample with observations $\{ (Y_i, \bX_i', Q_i)', i=1, \ldots, n \}$, the
nonparametric regression with known $s$ thresholds is specified as
\begin{eqnarray}
Y_{i} = \sum^{s+1}_{j=1} m_{\gamma_{j}}(\bX_{i}) I_{\gamma_{j}}(Q_{i}) +  e_{i} \label{Model}
\end{eqnarray}
where $Y_{i}$, $\bX_{i}$, and $Q_i$ are the $i$th sample observations of $Y$, $\bX$, and $Q$, respectively;
$e_{i}$ is the regression error. Note that the
threshold values satisfy $\gamma_{0} < \gamma_{1} < \ldots < \gamma_{s+1}$.

Given a $p$-dimensional product kernel function, $\bK(\bu)$, in which $\bK_{h}(\bu)$ is defined as
\begin{eqnarray*}
\bK_{h}(\bu) \equiv h^{-p} \bK(\bu/h),
\end{eqnarray*}
the sample kernel density estimators of $f_{\gamma_{j}}(y, \bx)$ and $f_{\gamma_{j}}(\bx)$ are
\begin{eqnarray}
\hat{f}_{\gamma_{j}}(y,\bx) &=& \frac{1}{n}\sum^{n}_{i=1} \bK_{h} (\bX_{i} - \bx)
I_{\gamma_{j}}(Q_{i}) K_{h} (Y_{i} - y) \label{eq:2-1}  \\
\hat{f}_{\gamma_{j}}(\bx)  &=& \frac{1}{n}\sum^{n}_{i=1} \bK_{h} (\bX_{i} - \bx) I_{\gamma_{j}}(Q_{i})  \label{eq:2-2}
\end{eqnarray}
Thus, the standard Nadaraya-Watson kernel regression estimator of $m_{\gamma_{j}}(\bx)$ is
\begin{eqnarray}
\hat{m}_{\gamma_{j}}(\bx) = \frac{\sum^{n}_{i=1} \bK_{h}(\bX_{i} - \bx)
I_{\gamma_{j}}(Q_{i})Y_{i}}{\sum^{n}_{i=1} \bK_{h}(\bX_{i} - \bx) I_{\gamma_{j}}(Q_{i})}.\label{eq:2-3}
\end{eqnarray}

\subsection{Assumptions}

To establish the asymptotic properties of 
the conditional mean estimator, $\hat{m}_{\gamma_{j}}(\bx)$,
and the density estimator,  $\hat{f}_{\gamma_{j}}(y,\bx)$, in the $j$th regime, 
as well as the convergence rate of the optimal bandwidth selector,
we make the following assumptions.

\medskip
\noindent
 {\bf Assumption 1}. The following assumptions are specified for the random variables under study.
 \begin{enumerate}
 \item [1-1.] $\bZ_{i} = (Y_i, \bX_i, Q_i)$ is strictly stationary, ergodic and $\beta$-mixing with $\beta$ coefficients
             for some fixed $\varepsilon > 0$,
             satisfying $\sum^{\infty}_{k=1} k^{2} [\beta(k)]^{\frac{\varepsilon}{1+\varepsilon}} < \infty$.
 \item [1-2.] The density $f(y, \bx,q)$ is bounded away from zero and globally integrable on the compact support $S$ of the weighting function $a(\cdot)$,
 where $a(\cdot)$ is defined in Section 3.1 when we construct 
 the proposed test statistic.
            Hence $\inf_{S}$ $f(\bx,q) \equiv b \geq 0$.
 \item [1-3.] The joint density $f_{1, 1+j}$ of $(\bZ_{1}, \bZ_{1+j})$ exists for all
             $j$ and is continuous on $(R  \times S)^{2}$.

             \item[1-4.] $\mbox{E}[e^{4}_{i}|\bX_{i}=\bx,Q_{i}=q]$ $\leq$ $\infty$ , $\mbox{E}(e^{2}_{i}|\bX_{i}=\bx,Q_{i}=q)
            = \sigma^{2}(\bx,q)$ and $\sigma^{2}(\bx,q)$ is square-integrable on $S$.
 \item[1.5.] $\int |m_{\gamma_{l}}(\bx_{i}) - m_{\gamma_{k}}(\bx_{i}) |  d\bx_{i} \neq 0 $ ,
            $l , k = 1, \ldots , s+1$ and $l \neq k$.

 \end{enumerate}

 \noindent
 {\bf Assumption 2}. The following assumptions are imposed on the kernel function. 
 \begin{enumerate}
 \item[2-1.] $\bK$ is a product kernel, $\bK = K_1 \times \cdots \times K_p = K^p$, given $K_i = K, \forall i$,
            and a  bounded function on $R^{p}$,  symmetric about 0,
            with $\int |K (z)| dz < \infty $, $\int K (u) du = 1 $ ,
            $\int u^{j} K (u) du = 0 , j = 1, \ldots , r-1$, and $\int u^{r} K(u) du < \infty$.
 \item[2.2.] The kernel $K$ is $r$th continuous differentiable with $r > 3p/4$.
 \end{enumerate}

 \noindent
 {\bf Assumption 3}. The following assumptions are assumed for the bandwidth selector.
 \begin{enumerate}
 \item[3-1.] As $n \to \infty$,$h \to 0$,$ nh^{p} \to  \infty$ and $nh^{p+2r+2} \to 0$.
 \item[3-2.] As $n \to \infty$, the bandwidth sequence $h = O(n^{-1/\delta})$ is such that
             $2p < \delta < 2r+ p/2$ and then
            $h \to 0$, $ nh^{p} \to = \infty$ and $nh^{p/2+2r} \to 0$.

 \end{enumerate}

\medskip
Assumptions 1-1 and 1-3 are similar to Assumption 7 in A\"{\i}t-Sahalia et al.~(2001), allowing for dependent observed data including macroeconomic or financial time-series data.
Assumptions 1-2 and 1-4 generalize Assumption 2 of A\"{\i}t-Sahalia et al.~(2001) to encompass the threshold models. Moreover, Assumptions 1-4 and 1-5 restrict the behaviors of the conditional moments and conditional mean functions across distinct thresholds.
Assumption 2.1 states the standard restrictions on the higher-order kernel functions, which are devices used to reduce bias ( cf.\ Li and Racine, 2007). Assumption 2.2, however, implies that there is no need to use a higher-order kernel $(r>2)$ unless the dimensionality of the covariate is greater than or equal to 3. Assumption 3 imposes  the joint restrictions on the bandwidth sequence $h$, order of the kernel $r$, dimensionality of the covariate $p$, and sample size $n$.
In particular, when $p=1$ and $r=2$, the restriction, $2p <\delta <2r+p/2$ which is also used by A\"{\i}t-Sahalia et al.~(2001), leads to $2 < \delta < 4.5$. In this study, when conducting Monte Carlo simulations, we impose $\delta = 4.25$, which suffices the nonparametric estimator valid asymptotic properties.

\subsection{Asymptotic Properties of the Estimators under Known Thresholds}

Assuming that the number of thresholds $s$ and corresponding threshold values $\gamma_j, j=1, \ldots, s+1$ are known already, the consistency and asymptotic normality of $\hat{f}_{\gamma_{j}}(\bx)$ are provided in Theorem \ref{Est-1} and the asymptotic properties of $\hat{m}_{\gamma_{j}}(\bx)$ are stated in Theorem \ref{Est-3}.

\begin{thm}\label{Est-1}
Suppose  that the assumptions in Assumptions 1,  2, and  3-1 hold.
The following results are established.

\noindent
a). The almost sure convergence rate of $\hat{f}_{\gamma_{j}}(\bx)$,
\begin{eqnarray*}
\sup | \hat{f}_{\gamma_{j}}(\bx) - f_{\gamma_{j}}(\bx) | = O_{p}(h^{r}+(\ln(n))^{1/2}/(nh^{p})^{1/2}),
\, j = 1 ,\ldots,s+1.
\end{eqnarray*}

\noindent
b). The asymptotic normality of $\hat{f}_{\gamma_{j}}(\bx)$,
\begin{eqnarray*}
(nh^{p})^{1/2} \left\{ \hat{f}_{\gamma_{j}}(\bx) - f_{\gamma_{j}}(\bx) -
\frac{1}{2} h^{2}\, C_{1} \sum^{p}_{l=1} f^{(2)}_{\gamma_{j},l}(\bx)  \right\}
\to N(0, C_{2} f_{\gamma_{j}}(\bx) )
\end{eqnarray*}
where
\begin{eqnarray*}
C_{1} = \int u^2 K (u) du,\, C_{2} = \left[\int K^{2}(u) du \right]^p, \,
f^{(2)}_{\gamma_{j},l}(\bx) = \frac{\partial^2 f_{\gamma_j} (\bx)}{\partial x_l^2}. \quad\square
\end{eqnarray*}
\end{thm}

\medskip
When the estimation is carried out at a single point $x$, 
we have the convergence rate 
$O_{p}(h^{r}+1/(nh^{p})^{1/2})$.
In empirical applications, 
multiple $\bx$ often appear 
and then the estimator has a slower uniform convergence rate
$O_{p}(h^{r}+(\ln(n))^{1/2}/(nh^{p})^{1/2})$.
Hence, from part $b)$, the kernel-smoothing density estimation is biased.
Given that a Gaussian product kernel is being used, 
we already know that $C_1 = 1$ and $C_{2} = 1/(2 \sqrt{\pi})^{p}$
according to A\"{\i}t-Sahalia  et al.~(2001).
Moreover, given that the number of thresholds $s$ and  corresponding threshold values
$\gamma_j, j=1, \ldots, s+1$ are known,
the consistency and asymptotic normality of $\hat{m}_{\gamma_{j}}(\bx)$ are provided as follows.

\begin{thm}\label{Est-3}
Suppose that the assumptions in Assumptions 1,  2 and  3-1 hold. The following results
are derived.

\noindent
a) The almost sure convergence rate of $\hat{m}_{\gamma_{j}}(\bx)$,
\begin{eqnarray*}
\sup | \hat{m}_{\gamma_{j}}(\bx) - m_{\gamma_{j}}(\bx) | = O_{p}(h^{r} + (\ln(n))^{1/2}/(nh^{p})^{1/2}),
\, j = 1 ,\ldots,s+1
\end{eqnarray*}

\noindent
b) The asymptotic normality of $\hat{m}_{\gamma_{j}}(\bx)$,
\begin{eqnarray*}
& &(nh^{p})^{1/2} \left[ \hat{m}_{\gamma_{j}}(\bx) - m_{\gamma_{j}}(\bx) - AB(\bx) \right]
\to N\left(0, C_{2} \frac{\sigma^{2}_{\gamma_{j}}(\bx)}{f_{\gamma_{j}}(\bx)} \right)
\end{eqnarray*}
where $AB(\bx)$ denotes the asymptotic bias,
\begin{eqnarray*}
AB(\bx) =  \frac{1}{2} h^{2} \,   C_{1}\,
  \sum_{l=1}^p \left[ m^{(2)}_{\gamma_{j},l}(\bx) f_{\gamma_{j}}(\bx) + 2 m^{(1)}_{\gamma_{j},l}(\bx)
 f^{(1)}_{\gamma_{j},l}(\bx) \right] /f_{\gamma_{j}}(\bx),
\end{eqnarray*}
$m^{(1)}_{\gamma_{j},l}(\bx) = \frac{\partial m_{\gamma_j} (\bx)}{\partial x_l}$ and $m^{(2)}_{\gamma_{j},l}(\bx) =
\frac{\partial^2 m_{\gamma_j} (\bx)}{\partial x_l^2}$ are the first- and second-order derivatives of the $j$th
regime's conditional mean with respect to the $l$th explanatory variable, respectively. \quad$\square$
\end{thm}

It is now clear that the sample estimator $\hat{m}_{\gamma_{j}}(\bx)$ is also asymptotically biased. However, this asymptotic bias could be reduced by using higher-order kernels.
Notice that the convergence rates and asymptotic results of $\hat{f}_{\gamma_{j}}(\bx)$ and
$\hat{m}_{\gamma_{j}}(\bx)$ are not affected by $s$, the number of thresholds.
In finite samples, the number of thresholds does affect the nonparametric estimation.
However, at the limit, the convergence rate does not depend on $s$.
Our results are therefore similar to those presented by Li and Racine (2007).

\subsection{Optimal Bandwidth Selector}

In nonparametric regressions,  bandwidth plays a crucial role in the estimation.
Different bandwidth selection rules have been suggested in the literature. Among the selectors, the
optimal bandwidth selector is the most comprehensively studied and is obtained by minimizing the mean integrated squared error (MISE).
That is, for a model with $s$ thresholds,
the corresponding MISE is defined as
\begin{eqnarray}
\mbox{MISE} (h) = \int \int \mbox{E}
\left[ \sum^{s+1}_{j=1} \left(\hat{m}_{\gamma_{j}}(\bx) - m_{\gamma_{j}}(\bx)\right)
I_{\gamma_{j}}(q) \right]^{2} w(\bx) d\bx dq \label{Opt}
\end{eqnarray}
and then the optimal bandwidth selector is obtained from
\begin{eqnarray*}
h_{opt} = \arg\min_h \mbox{MISE} (h).
\end{eqnarray*}
The weighting function $w(\bx)$ is an indicator function selecting a particular $x$-region of interest, and this depends generally on empirical studies.
Since the threshold variable $q$ does not affect the convergence rate of the proposed estimator,  we construct the  weighting function without including the threshold variable.
The convergence rate of $h_{opt}$ is derived and summarized in the following theorem.

\begin{thm}\label{optband}
Under Assumptions 1, 2, and 3, the convergence rate of the optimal bandwidth selector is $h_{opt} = O(n^{- \frac{1}{\delta}})$
in which $\delta = p + 2r$. \quad$\square$
\end{thm}

This result shows that the convergence rate of the
optimal bandwidth selector depends on 
the number of covariates $p$ 
and 
order of continuous differentiability of the kernel function,
but that the convergence rate is not affected by the number of thresholds.
In other words, 
the additional thresholds do not worsen the curse-of-dimensionality problem.

\section{Determining the Number of Thresholds}

The number of thresholds and  corresponding threshold values 
are typically unknown in practice.
In this section, 
we thus present a procedure for determining the unknown number of thresholds
and estimating the threshold values.
In linear regressions with thresholds, 
the number of thresholds is commonly determined by carrying out a sequential significance test (see Hansen, 1997). 
This sequential test is conducted by comparing the estimated
sum of the squared errors from a model with $s$ thresholds (under the null hypothesis) with that from a model
with $s + 1$ thresholds (under the alternative) sequentially.
The number of thresholds is determined as $s$ when the null of $s - 1$ thresholds versus the alternative of $s$
thresholds is rejected, whereas the null of $s$ thresholds versus the alternative of $s + 1$ thresholds is not rejected.
Similarly, we determine the number of thresholds in nonparametric regressions based on sequential tests in this study.
Instead of comparing the estimated error sum of squares from the linear regressions, 
however, we use the significance test suggested by
A\"{\i}t-Sahalia  et al.~(2001) for the nonparametric regressions as the basis in the sequential tests.
The test statistic for the null of $s+1$ thresholds to $s$ thresholds is constructed
and its asymptotic distribution is established as follows.

The test of A\"{\i}t-Sahalia et al.~(2001) is constructed
to test the significance of a subset of covariates in a nonparametric regression. 
The intuition behind the test is to check the difference between
the nonparametric regression estimates of unconstrained and constrained conditional means.
That is, the null of the significance test is written as
\begin{eqnarray}
H_{0} : Pr[m(\bW, \bV) - m(\bW)] = 1  \label{Goodness} 
\end{eqnarray}
where $\bW$ represents the $p$-dimensional explanatory variables,
$\bV$ is the $q$-dimensional explanatory variables under testing, $m(\bw, \bv)$ and $m(\bw)$ denote
the conditional means under  the alternative and null hypotheses, 
and $f(\bw, \bv)$ and $f(\bw)$
are the joint probability density functions of $(\bw, \bv)$ and $\bw$, respectively.

To test the null of $s$ thresholds versus the alternative of $s+1$, this test can be modified
by taking $\bW$ as the $p \times (s+1)$ independent variables in the regression with $s$ thresholds
and  $\bV$ as the extra $p$ independent variables in the regression with $s+1$ thresholds.
The significance of $\bV$ implies that the regression with $s+1$ thresholds must be considered.
However, the regression remains with $s$ thresholds if $\bV$ is not significant. The details
are discussed as follows. First, we construct the test for detecting whether 
an extra threshold (known at value, $\tau_j$) exists in the $j$th regime. Second,
since the threshold value $\tau_j$ is unknown in general, the test is extended to test
whether  an extra unknown threshold exists in the $j$th regime.

\subsection{Testing for the Existence of an Extra Threshold}

Given a regression with $s$ thresholds expressed as (\ref{Model}),
a new threshold $\tau_{j}$ is suspected to exist in the $j$th regime $[\gamma_{j-1},\gamma_{j})$.
Then, the conditional mean for the regime $[\gamma_{j-1},\gamma_{j})$ is split into two parts:
$m_{\gamma_{j-1},\tau_{j}}(\bX_i)I_{\gamma_{j-1},\tau_{j}}(Q_i)$ in the regime $[\gamma_{j-1},\tau_j)$
and $m_{\tau_{j},\gamma_{j}}(\bX_i)I_{\tau_{j},\gamma_{j}}(Q_i)$ in the regime $[\tau_j, \gamma_j)$,
 where
\begin{eqnarray*}
I_{\gamma_{j-1},\tau_{j}}(Q_{i})
 = \left\{\begin{array}{cl}
      1, & Q_{i} \in [ \gamma_{j-1} , \tau_{j} ) , \\
      0, & else,
      \end{array}\right. \, , \hspace{3mm}
I_{\tau_{j},\gamma_{j}}(Q_{i})
=  \left\{\begin{array}{cl}
     1, & Q_{i} \in [ \tau_{j},\gamma_{j} ) , \\
     0, & else,
     \end{array}\right.,
\end{eqnarray*}
and $m_{\gamma_{j-1},\tau_{j}}(\bx)$ is defined as
\begin{eqnarray*}
f_{\gamma_{j-1},\tau_{j}}(\bx,y) &=& \int I_{\gamma_{j-1},\tau_{j}}(q) f(\bx,y,q)\, dq \\
f_{\gamma_{j-1},\tau_{j}}(\bx) &=& \int I_{\gamma_{j-1},\tau_{j}}(q) f(\bx,q)\, dq \\
m_{\gamma_{j-1},\tau_{j}}(\bx) &=& \mbox{E}(Y_{i}| \bX_{i} = \bx , I_{\gamma_{j-1},\tau_{j}}(Q_{i}) = 1 ) \\
                &=& \int y \frac{f_{\gamma_{j-1},\tau_{j}}(y,\bx)}{f_{\gamma_{j-1,\tau_{j}}}(\bx)}\, d\bx
\end{eqnarray*}
and $m_{\tau_{j},\gamma_{j}}(\bx)$ is defined similarly to $m_{\gamma_{j-1},\tau_{j}}(\bx)$.

Denote $E(Y|\bX,Q; \gamma_{1}, \ldots, \gamma_{s})$ as the conditional mean with $s$ thresholds under the null
and $E(Y|\bX,Q; \gamma_{1}, \ldots, \gamma_{j-1}, \tau_{j}, \gamma_{j}, \ldots, \gamma_{s})$ as
the conditional mean function with $s+1$ thresholds under the alternative.
Then,
the null hypothesis for testing whether an extra threshold exists in the regime
$[\gamma_{j-1},\gamma_{j})$ can be written as
\begin{eqnarray*}
H_0: Pr [E (Y|\bX,Q; \gamma_{1}, \ldots, \gamma_{s})
    = E (Y|\bX,Q; \gamma_{1}, \ldots, \gamma_{j-1}, \tau_{j}, \gamma_{j}, \ldots, \gamma_{s}) ] = 1.
\end{eqnarray*}

The sample statistic analogous to the test $\Gamma(\tau_{j})$ 
in A\"{\i}t-Sahalia  et al.~(2001) is constructed as 
\begin{eqnarray}
\tilde{\Gamma}(\tau_{j})
&=& \frac{1}{n}\sum^{n}_{i=1} \left\{ \hat{m}_{\gamma_{j}}(\bX_{i})I_{\hat{\gamma}_{j}}(Q_{i}) -
     \hat{m}_{\gamma_{j-1},\tau_{j}}(\bX_{i})I_{\gamma_{j-1},\tau_{j}}(Q_{i}) \right. \nonumber \\
& & \left. \hspace{15mm} - \hat{m}_{\tau_{j},\gamma_{j}}(\bX_{i})
   I_{\tau_{j},\gamma_{j}}(Q_{i})   \right\}^{2} a(\bX_{i}), \label{gamma1}
\end{eqnarray}
where $\hat{m}_{\gamma_{j}}(\bx)$, $\hat{m}_{\gamma_{j-1},\tau_{j}}(\bx)$, and
$\hat{m}_{\tau_{j},\gamma_{j}}(\bx)$ are the sample estimates of $m_{\gamma_{j}}(\bx)$,
$m_{\gamma_{j-1},\tau_{j}}(\bx)$, and $m_{\tau_{j},\gamma_{j}}(\bx)$, respectively,
and $a(\bX_i)$ is a weighting function.
Specifically,
\begin{eqnarray*}
a(\bX) &=& \left\{ \begin{array}{ll}
                                   1 & \ \bX \in \bC , \ \ \mbox{where} \ \bC \in \mathcal{R}^{p} \\
                                   0 & \ \mbox{otherwise}.
                                   \end{array} \right.
\end{eqnarray*}
The choice of $\bC$ is application-dependent.
For example, 
in an empirical analysis of options prices,
 $a(\bX)$ can be set to exclude those in-the-money options with price biases.
Similarly, it can be set by using prior information to tackle boundary effects so that the density is bounded away from zero.
Since $\tilde{\Gamma}(\tau_{j})$ is the weighted sum of the squares of the
differences from $\hat{m}_{\gamma_{j}}(\bx)$ to $\hat{m}_{\gamma_{j-1},\tau_{j}}(\bx)$ and to
$\hat{m}_{\tau_{j},\gamma_{j}}(\bx)$,
the null hypothesis,  $\Gamma(\tau_{j})=0$, is not rejected when $\tilde{\Gamma}(\tau_{j})$ is insufficiently large and is rejected when
$\tilde{\Gamma}(\tau_{j})$ is sufficiently large. Therefore, this inference is a right-tailed test.
The asymptotic distribution of $\tilde{\Gamma}(\tau_{j})$ is constructed as follows.

\begin{thm}\label{gammatest}
Under the null hypothesis and according to Assumptions 1, 2, and 3,
 the asymptotic normality of the statistic $\tilde{\Gamma}(\tau_{j})$ is represented as
\begin{eqnarray}
\sigma^{-1}(\tau_{j}) \{ nh^{p/2}\, \tilde{\Gamma}(\tau_{j}) - h^{-p/2}\, \xi (\tau_{j})  \} 
\stackrel{d}{\longrightarrow}  N(0,1), \label{testsat}
\end{eqnarray}
where $\xi(\tau_{j})$ and $\sigma^{2} (\tau_{j})$ denote the bias and variance terms, respectively, and
where the bias term is
\begin{eqnarray*}
\xi(\tau_{j}) = C_{2} [\xi_{1}(\tau_{j}) +  \xi_{2}(\tau_{j})]
\end{eqnarray*}
with
\begin{eqnarray*}
\xi_{1}(\tau_{j}) &=& \int_{\bx} \sigma^{2}_{\gamma_{j}}(\bx)\, a(\bx)\, d\bx \\
\xi_{2}(\tau_{j}) &=& \int_{\bx} \left(1 - 2 \frac{f_{\gamma_{j-1}, \tau_{j}}(\bx) }{f_{\gamma_{j}}(\bx)} \right)
                       \sigma^{2}_{\gamma_{j-1}, \tau_{j}}(\bx)\, a(\bx)\, d\bx \\
             & & + \int_{\bx} \left(1 - 2 \frac{f_{\tau_{j}, \gamma_{j}}(\bx) }{f_{\gamma_{j}}(\bx)} \right)
              \sigma^{2}_{\tau_{j}, \gamma_{j}}(\bx)\, a(\bx)\, d\bx.
\end{eqnarray*}
$C_2$ was defined in Theorem 1, 
and the variance term is
\begin{eqnarray*}
\sigma^{2} (\tau_{j}) = 2C_{3} [ \sigma^{2}_{1}(\tau_{j}) + \sigma^{2}_{2}(\tau_{j}) ]
\end{eqnarray*}
with
\begin{eqnarray*}
\sigma^{2}_{1}(\tau_{j}) &=& \int_{\bx} \sigma^{4}_{\gamma_{j}}(\bx)\, a^{2}(\bx)\,  d\bx \\
\sigma^{2}_{2}(\tau_{j}) &=& \int_{\bx} \left(1 - 2 \frac{f_{\gamma_{j-1}, \tau_{j}}(\bx) }{f_{\gamma_{j}}(\bx)} \right)
                \sigma^{4}_{\gamma_{j-1}, \tau_{j}}(\bx)\, a^{2}(\bx)\, d\bx \\
           & &   + \int_{\bx} \left(1 - 2 \frac{f_{\tau_{j}, \gamma_{j}}(\bx) }{f_{\gamma_{j}}(\bx)}\right)
              \sigma^{4}_{\tau_{j}, \gamma_{j}}(\bx)\, a^{2}(\bx)\, d\bx
\end{eqnarray*}
where $\sigma^{2}_{\gamma_{j}}(\bx)$, $\sigma^{2}_{\gamma_{j-1},\tau_{j}}(\bx)$, and
$\sigma^{2}_{\tau_{j},\gamma_{j}}(\bx)$ are
\begin{eqnarray*}
\sigma^{2}_{\gamma_{j}}(\bx)
&=& \int \left[y-m_{\gamma_{j}}(\bx)\right]^{2} \frac{f_{\gamma_{j}}(y,\bx)}{f_{\gamma_{j}}(\bx)}\, dy
                            = \int \sigma^{2}(\bx,q)  I_{\gamma_{j}}(q) \frac{f(\bx,q)}{f_{\gamma_{j}}(\bx)}\, dq \\
\sigma^{2}_{\gamma_{j-1},\tau_{j}}(\bx)
&=& \int \left[y-m_{\gamma_{j-1},\tau_{j}}(\bx) \right]^{2}\,
     \frac{f_{\gamma_{j-1},\tau_{j}}(y,\bx)}{f_{\gamma_{j-1},\tau_{j}}(\bx)}\, dy \\
&=& \int \sigma^{2}(\bx,q)  I_{\gamma_{j-1},\tau_{j}}(q)\, \frac{f(\bx,q)}{f_{\gamma_{j-1},\tau_{j}}(\bx)}\, dq \\
\sigma^{2}_{\tau_{j},\gamma_{j}}(\bx)
&=& \int \left[y-m_{\tau_{j},\gamma_{j}}(\bx)\right]^{2}\,
    \frac{f_{\tau_{j},\gamma_{j}}(y,\bx)}{f_{\tau_{j},\gamma_{j}}(\bx)}\, dy \\
&=& \int \sigma^{2}(\bx,q)  I_{\tau_{j},\gamma_{j}}(q)\, \frac{f(\bx,q)}{f_{\tau_{j},\gamma_{j}}(\bx)}\, dq
\end{eqnarray*}
and
\begin{eqnarray*}
C_{3} = \int_w \left\{ \int_{u}  K(u) K(u +w)\, du \right\}^{2} dw. \quad\square
\end{eqnarray*}
\end{thm}

\bigskip
Note that A\"{\i}t-Sahalia  et al.~(2001) also show that $C_{3} = 1/(2 \sqrt{2 \pi})^{p}$ when
the Gaussian product kernel is used. Given the result in Theorem \ref{gammatest},
we denote
\begin{eqnarray*}
\delta (\tau_{j}) = \sigma^{-1}(\tau_{j}) \left[ nh^{p/2}\, \tilde{\Gamma}(\tau_{j}) -
h^{-p/2}\, \xi(\tau_{j})  \right]
\end{eqnarray*}
and then the test statistic for the null of having an extra threshold $\tau_j$ in
the $j$th regime can be considered to be
\begin{eqnarray*}
\hat\delta (\tau_{j}) = \hat\sigma^{-1}(\tau_{j}) \left[ nh^{p/2}\, \tilde{\Gamma}(\tau_{j}) -
h^{-p/2}\, \hat\xi(\tau_{j})  \right],
\end{eqnarray*}
where  $\hat{\sigma}^{2}$ and $\hat{\xi}$ are  the consistent estimators for
$\sigma^2$ and $\xi$, respectively. The limiting distribution of $\hat\delta (\tau_{j})$ is $N(0, 1)$.
The power property of $\hat\delta (\tau_{j})$ is investigated in Section 3.4;
Consequently it is a consistent test.
We describe the consistent estimation of 
$\sigma^2$ and $\xi$ in the following subsections.

\subsection{Testing for an Extra Unknown Threshold}

In practice,  $\tau_{j}$ is unknown a priori and 
there are, in principle, infinite many of $\tau_j$s 
in the regime $[\gamma_{j-1}, \gamma_{j})$. 
To make the test implementable, instead of infinite many of $\tau_j$s,
we only consider the
$m$ candidate threshold values within the regime
$[\gamma_{j-1},\gamma_{j})$, i.e., $\gamma_{j-1} < \tau_{j,1} < \tau_{j,2} < \ldots < \tau_{j,m} < \gamma_{j}$,
where $\tau_{j, 1} - \gamma_{j-1} = \tau_{j, 2} - \tau_{j, 1} = \cdots = \gamma_j - \tau_{j, m} = (\gamma_j - \gamma_{j-1})/m$.
Given the suspected $m$ pseudo thresholds, $\tau_{j,1}, \tau_{j,2}, \ldots,  \tau_{j,m}$,
the null of an extra unknown threshold can be written as
\begin{eqnarray}
H_{0}:
Pr\left(
\begin{array}{c}
\Gamma(\tau_{j,1}) = 0 \\
\Gamma(\tau_{j,2}) = 0\\
\vdots \\
\Gamma(\tau_{j,m}) = 0
\end{array} \right)
=1.  \label{Quention}
\end{eqnarray}

Given the sample counterpart $\tilde{\Gamma}(\tau_{j,i})$ of 
$\Gamma(\tau_{j,i}), i=1, \ldots, m,$
as defined in (\ref{gamma1}), 
the following theorem reports
the joint asymptotic distribution of the $m$ statistics.

\begin{thm}\label{D-1}
Given that the assumptions in Assumptions 1,  2, and  3 hold,
$\mbox{E}(e^{2}_{i}|\bX_{i}=\bx,Q_{i}=q) = \sigma^{2}(\bx,q)$, and under the null,
\begin{eqnarray*}
\left(
\begin{array}{c}
\delta^{*} (\tau_{j,1}) \\
\delta^{*} (\tau_{j,2}) \\
\vdots \\
\delta^{*} (\tau_{j,m})
\end{array} \right)
=
\Sigma^{-1/2} \left(
\begin{array}{c}
\delta (\tau_{j,1}) \\
\delta (\tau_{j,2}) \\
\vdots \\
\delta (\tau_{j,m})
\end{array} \right)
\stackrel{d}{\longrightarrow} N(0,I)
\end{eqnarray*}
where
\begin{eqnarray*}
\delta (\tau_{j, k}) = \sigma^{-1}(\tau_{j,k}) \left[ nh^{p/2}\, \tilde{\Gamma}(\tau_{j,k}) -
h^{-p/2}\, \xi(\tau_{j,k})  \right]
\end{eqnarray*}
and $\Sigma$ is the variance-covariance matrix of $\delta (\tau_{j,1}), \ldots, \delta (\tau_{j,m})$.
The $(l,k)$-element in the variance-covariance matrix $\Sigma$, assuming $\tau_{j,l} < \tau_{j,k}$, is
 \begin{eqnarray*}
 \lefteqn{\COV(\delta (\tau_{j,l}),\delta (\tau_{j,k}))} \\
 &=&[\sigma^{2}_{1}(\tau_{j,l}) + \sigma^{2}_{2}(\tau_{j,l})]^{-1/2}\,
   [\sigma^{2}_{1}(\tau_{j,k}) + \sigma^{2}_{2}(\tau_{j,k})]^{-1/2} \\
 & &   \times \varphi(\tau_{j,l},\tau_{j,k})
 \end{eqnarray*}
 where
 $\varphi(\tau_{j,l},\tau_{j,k})$ is defined in the Appendix because of its complex form. \quad$\square$
\end{thm}

Theorem \ref{D-1} is applicable to nonparametric regressions with
heteroskedastic errors whose variances depend on the values of $\bX_i$ and $Q_i$,
i.e, $\mbox{E}(e^{2}_{i}|\bX_{i}=\bx,Q_{i}=q) = \sigma^{2}(\bx,q)$.\footnote{ 
For the two restricted cases with
heteroskedastic errors whose variances
depend on the values of $\bX_i$ but not on those of $Q_i$, i.e., $\mbox{E}(e^{2}_{i}|\bX_{i}=\bx,Q_{i}=q)^{2}
= \sigma^{2}(\bx)$, when $\bX$ and $Q$ are either dependent or independent,
the joint asymptotic distribution of the $m$ statistics is also derived but not provided in this paper. The detailed results and proofs of the corresponding asymptotic distributions are available from the authors upon request.}
By replacing $\sigma^{2}$, $\xi$, and $\Sigma$ in Theorem \ref{D-1} with consistent estimates, namely $\hat{\sigma}^{2}$,
$\hat{\xi}$, and $\hat{\Sigma}$, respectively, we have
\begin{eqnarray}
\left(
\begin{array}{c}
\hat{\delta}^{*} (\tau_{j,1}) \\
\hat{\delta}^{*} (\tau_{j,2}) \\
\vdots \\
\hat{\delta}^{*} (\tau_{j,m})
\end{array} \right)
=
\hat{\Sigma}^{-1/2}\left(
\begin{array}{c}
\hat{\delta} (\tau_{j,1}) \\
\hat{\delta} (\tau_{j,2}) \\
\vdots \\
\hat{\delta} (\tau_{j,m})
\end{array} \right) \stackrel{d}{\longrightarrow} \bN (\bzero, \bI_m), \label{EstD-1}
\end{eqnarray}
where
\begin{eqnarray*}
\hat{\delta} (\tau_{j,k}) = \hat{\sigma}^{-1} (\tau_{j,k})\, \left[ n h^{p/2}\, \tilde{\Gamma}(\tau_{j}) - h^{-p/2}\,
\hat{\xi} (\tau_{j,k})\right].
\end{eqnarray*}

\subsection{Estimation of the Nuisance Parameters}

Given the asymptotic normality of the test statistic $\tilde{\Gamma}(\tau_{j})$,
the nuisance parameters must be estimated consistently. 
First, the parameter $\sigma^{2}_{\gamma_{j}}(\bx)$
can be estimated by  using the Nadaraya-Watson estimator as follows:
\begin{eqnarray}
\hat{\sigma}^{2}_{\gamma_{j}}(\bx)
&=& \frac{\sum^{n}_{i=1} \bK_{h}(\bX_{i} - \bx)I_{\gamma_{j}}(Q_{i})Y^{2}_{i}}
         {\sum^{n}_{i=1} \bK_{h}(\bX_{i} - \bx)I_{\gamma_{j}}(Q_{i})} - \hat{m}^{2}_{\gamma_{j}}(\bx) \label{hatsigma}
\end{eqnarray}
Thus, $\sigma^{2}$, $\xi$, and $\Sigma$ can be estimated as
\begin{eqnarray*}
\hat{\xi} (\tau_{j,k})
&=& C_{2} (\hat{\xi}_{1}(\tau_{j,k}) +  \hat{\xi}_{2}(\tau_{j,k})) \\
\hat{\xi}_{1}(\tau_{j,k})
&=&  \frac{1}{n} \sum^{n}_{i=1} \frac{\hat{\sigma}^{2}_{\gamma_{j}}(\bX_{i}) a(\bX_{i})}{\hat{f}(\bX_{i})}  \\
\hat{\xi}_{2}(\tau_{j,k})
&=& \frac{1}{n} \sum^{n}_{i=1} \left(1 - 2\, \frac{\hat{f}_{\gamma_{j-1}, \tau_{j,k}}(\bX_{i}) }
                                                  {\hat{f}_{\gamma_{j}}(\bX_{i})}\right)
     \frac{\hat{\sigma}^{2}_{\gamma_{j-1}, \tau_{j,k}}(\bX_{i})a(\bX_{i})}{\hat{f}(\bX_{i})}  \\
& &  + \frac{1}{n} \sum^{n}_{i=1}  \left(1 - 2\, \frac{\hat{f}_{\tau_{j,k}, \gamma_{j}}(\bX_{i}) }
                                                    {\hat{f}_{\gamma_{j}}(\bX_{i})} \right)
            \frac{\hat{\sigma}^{2}_{\tau_{j,k}, \gamma_{j}}(\bX_{i})a(\bX_{i})}{\hat{f}(\bX_{i})}
\end{eqnarray*}
and
\begin{eqnarray*}
\hat{\sigma}^{2} (\tau_{j,k})
&=& 2C_{3}( \hat{\sigma}^{2}_{1}(\tau_{j,k}) + \hat{\sigma}^{2}_{2}(\tau_{j,k}) ) \\
\hat{\sigma}^{2}_{1}(\tau_{j,k})
&=& \frac{1}{n} \sum^{n}_{i=1} \frac{\hat{\sigma}^{4}_{\gamma_{j}}(\bX_{i})a(\bX_{i})}{\hat{f}(\bX_{i})} \\
\hat{\sigma}^{2}_{2}(\tau_{j,k})
&=& \frac{1}{n}  \sum^{n}_{i=1} (1 - 2\, \frac{\hat{f}_{\gamma_{j-1}, \tau_{j,k}}(\bX_{i}) }{\hat{f}_{\gamma_{j}}(\bX_{i})})
    \frac{\hat{\sigma}^{4}_{\gamma_{j-1}, \tau_{j,k}}(\bX_{i}) a^{2}(\bX_{i})}{\hat{f}(\bX_{i})} \\
& & + \frac{1}{n}  \sum^{n}_{i=1} (1 - 2\, \frac{\hat{f}_{\tau_{j,k}, \gamma_{j}}(\bX_{i}) }{\hat{f}_{\gamma_{j}}(\bX_{i})})
              \frac{\hat{\sigma}^{4}_{\tau_{j,k}, \gamma_{j}}(\bX_{i}) a^{2}(\bX_{i})}{\hat{f}(\bX_{i})}.
\end{eqnarray*}
Further, the $(i,j)$th elements of $\Sigma$ can be estimated as
\begin{eqnarray*}
\widehat{\COV}(\delta (\tau_{j,l}),\delta (\tau_{j,k})) =&[\hat{\sigma}^{2}_{1}(\tau_{j,l}) + \hat{\sigma}^{2}_{2}(\tau_{j,l})]^{-1/2}\,
   [\hat{\sigma}^{2}_{1}(\tau_{j,k}) + \hat{\sigma}^{2}_{2}(\tau_{j,k})]^{-1/2} \\
 &   \times ( \hat{c}_{1} + \hat{c}_{2} + \hat{c}_{3} + \hat{c}_{4} + \hat{c}_{5} + \hat{c}_{6} + \hat{c}_{7} + \hat{c}_{8} + \hat{c}_{9}),
\end{eqnarray*}
where the terms $\hat{c}_{1}$ to $\hat{c}_{9}$ are

\begin{eqnarray*}
\hat{c}_{1}
&=& \frac{1}{n} \sum^{n}_{i=1}  \frac{\hat{\sigma}^{4}_{\gamma_{j}}(\bX_{i})}{\hat{f}(\bX_{i})} a^{2}(\bX_{i}) \\
\hat{c}_{2}
&=& -2 \left\{ \frac{1}{n} \sum^{n}_{i=1}  \frac{\hat{\sigma}^{2}_{\gamma_{j}}(\bX_{i})\hat{\sigma}^{2}_{\gamma_{j-1},\tau_{j,k}}(\bX_{i})}{\hat{f}(\bX_{i})}
    \frac{\hat{f}_{\gamma_{j-1},\tau_{j,k}}(\bX_{i})}{\hat{f}_{\gamma_{j}}(\bX_{i})} a^{2}(\bX_{i}) \right. \\
& & \left. + \frac{1}{n} \sum^{n}_{i=1} \frac{\hat{\sigma}^{2}_{\gamma_{j}}(\bX_{i}) \hat{\sigma}^{2}_{\tau_{j,k},\gamma_{j}}(\bX_{i})}{\hat{f}(\bX_{i})}
          \frac{\hat{f}_{\tau_{j,k},\gamma_{j}}(\bX_{i})}{\hat{f}_{\gamma_{j}}(\bX_{i})} a^{2}(\bX_{i}) \right\}  \
\end{eqnarray*}
\begin{eqnarray*}        
\hat{c}_{3}
&=& \frac{1}{n} \sum^{n}_{i=1} \frac{\hat{\sigma}^{4}_{\gamma_{j-1},\tau_{j,k}}(\bX_{i})}{\hat{f}(\bX_{i})} \frac{\hat{f}_{\gamma_{j-1},\tau_{j,k}}(\bX_{i})}{\hat{f}_{\gamma_{j}}(\bX_{i})} a^{2}(\bX_{i})
   + \frac{1}{n} \sum^{n}_{i=1} \frac{\hat{\sigma}^{4}_{\tau_{j,k},\gamma_{j}}(\bX_{i})}{\hat{f}(\bX_{i})} \frac{\hat{f}_{\tau_{j,k},\gamma_{j}}(\bX_{i})}{\hat{f}_{\gamma_{j}}(\bX_{i})} a^{2}(\bX_{i}) \\ 
\hat{c}_{4}
&=& -2 \left\{ \frac{1}{n} \sum^{n}_{i=1} \frac{\hat{\sigma}^{2}_{\gamma_{j}}(\bX_{i}) \hat{\sigma}^{2}_{\gamma_{j-1},\tau_{j,l}}}{\hat{f}(\bX_{i})}
                \frac{\hat{f}_{\gamma_{j-1},\tau_{j,l}}(\bX_{i})}{\hat{f}_{\gamma_{j}}(\bX_{i})} a^{2}(\bX_{i}) \right.  \\
& & \left.  + \frac{1}{n} \sum^{n}_{i=1} \frac{\hat{\sigma}^{2}_{\gamma_{j}}(\bX_{i}) \hat{\sigma}^{2}_{\tau_{j,l},\gamma_{j}}(\bX_{i})}{\hat{f}(\bX_{i})}
    \frac{\hat{f}_{\tau_{j,l},\gamma_{j}}(\bX_{i})}{\hat{f}_{\gamma_{j}}(\bX_{i})} a^{2}(\bX_{i})  \right\} 
\end{eqnarray*}
\begin{eqnarray*}   
\hat{c}_{5}
&=& 4  \left\{ \frac{1}{n} \sum^{n}_{i=1} \frac{\hat{\sigma}^{2}_{\gamma_{j}}(\bX_{i}) \hat{\sigma}^{2}_{\gamma_{j-1},\tau_{j,l}}(\bX_{i})}{\hat{f}(\bX_{i})}
                          \frac{\hat{f}_{\gamma_{j-1},\tau_{j,l}}(\bX_{i})}{\hat{f}(\bX_{i})} a^{2}(\bX_{i}) \right.   \\
      & &             +  \frac{1}{n} \sum^{n}_{i=1} \frac{\hat{\sigma}^{2}_{\gamma_{j}}(\bX_{i}) \hat{\sigma}^{2}_{\tau_{j,l},\tau_{j,k}}(\bX_{i})}{\hat{f}(\bX_{i})}
                          \frac{\hat{f}_{\tau_{j,l},\tau_{j,k}}(\bX_{i})}{\hat{f}_{\gamma_{j}}(\bX_{i})} a^{2}(\bX_{i})    \\
      & &    \left.         +  \frac{1}{n} \sum^{n}_{i=1} \frac{\hat{\sigma}^{2}_{\gamma_{j}}(\bX_{i}) \hat{\sigma}^{2}_{\tau_{j,k},\gamma_{j}}(\bX_{i})}{\hat{f}(\bX_{i})}
                          \frac{\hat{f}_{\tau_{j,k},\gamma_{j}}(\bX_{i})}{\hat{f}_{\gamma_{j}}(\bX_{i})} a^{2}(\bX_{i})    \right\} \\
\hat{c}_{6} &=& -2 \left\{ \frac{1}{n} \sum^{n}_{i=1} \frac{\hat{\sigma}^{2}_{\gamma_{j-1},\tau_{j,k}}(\bX_{i}) \hat{\sigma}^{2}_{\gamma_{j-1},\tau_{j,l}}(\bX_{i})}{\hat{f}(\bX_{i})}
                          \frac{\hat{f}_{\gamma_{j-1},\tau_{j,l}}(\bX_{i})}{\hat{f}(\bX_{i})} a^{2}(\bX_{i})  \right.  \\
      & &             +  \frac{1}{n} \sum^{n}_{i=1} \frac{\hat{\sigma}^{2}_{\gamma_{j-1},\tau_{j,k}}(\bX_{i}) \hat{\sigma}^{2}_{\tau_{j,l},\tau_{j,k}}(\bX_{i})}{\hat{f}(\bX_{i})}
                          \frac{\hat{f}_{\tau_{j,l},\tau_{j,k}}(\bX_{i})}{\hat{f}_{\gamma_{j}}(\bX_{i})} a^{2}(\bX_{i})    \\
      & &     \left.        +  \frac{1}{n} \sum^{n}_{i=1} \frac{\hat{\sigma}^{4}_{\tau_{j,k},\gamma_{j}}(\bX_{i})}{\hat{f}(\bX_{i})}
                          \frac{\hat{f}_{\tau_{j,k},\gamma_{j}}(\bX_{i})}{\hat{f}_{\gamma_{j}}(\bX_{i})} a^{2}(\bX_{i}) \right\}  
\end{eqnarray*}
\begin{eqnarray*}                                                                        
\hat{c}_{7} &=& \left\{ \frac{1}{n} \sum^{n}_{i=1} \frac{\hat{\sigma}^{4}_{\gamma_{j-1},\tau_{j,l}}(\bX_{i})}{\hat{f}(\bX_{i})} \frac{\hat{f}_{\gamma_{j-1},\tau_{j,l}}(\bX_{i})}{\hat{f}_{\gamma_{j}}(\bX_{i})} a^{2}(\bX_{i})
            + \frac{1}{n} \sum^{n}_{i=1} \frac{\hat{\sigma}^{4}_{\tau_{j,l},\gamma_{j}}(\bX_{i})}{\hat{f}(\bX_{i})} \frac{\hat{f}_{\tau_{j,l},\gamma_{j}}(\bX_{i})}{\hat{f}_{\gamma_{j}}(\bX_{i})} a^{2}(\bX_{i}) \right\} \\  
\hat{c}_{8} &=& -2  \left\{ \frac{1}{n} \sum^{n}_{i=1} \frac{\hat{\sigma}^{4}_{\gamma_{j-1},\tau_{j,l}}(\bX_{i})}{\hat{f}(\bX_{i})}
                          \frac{\hat{f}_{\gamma_{j-1},\tau_{j,l}}(\bX_{i})}{\hat{f}(\bX_{i})} a^{2}(\bX_{i})  \right.  \\
      & &              +  \frac{1}{n} \sum^{n}_{i=1} \frac{\hat{\sigma}^{2}_{\tau_{j,k},\gamma_{j}}(\bX_{i}) \hat{\sigma}^{2}_{\tau_{j,l},\tau_{j,k}}(\bX_{i})}{\hat{f}(\bX_{i})}
                          \frac{\hat{f}_{\tau_{j,l},\tau_{j,k}}(\bX_{i})}{\hat{f}_{\gamma_{j}}(\bX_{i})} a^{2}(\bX_{i})    \\
      & &     \left.        +  \frac{1}{n} \sum^{n}_{i=1} \frac{\hat{\sigma}^{2}_{\tau_{j,l},\gamma_{j}}(\bX_{i}) \hat{\sigma}^{2}_{\tau_{j,k},\gamma_{j}}(\bX_{i})}{\hat{f}(\bX_{i})}
                          \frac{\hat{f}_{\tau_{j,k},\gamma_{j}}(\bX_{i})}{\hat{f}_{\gamma_{j}}(\bX_{i})} a^{2}(\bX_{i})   \right\} 
\end{eqnarray*}
\begin{eqnarray*}                             
\hat{c}_{9} &=&    \frac{1}{n} \sum^{n}_{i=1} \frac{\hat{\sigma}^{4}_{\gamma_{j-1},\tau_{j,l}}(\bX_{i})}{\hat{f}(\bX_{i})}
                          \frac{\hat{f}_{\gamma_{j-1},\tau_{j,l}}(\bX_{i})}{\hat{f}_{\gamma_{j-1},\tau_{j,k}}(\bX_{i})} a^{2}(\bX_{i})   \\
      & &             +  \frac{1}{n} \sum^{n}_{i=1} \frac{\hat{\sigma}^{4}_{\tau_{j,l},\tau_{j,k}}(\bX_{i})}{\hat{f}(\bX_{i})}
                          \frac{\left\{\hat{f}_{\tau_{j,l},\tau_{j,k}}(\bX_{i})\right\}^{2}}{\hat{f}_{\gamma_{j-1},\tau_{j,k}}(\bX_{i}) \hat{f}_{\tau_{j,l},\gamma_{j}}(\bX_{i})} a^{2}(\bX_{i})   \\
      & &            +  \frac{1}{n} \sum^{n}_{i=1} \frac{\hat{\sigma}^{4}_{\tau_{j,k},\gamma_{j}}(\bX_{i})}{\hat{f}(\bX_{i})}
                          \frac{\hat{f}_{\tau_{j,k},\gamma_{j}}(\bX_{i})}{\hat{f}_{\tau_{j,l},\gamma_{j}}(\bX_{i})} a^{2}(\bX_{i}).
\end{eqnarray*}

\medskip
Given Lemma 6 , Theorems $\ref{Est-1}$ and $\ref{Est-3}$,  and Assumptions 1, 2 and 3,
we have the following results as in A\"{\i}t-Sahalia  et al.~(2001):
\begin{eqnarray*}
\hat{\xi}_{1}(\tau_{j,k}) - \xi_{1}(\tau_{j,k}) &=& o_{p}(h^{p/2}) \\
\hat{\xi}_{2}(\tau_{j,k}) - \xi_{2}(\tau_{j,k}) &=& o_{p}(h^{p/2}) 
\end{eqnarray*}
and
\begin{eqnarray*}
\hat{\sigma}^{2}_{1}(\tau_{j,k}) - \sigma^{2}_{1}(\tau_{j,k}) &=& o_{p}(1) \\
\hat{\sigma}^{2}_{2}(\tau_{j,k}) - \sigma^{2}_{2}(\tau_{j,k}) &=& o_{p}(1).
\end{eqnarray*}
That is, $\hat{\xi}_{1}(\tau_{j,k})$, $\hat{\xi}_{2}(\tau_{j,k})$, $\hat{\sigma}^{2}_{1}(\tau_{j,k})$, and
$\hat{\sigma}^{2}_{2}(\tau_{j,k})$ are the consistent estimators of $\xi_{1}(\tau_{j,k})$, $\xi_{2}(\tau_{j,k})$,
$\sigma^{2}_{1}(\tau_{j,k})$, and  $\sigma^{2}_{2}(\tau_{j,k})$, respectively.
For $C_{2}$ and $C_{3}$, A\"{\i}t-Sahalia  et al.~(2001) show that
\begin{eqnarray*}
C_{2} &=& 1/(2 \sqrt{\pi})^{p}, \\
C_{3} &=& 1/(2 \sqrt{2 \pi})^{p}.
\end{eqnarray*}

In light of the results in (\ref{EstD-1}), the following test statistics are suggested to test the null of no
extra unknown threshold existing in the regime $[\gamma_{j-1}, \gamma_{j})$:
\begin{eqnarray}
Z_{\gamma_{j}} = \frac{1}{\sqrt{m}} \sum^{m}_{i=1} \hat\delta^{*} (\tau_{j,i}).
\end{eqnarray}
Furthermore, 
we know that $\hat\delta^{*} (\tau_{j,i})$
converge to the standard
normal distribution. Therefore, the distribution in the limit of $Z_{\gamma_{j}}$ is also standard normally
distributed, i.e.,
\begin{eqnarray}
Z_{\gamma_{j}} \sim N(0,1).  \label{Normdistribution}
\end{eqnarray}

\subsection{Local Alternative Power}
In this subsection, 
we study the consistency of the test.
We then examine  its power, that is, the probability of rejecting a false hypothesis
against the sequences of alternatives that approach the null as $n\rightarrow\infty$.
Given an extra threshold existing in $[\gamma_{j-1},\gamma_{j})$ and being neglected,
\begin{eqnarray}
m_{\gamma_{j}}(\bx) I_{\gamma_{j}}(q) - m_{\gamma_{j-1},\tau_{j}}(\bx) I_{\gamma_{j-1},\tau_{j}}(q) -
m_{\tau_{j},\gamma_{j}}(\bx) I_{\tau_{j},\gamma_{j}}(q)
\neq 0 \label{Localalt}
\end{eqnarray}
for $q\in[\gamma_{j-1},\gamma_{j})$.
Suppose an extra threshold does exist in $[\gamma_{j-1},\gamma_{j})$ under the alternative and denote
the sequence of densities as $f^{[n]}_{\gamma_{j}}$, $f^{[n]}_{\gamma_{j-1},\tau_{j}}$
and $f^{[n]}_{\tau_{j},\gamma_{j}}$.  The superscript $[n]$ is specified to show that these densities are dependent
on $n$ since the value of the extra threshold is unknown.
The local alternatives can be specified as
\begin{eqnarray*}
H_{1n}&:& \sup[ m^{[n]}_{\gamma_{j}}(\bx)I_{\gamma_{j}}(q) - m^{[n]}_{\gamma_{j-1},\tau_{j}}(\bx) I_{\gamma_{j-1},\tau_{j}}(q)
       - m^{[n]}_{\tau_{j},\gamma_{j}}(\bx) I_{\tau_{j},\gamma_{j}}(q) \\
       & & \hspace{20mm}- \epsilon_{n} \lambda_{\tau^{*},\tau_{j}}(\bx,q) |: \bx,q \in S ]
      = o(\epsilon_{n})
\end{eqnarray*}
where
\begin{eqnarray*}
& & ||f^{[n]}_{\gamma_{j}} - f_{\gamma_{j}}||_{\infty} = o(n^{-1}h^{-p/2}) \\
& & ||f^{[n]}_{\gamma_{j-1},\tau_{j}} - f_{\gamma_{j-1},\tau_{j}}||_{\infty} = o(n^{-1}h^{-p/2}) \\
& & ||f^{[n]}_{\tau_{j},\gamma_{j}} - f_{\tau_{j},\gamma_{j}}||_{\infty} = o(n^{-1}h^{-p/2})
\end{eqnarray*}
and $\lambda_{\tau^{*},\tau_{j}}(\bx,q)$ satisfies
\begin{eqnarray*}
\int \lambda_{\tau^{*},\tau_{j}}(\bx,q) f(\bx,q) dq = 0
\end{eqnarray*}
and
\begin{eqnarray*}
\Lambda_{\tau^{*},\tau_{j}} \equiv \int \int \lambda^{2}_{\tau^{*},\tau_{j}}(\bx,q) f(\bx,q) d\bx dq < \infty
\end{eqnarray*}
It is clear that the alternative $H_{1n}$ converges to the null $H_{0}$ at speed $n^{-1/2} h^{-p/4}$
(i.e., $\epsilon_n = n^{-1/2} h^{-p/4}$).

\begin{thm}
Under Assumptions 1, 2, and 3, the asymptotic power of the test is
\begin{eqnarray*}
P(\hat{\delta} (\tau_{j}) \geq z_{\alpha}|H_{1n}) \to 1 - \Phi(z_{\alpha} -
\Lambda_{\tau^{*},\tau_{j}}/\sigma_{1}(\tau_{j})),
\end{eqnarray*}
where $\Phi(z_{\alpha}) = 1 - \alpha$ with $\Phi(\cdot)$, the CDF function of a standard normal random variable. \quad$\square$
\end{thm}

\subsection{Identifying the Number of Thresholds}

The test statistic, the average norm $Z_{\gamma_{j}}$,
is suggested to check whether an extra threshold exists in the regime
$[\gamma_{j-1}, \gamma_{j})$ given that the $s$ threshold values $\gamma_{1}, \ldots, \gamma_s$ are already known.
Logically, the test can be applied to check for an extra threshold existing in the regime
$[\gamma_{j-1}, \gamma_{j})$ for $j=1, \ldots, s$ jointly. This thus ends up being the test for
whether there is an extra threshold in a given $s$ threshold regression. 
Accordingly, we construct, in what follows, 
the test
for the null of $s$ thresholds against the alternative of $s+1$ thresholds.

Since the indicator functions are independent, i.e., $I_{\gamma_{i}}(Q_{i}) \times I_{\gamma_{j}}
(Q_{i}) = 0 , i \neq j$, the covariance of $\delta (\tau_{i})$ and $\delta (\tau_{j})$ for $i \neq j$ is zero.
That is
\begin{eqnarray*}
E[\delta (\tau_{i}) \delta (\tau_{j})] = 0, \, \mbox{for}\, i\neq j
\end{eqnarray*}
This fact implies that  $Z_{\gamma_{j}}$ and $Z_{\gamma_{l}}$($j \neq l$) are  asymptotically independent. 
The test statistic for the null $s$ thresholds against $s+1$ thresholds is constructed
as characterized in the following theorem.

\begin{thm}{\label{t-2}}
Under the same assumptions as for Theorem 5, the test statistic for the null $s$ thresholds against $s+1$ thresholds is constructed as
\begin{eqnarray*}
F_{n}(s+1|s) = \max_{1 \leq j \leq s+1} Z_{\gamma_{j}},
\end{eqnarray*}
with $\lim_{n \to \infty} P( F_{n} (s+1|s) \leq  x) = \Phi^{s+1}(x)$, where
$\Phi(x)$ is the CDF of a standard normal distribution
and $Z_{\gamma_j}$ is defined in  equation (12). \quad$\square$
\end{thm}

Table 1 presents he critical values of the test statistic $F_{n}(s+1|s)$ for $s+1 = 1, 2, 3, 4, 5$ at 1\%, 5\% and 10\%.

\begin{table}
\begin{center}
\caption{Critical values of $F_{n}(s+1|s)$}
\medskip
\begin{tabular}[b]{|l|ccc|} \hline
$s+1$ & 10\% & 5\% & 1\%  \\ \hline
1  & 1.281552 & 1.644854 & 2.326348  \\
2  &  1.632219 & 1.954508 & 2.574961 \\
3  & 1.818281 &  2.121201 & 2.711943 \\
4  & 1.943196  & 2.234002 & 2.805821 \\
5  & 2.036469 & 2.318679 & 2.876895 \\ \hline
\end{tabular}
\end{center}
\end{table}

Given the test statistic for testing $s$ thresholds against $s+1$ thresholds in Theorem
\ref{t-2}, the number of thresholds can be determined by conducting these tests sequentially
for $s = 0, 1, \ldots$ and so on. The number of thresholds is determined by sequential inferences until the not rejection result is obtained. 
In other words, the number of thresholds is
$s$ when the null of $s$ thresholds against
$s+1$ thresholds is not rejected. 
When the number of thresholds is determined, 
we estimate the corresponding threshold values
by using the methods discussed in the next section.

\section{Statistical Properties of the Threshold Estimators}

In the preceding discussions on testing an extra unknown threshold in a certain regime and testing the null of $s$ thresholds against $s+1$ thresholds, 
the threshold values under the null are assumed to be known already.
In applied research, 
the threshold values are unknown and need to be estimated by using a valid procedure.
In the framework of linear regressions, 
Bai~(1997) and Bai and Perron~(1998) determine the number of structural changes
by using a sequential test and estimate the breakpoints by looking up the sums of the squared errors at which the minimization is obtained. Hansen~(1999) discusses the determination of the number of thresholds and  estimation of threshold values in linear regressions by using similar procedures. 
We thus extend these procedures to the framework of nonparametric regressions.

\subsection{Added Assumptions}
To derive the statistical properties of the threshold value estimators, 
we need the following assumptions.

\medskip
\noindent
 {\bf Assumption 4}. 
 \begin{enumerate}
 \item [4-1.] $f_{q}(q)$, $\mbox{E}(c^{2}_{l.k}(\bX)|q)$, and
 $\mbox{E}(c^{2}_{l.k}(\bX)e^{2}|q)$ exist
              and are continuous at $q = \gamma_{1}, \ldots, \gamma_{s}$,
              where $c_{l.k}(\bX_{i}) := m_{\gamma_{l}}(\bX_i) - m_{\gamma_{k}}(\bX_i)$.
 \item [4-2.] $\max_{l,k \in [1, \ldots , s]], l \neq k} \mbox{E}|c_{l.k}(\bX_{i})|^{4} < \infty$ ,
              $\mbox{E}|c_{l.k}(\bX_{i}) e_{i}|^{4} < \infty$.
 \item [4-3.] $\forall \gamma \in R$ , $\mbox{E}(| c^{4}_{l.k}(\bX_{i}) e^{4}_{i} | | Q_{i} = \gamma) < D  $ ,
             $ \mbox{E}(|c^{4}_{l.k}(\bX_{i}) | |Q_{i} = \gamma) < D  $ for some $D \leq \infty$ ,
             and $f_{q}(\gamma) \leq \bar{f} \leq \infty$.
  \item [4-4.] $\delta_{n,l,k}(\bX_{i}) = n^{- \alpha} c_{l,k}^*(\bX_{i})$ , $\int |c(\bx_{i})| d\bx_{i} \neq 0$
         and $0< \alpha < 1/2$.
 \item [4-5.] $n h^{2/p + 2r} \to 0$ and $ [(\ln(n))^{1/2}n^{\alpha}]/[n^{1/2}h^{p/2}] \to 0$, where $0<\alpha<1/2$.
 \end{enumerate}

\smallskip
Assumptions 4-1, 4-2, and 4-3 are standard in proving the consistency of the threshold estimators. Assumptions 4-4 and 4-5 relate to a condition called the small effect, $\delta_{n, l, k}(\cdot)$, which is needed when we derive the asymptotic property of the threshold value estimator; see the proofs of Lemma 7 and Theorem 9. 
The small effect can approach zero when the sample size is sufficiently large;
therefore, it depends on $n$.
$c_{l,k}^*(\bX_{i})$ is the remainder of the difference
between $m_{\gamma_{l}}(\bX_i)$ and  $m_{\gamma_{k}}(\bX_i)$
when we extract the effect of the sample size, $n^{-\alpha}$,
from $c_{l,k}(\bX_{i})$.

\subsection{Asymptotic Properties of the Threshold Value Estimators}
Given that the number of thresholds $s$ is known, 
the estimator of the threshold values can be defined in a manner
similar to that in Proposition 5 of Bai and Perron~(1998):
\begin{eqnarray*}
[\hat{\gamma}_{1}, \ldots \hat{\gamma}_{s}]
= \arg \min  \sum^{n}_{i=1} \left[Y_{i} - \sum^{s+1}_{j=1} \hat{m}_{\gamma_{j}}(\bX_{i})
              I_{\gamma_{j}}(Q_{j}) \right]^{2}.
\end{eqnarray*}
Clearly, $\hat{\gamma}_{1}, \ldots \hat{\gamma}_{s}$ are determined simultaneously 
by global minimization. In practice, the estimation is implemented by an algorithm based on the principle of dynamic programming.
Under Assumptions 1, 2, 3, and 4, 
the following theorem establishes
the consistency of $\hat{\gamma}_{j}, j=1, \ldots, s$.

\noindent
\begin{thm} \label{thm:9}
For $j = 1, \ldots, s$,
\begin{enumerate}
\item[a)]
\begin{eqnarray*}
\hat{\gamma}_{j} \stackrel{p}{\rightarrow} \gamma_{j}
\end{eqnarray*}
\item[b)]
\begin{eqnarray*}
 n (\hat{\gamma}_{j} - \gamma_{j} ) = O_{p}(1). \quad\square
\end{eqnarray*} 
\end{enumerate}
\end{thm}

The convergency rate of $\hat{\gamma}_{j}$ is $n$,
which is a common result in the literature on structural changes and threshold models within the framework of linear regressions and linear quantile regressions
(cf.\ Chen, 2008). The limiting distribution of the threshold value estimator is provided by Chan~(1998) for linear models. On the contrary, Hansen~(2000) and Bai and Perron~(2003) introduce the existence of the small effect to obtain the limiting distribution without the nuisance parameters of the threshold value estimation. That is,
denote
\begin{eqnarray*}
\delta_{n,l,k}(\bX_{i}) = m_{\gamma_{l}}(\bX_{i}) - m_{\gamma_{k}}(\bX_{i}) = n^{- \alpha}c^*_{l,k}(\bX_{i}).
\end{eqnarray*}
Under the assumption of $\delta_{n,l,k}(\bX_{i}) \to 0$, which is called the small effect,
we then obtain the asymptotic property of $\hat{\gamma}_{j}$:
\begin{thm}
\begin{eqnarray*}
n^{1 - 2\alpha} (\hat{\gamma}_{j} - \gamma_{j} ) \stackrel{d}{\longrightarrow} Q_{j}, j=1, \ldots, s,
\end{eqnarray*}
where
\begin{eqnarray*}
Q_{j} &=& \arg\max_{ - \infty < v < \infty } \omega_{j} P_{j}(v) \\
P_{j}(v) &=& \left\{ \begin{array}{ll}
                B_{2,j}(-v), & v < 0 \\
                0,       & v = 0 \\
                B_{1,j}(v), & v > 0,
                \end{array} \right.
\end{eqnarray*}
where
\begin{eqnarray*}
\omega_{j} = \frac{\mbox{E}(c^{*2}_{j,j+1}(\bX_{i}) e^{2}_{i} | q_{i}
           = \gamma_{j} )}{[\mbox{E}(c^{*2}_{j,j+1}(\bX_{i}) | q_{i} = \gamma_{j} )]^{2} f(\gamma_{j})} 
\end{eqnarray*}
and $B_{1,j}(\cdot)$ and $B_{2,j}(\cdot)$ are two independent Brownian motions. \quad$\square$
\end{thm}
Note that the convergence rate of $\hat{\gamma}_{j}$ under the existence of the small effect is slower
than the rate in the case in which no small effect  is assumed.
The CDF of $Q_j$ can be obtained from Bhattacharya and Brockwell (1976), i.e.,
for $a \geq 0$,
\begin{eqnarray*}
P(Q_j \leq a) = 1 + \sqrt{\frac{a}{2 \pi}} e^{-\frac{a}{8}}
 + \frac{3}{2} e^{a} \Phi \left(-\frac{3 \sqrt{a}}{2}\right) - \left(\frac{x+5}{2} \right) \Phi \left(- \frac{\sqrt{x}}{2} \right)
\end{eqnarray*}
and for $a \leq 0$, $P(Q_j \leq x) = 1 - P(Q_j \leq -x)$, where $\Phi(x)$ is the CDF of a standard normal random variable.

\subsection{Sequential Method}
Instead of using a global minimization algorithm in the threshold value estimations,
the sequential method can  be adopted. Bai~(1997) proposes the sequential method for estimating the change points in a linear regression with multiple structural changes and provides the proof of the consistency of his estimator without knowing the number of breaks. Bai and Perron~(1998) also suggest using the sequential method to estimate the change points in linear regressions,
while Hansen~(1998) applies the sequential method to estimate the threshold values
for nondynamic panel threshold models. 
Following the literature, we thus use the sequential method to estimate the threshold values in the nonparametric regressions. Without loss of generality, a nonparametric regression with three thresholds is considered.
The model under consideration is, for $s = 3$,
\begin{eqnarray*}
Y_{i} = \sum^{4}_{j=1} m_{\gamma_{j}}(\bX_{i}) I_{\gamma_{j}}(Q_{i}) + e_{i}.
\end{eqnarray*}
The true threshold values implied by this model are $\gamma_1, \gamma_2$, and $\gamma_3$, while $\gamma_0$ and $\gamma_4$
are the lower and upper bounds of the threshold values. 
However, a nonparametric regression is mis-specified
when a model  with one threshold  is estimated as
\begin{eqnarray*}
\hat{Y}_{i} = \hat{m}_{\gamma}(\bX_{i}) I_{\gamma}(Q_{i}) + \hat{m}_{\gamma}^* (\bX_{i}) [1-I_{\gamma}(Q_{i})],
\label{1threshold}
\end{eqnarray*}
where $\hat{m}_{\gamma}(\bX_{i})$ and $\hat{m}_{\gamma}^* (\bX_{i})$ denote the kernel estimations from the sample observations 
$Q_i \in (-\infty, \gamma]$ and $Q_i \in[\gamma, \infty)$, respectively. 
The indicator function
$I_{\gamma}(Q_{i}) = 1$ for $Q_i \in (-\infty, \gamma]$ and 0 otherwise.

Denote $SSR(\gamma)$ as the sum of the squared residuals from the nonparametric
regression with the threshold value $\gamma$. That is,
\begin{eqnarray*}
SSR(\gamma) = \frac{1}{n} \sum^{n}_{i=1} \{ Y_{i} - \hat{m}_{\gamma}(\bX_{i}) I_{\gamma}(Q_{i})
                  - \hat{m}_{\gamma}^* (\bX_{i}) [1 - I_{\gamma}(Q_{i})] \}^{2}.
\end{eqnarray*}

\begin{thm}\label{Sequential-1}
Given a threshold value specified at $\gamma$ in a mis-specified nonparametric regression with one threshold,
the model mis-specification error is
\begin{eqnarray*}
SSR(\gamma) \stackrel{p}{\rightarrow}  S(\gamma) = \sum^{4}_{j=1} b_{j}(\gamma) I_{\gamma_{j}}(\gamma),
\end{eqnarray*}
where $b_{j}(\gamma)$  and $I_{\gamma_{j}}(\gamma)$ for $j=1, \ldots, 4$ are defined in the Appendix.
\quad$\square$
\end{thm}

Given the three true threshold values $\gamma_1$, $\gamma_2$, and $\gamma_3$, the threshold value $\gamma$ of a
mis-specified nonparametric regression with one threshold may be 
 in $[\gamma_0, \gamma_1)$, 
in $(\gamma_1, \gamma_2)$, 
in $(\gamma_2, \gamma_3)$, 
or in $(\gamma_3, \gamma_4]$. 
The model mis-specification error of the whole sample is 
$b_1(\gamma)$,  $b_2(\gamma)$,  $b_3(\gamma)$, or $b_4(\gamma)$ 
if the threshold value is mis-specified at the regime 
$[\gamma_0, \gamma_1)$, $(\gamma_1, \gamma_2)$,  $(\gamma_2, \gamma_3)$, 
or $(\gamma_3, \gamma_4]$, respectively.
In the Appendix, we describe the foregoing results in detail.

\begin{thm}
Let $S(\gamma_1) = \min(S(\gamma_{1}),S(\gamma_{2}),S(\gamma_{3}))$. $S(\gamma_1)$ is the smallest model
mis-specification error among  all $\gamma \in [\gamma_0, \gamma_4]$.
The exact expression of $S(\cdot)$ can be found in the Appendix. \quad$\square$
\end{thm}

$S(\gamma_{1}), S(\gamma_{2})$, and $S(\gamma_{3})$ are three smallest model mis-specification errors among all $\gamma \in [\gamma_0, \gamma_4]$. Moreover,
since $S(\gamma)$ is the limit of $SSR(\gamma)$ in probability and, without loss of generality, $\min(S(\gamma_{1}),S(\gamma_{2}),S(\gamma_{3})) = S(\gamma_{1})$ is assumed, we have the following theorem to prove $S(\gamma_{1})$ is global minimization. That is, Theorem 12 is sufficient to justify the sequential procedures discussed.

\begin{thm}\label{Mult}
Assume that the true model is a nonparametric regression with  three threshold values, namely $\gamma_1$, $\gamma_2$, and $\gamma_3$, and that
a nonparametric regression with one threshold is mis-specified and estimated 
via 
\begin{eqnarray*}
\hat{\gamma} = \arg \min  \frac{1}{n} \sum^{n}_{i=1} \{ Y_{i} - \hat{m}_{\gamma}(\bX_{i}) I_{\gamma}(Q_{i})
                  - \hat{m}_{\gamma}^* (\bX_{i}) [1 - I_{\gamma}(Q_{i})] \}^{2}.
\end{eqnarray*}
We then have
\begin{enumerate}
\item[a).] If $S(\gamma_1) = \min(S(\gamma_{1}),S(\gamma_{2}),S(\gamma_{3}))$,  $S(\gamma_1)$ is the smallest model
      mis-specification error among  all $\gamma \in [\gamma_0, \gamma_4]$
\item[b).] $SSR(\hat{\gamma}) \to S(\gamma_{1})$.
\item[c).] $\hat{\gamma}$ will, with probability one, converge to $\gamma_{1}$. \quad$\square$
\end{enumerate}
\end{thm}

According to Theorem \ref{Mult}, even if the nonparametric regression is mis-specified
and a threshold value is mis-estimated at which the sum of the squared errors is smallest, the mis-estimated threshold value converges to the true threshold value at which the model mis-specification error is the smallest. The result of Theorem  \ref{Mult} is thus similar to those in the study by Bai and Perron~(1998) for the estimation of the change points in a linear regression with multiple structural changes. 
To the best of our knowledge, this is the first theorem that ensures the consistency of the estimators obtained from using a sequential method in nonparametric regressions.

Note that the assumption $\min(S(\gamma_{1}),S(\gamma_{2}),S(\gamma_{3})) = S(\gamma_{1})$ indicates that the threshold value $\gamma_1$ has the largest influence on the regression.Theorem \ref{Mult} can be extended to a mis-specified regression model with two threshold values, and then the two estimated threshold values will be consistent with the two true threshold values that have a larger impact on the regression. Based on Theorem $\ref{Mult}$, the determination of the number of thresholds and estimation of the threshold values can be obtained by using the following sequential procedure.
\begin{enumerate}
\item Implement the test for the null of $s=0$ against $s=1$. That is, run the test to check whether an extra threshold exists in $(\gamma_{\min}, \gamma_{\max})$. If the null is not rejected, it is inferred that the regression has no threshold. If the null is rejected, move onto the next step.
\item Specify $s=1$ and estimate the threshold value as $\hat\gamma_1$. 
		Given $\hat\gamma_1$, carry out the test
      for the null of $s=1$ against $s=2$. That is, run the test to check whether an extra threshold exists in regimes $(\gamma_{\min}, \hat\gamma_1]$ and $(\hat\gamma_1, \gamma_{\max})$.
      If the null is not rejected, it is inferred that the regression has one threshold.
      If the null is rejected, move onto the next step.
\item Specify $s=2$ and estimate the extra threshold value from regimes $(\gamma_{\min}, \hat\gamma_1]$
      and $(\hat\gamma_1, \gamma_{\max})$as $\hat\gamma_2$. Pick up the estimation of the threshold values,
      $\hat\gamma_2$, which has a smaller sum of squared errors. Given $\hat\gamma_1$ and $\hat\gamma_2$, carry out the test for the null of $s=2$ against $s=3$. That is, run the test to check whether an extra threshold exists in regimes $(\gamma_{\min}, \hat\gamma_1]$, $(\hat\gamma_1, \hat\gamma_2]$, and $(\hat\gamma_2, \gamma_{\max})$ if $\hat\gamma_2 > \hat\gamma_1$.
      If the null is not rejected, it is inferred that the regression has two thresholds.
      If the null is rejected, repeat the above test until  the null of $s$ against $s+1$ thresholds is not rejected.
\end{enumerate}
When the procedure is conducted to the end such that the null of $s$ thresholds against $s+1$ thresholds is not rejected, we then pin down a nonparametric regression with $s$ thresholds. Along with this procedure, the estimates of the $s$ threshold values, $\hat\gamma_1, \hat\gamma_2, \ldots, \hat\gamma_s$, are obtained as a byproduct. Following Theorem \ref{Mult}, the consistency of $\hat\gamma_1, \hat\gamma_2, \ldots, \hat\gamma_s$, is obtained consequently.

As mentioned in Proposition 8 of Bai and Perron~(1998), 
the drawback of the previously described sequential method is that the determined number of thresholds is larger than the true number of thresholds with a nonzero probability value. Therefore, Bai and Perron~(1998) recommend applying the sequential method with a certain Type I error that converges to zero at a slower rate with the sample size. By doing so, the determined number of thresholds converges to the true number of thresholds.

\section{Monte Carlo Studies}

In this section, Monte Carlo studies are conducted to evaluate the performance
of the proposed test statistic,  $F_{n}(s+1|s)$.
We also conduct simulations to assess the finite sample performance of the sequential method for estimating the threshold values.

\subsection{Empirical Performance of the Test Statistic}

Monte Carlo simulations are designed to evaluate the empirical size and power
performances of the tests to identify the number of thresholds. 
Our experimental design is mainly based on the data-generating process (DGP)
considered in A\"{\i}t-Sahalia  et al.~(2001).
We consider the null of no threshold against the alternative with one threshold.
The DGP under the null is specified  as
\begin{eqnarray*}
Y_{i} &=& e^{-0.25X_{i}} + \sqrt{ e^{-0.2\, (X_{i}+Q_{i})^2}} \cdot\epsilon_i \\
X_{i} & \stackrel{i.i.d.}{\sim} &  \sqrt{0.2}\, Q_{i} + \sqrt{0.8} u_{i} \\
Q_{i} &\stackrel{i.i.d.}{\sim} &  N(0,1), \, 
u_{i} \stackrel{i.i.d.}{\sim} N(0,1), \, \epsilon_{i} \stackrel{i.i.d.}{\sim} N(0,1).
\end{eqnarray*}
In this DGP, 
the random variable $X$ is dependent on the threshold variable $Q$ 
and the heteroskedasticity of the regression depends on $X$ and $Q$. 
By using a univariate normal kernel function,
we compute the bandwidth as $h = c \cdot \sigma \cdot n^{-1/\delta} =  n^{-1/\delta}$,
where $\delta = 4.25$ (cf.\ A\"{\i}t-Sahalia  et al., 2001, p.383),
$c=1$, and $\sigma$ is set to  one in our simulation.
We  also conduct robustness checks on the bandwidth selection. Since $s = 0$ under the null, the critical values of the test statistic $F_{n}(s+1|s)$ in Theorem $\ref{t-2}$ are 1.282,  1.645, and 2.326 for Type I errors at 1\%, 5\%, and 10\%, respectively.

We conduct simulations with sample sizes of 500, 1000 and 2000. Throughout our simulations, the numbers of replications and partitions $m$ are set to be 1000 and 7, respectively. Table 2 presents the empirical sizes of $F_{n}(s+1|s)$ at 1\%, 5\%, and 10\%, showing that the proposed test performs well with decent empirical sizes.

\begin{table}
\begin{center}
\caption{Empirical Sizes of  $F_{n}(s+1|s): \ h= c\cdot \sigma \cdot n^{-1/4.25}$}
\medskip
\begin{tabular}{lccc}
 \hline
$F_{n}(s+1|s)$       &   & \hspace{-0.01cm}$c = 1$  \\     \hline
$n$      &  500      & 1000   & 2000     \\   \hline
1\%    &    0.021     & 0.017  & 0.011   \\
5\%     &    0.045    & 0.051  &  0.051  \\
10\%     &   0.076    & 0.084  &  0.086  \\ \hline\hline
\end{tabular}\par
\medskip
\parbox{7cm}{\scriptsize \hspace{0.36cm} Note: Heterogeneity depends on $X$ and $Q$.}
\end{center}
\end{table}

Table 3 shows the corresponding Monte Carlo results with the robustness checks on the choice of bandwidth. The proposed test copes well with decent sizes across the distinct bandwidth values. 

\begin{table}
\begin{center}
\caption{Empirical Sizes of  $F_{n}(s+1|s): \ h= c\cdot \sigma \cdot n^{-1/4.25}$}
\medskip
\begin{tabular}[b]{l  c || c || c} \hline
$F_{n}(s+1|s)$      &  $c = 1.24$         &  $c = 1.30$       &  $c = 1.37$ \\ \hline
$n$              & 2000                   & 2000         & 2000 \\ \hline
1\%             & 0.018                & 0.011         & 0.018\\
5\%            &  0.050                  &  0.044      &  0.056\\
10\%           &  0.087               &  0.090        &  0.086\\ \hline \hline
\end{tabular}
\parbox{7cm}{\scriptsize \hspace{-0.36cm}Note: Heterogeneity depends on $X$ and $Q$.}
\end{center}
\end{table}

\subsection{Finite-sample Performance of the Sequential Method}

To assess the accuracy of the sequential method for estimating the threshold values, 
we consider the following DGP in the Monte Carlo studies,
which are similar to those in A\"{\i}t-Sahalia  et al.~(2001, p.383):
\begin{eqnarray*}
Y_{i} &=& e^{-0.25*X_{t}} I_{\gamma_{1}}(Q_{i}) + ( 1 + e^{-0.5\,X_{i}} ) I_{\gamma_{2}}(Q_{i})
         + ( 2 + e^{-0.1\, X_{i}} ) I_{\gamma_{3}}(Q_{i}) \\
      & & \quad + ( 0.5 + e^{-0.8\, X_{i}} ) I_{\gamma_{4}}(Q_{i}) + \sqrt{0.5625\, e^{-X^{2}_{i}}} 
      \cdot \epsilon_i,  \\
X_{i} &\stackrel{i.i.d.}{\sim} & N(0,1), \\
Q_{i} &\stackrel{i.i.d.}{\sim} &  N(0,1), \epsilon \stackrel{i.i.d.}{\sim} N(0,1). \\
 &  & \hspace{-1.75cm}\mbox{Thresholds}: \ \gamma_{1} = -0.7 , \ \ \gamma_{2}  =  0.15 , \ \ \gamma_{3} = 0.5.
\end{eqnarray*}

Let $\hat{\gamma}_{1,i}$ denote
the threshold value estimate in the first-round identification from
the $i$th replication of the DGP.
Then, the mean, standard error, and MSE (mean square error) 
from all the $nr$ replications are computed by
\begin{eqnarray*}
\bar{\hat{\gamma}}_{1} &=& \frac{1}{nr} \sum^{nr}_{i=1} \hat{\gamma}_{1,i} \\
se(\hat{\gamma}_{1}) &=& \left[\frac{1}{nr-1} \sum^{nr}_{i=1}
  (\hat{\gamma}_{1,i} - \bar{\hat{\gamma}}_{1} )^{2} \right]^{1/2} \\
MSE(\hat{\gamma}_{1}) &=& (\bar{\hat{\gamma}}_{1} - \gamma_{1})^{2} + [sd(\hat{\gamma}_{1}) ]^{2}.
\end{eqnarray*}

\begin{table}
\begin{center}
\caption{Performance of the Threshold Estimations}
{\footnotesize
\begin{tabular}{|l|ccc|ccc|ccc|} \hline
    &\multicolumn{3}{c|}{$\hat{\gamma}_{3}$}&\multicolumn{3}{c|}{$\hat{\gamma}_{2}$} &\multicolumn{3}{c|}{$\hat{\gamma}_{1}$} \\ \cline{2-10}
$n$ & $\bar{\hat\gamma}_{3}$ & $se(\hat\gamma_{3})$ & MSE($\hat\gamma_{3}$)  & $\bar{\hat\gamma}_{2}$ & $se(\hat\gamma_{2})$ & MSE($\hat{\gamma}_{2}$) & $\bar{\hat{\gamma}}_{1}$ & $se(\hat{\gamma}_{1})$ & MSE($\hat{\gamma}_{1}$) \\ \hline
500 & 0.4227  & 0.2542 & 0.0705 & 0.1775 & 0.0960 & 0.0100 & -0.6523 & 0.2407 & 0.0602 \\
1000& 0.4867  & 0.1245 & 0.0160 & 0.1529  & 0.0320 & 0.0010 & -0.6894 & 0.1198 & 0.0140 \\
3000& 0.5025 & 0.0079  & 6.9 $\times 10^{-6}$     & 0.1498     & 0.0077     & 5.9$\times 10^{-5}$ & -0.7029 & 0.0040 & 2.4$\times 10^{-5}$ \\ \hline
\end{tabular}
}
\end{center}
\end{table}

Given $n=500, 1000, 3000$ and 1000 replications, Table 4 shows the Monte Carlo results. We can draw the following conclusions from the simulation results. The standard error and MSE of the estimated threshold values decrease as the sample size increases. The sequential method consistently estimates the unknown threshold values. In particular, the mean and standard error of the first estimated threshold values are 0.5029813 and 0.0107737, respectively. The mean value is close to $\gamma_3 = 0.5$.  For the second estimated threshold values, the mean is 0.152506, which is close to $\gamma_2 = 0.15$. The mean of the third estimated threshold values is -0.6966892,  which is close to $\gamma_1 = -0.7$. 
These simulated results indicate the accuracy of the sequential method for estimating the threshold values.  Given the good performance of the simulations,
and based on Theorems 8 and 9,
the threshold value estimators are super-consistent, as we see in Hansen (2000).

\section{An Empirical Application: the 401(K) Retirement Savings Plan with Income Thresholds}

Examining the effects of 401(k) plans on savings 
is an issue of  long-standing empirical interest
(see Chernozhukov and Hansen (2004) and the references cited therein).
Intuitively, because different income groups face distinct resource constraints, income thresholds should play an important role in the analysis of individual savings for retirement.
Chernozhukov and Hansen (2013) study the effect of 401(k) eligibility on  total wealth by using high-dimensional methods that allow for flexible functional forms.
By using a sample of 9915, they generate 10,763 technical variables through a spline basis and polynomial basis and then select a few important variables out of the technical variables by using a LASSO-based double selection procedure.
The selected few important variables include $\max (0, \ income - 0.33)$,
where the $income$ variable is normalized on the $[0, 1]$ interval.
Their result suggests that the income threshold exists in the 401(k) study.
In the literature, however, no test procedures have thus far been implemented to investigate the relevant income threshold values in 401(k) applications. 
In this section, we use our testing procedure to show that income thresholds indeed exist in 401(k) applications, and confirm that this finding is robust to functional form specifications.

To illustrate the testing procedure proposed in the preceding sections,
we consider the estimation and inference of the thresholds 
associated with the effect of 401(k) eligibility on total wealth.
401(k) eligibility, the variable of interest,  
is an indicator of being eligible to enroll in a 401(k) plan (i.e., whether individual $i$ is working for a firm that offers access to a 401(k) plan).
Poterba et al.~(1994a, 1994b) and Chernozhukov et al.~(2016) argue that
401(k) eligibility may be taken as exogenous conditional on income.
Following Chernozhukov et al.~(2016) and 
by using the data set in Chernozhukov and Hansen (2004),
we thus construct both our outcome variable and the explanatory variable of interest 
after partialling out the effects of the other variables including
the dummies for age, education, marital status, family size, and homeownership. 
The sample size is 9915.
In the example presented herein, we consider the following
nonparametric regression with $s$ thresholds:
\begin{align*}
Y_{po} = \sum_{j=1}^s m_{\gamma_j}(D_{po}) + e_i,
\end{align*}
where the threshold variable is income,
while $Y_{po}$ and $D_{po}$ are the partialled out total wealth
and partialled out 401(k) eligibility, respectively.

We implement the test $F_n(s+1|s)$ in Theorem 7 to
determine the number of thresholds and then 
estimate the corresponding threshold values by using the sequential method.
The weighting function is constructed as $A(d)= \{d\in [-0.5, 0.5] \}$,
and the bandwidth $h= c\cdot \hat{\sigma}  \times (9915)^{-1/4.25}$,
where $\hat{\sigma} = 0.46$ and $c$ is set to 1.
We first conduct a test for the null hypothesis that $s=0$ versus $s=1$. 
We find that the value of the test statistic is 50.46, thereby rejecting the null.
The first-round estimated threshold value $\hat{\gamma}_1 = 75,000.3$ (92nd percentile). Since there are a small number of observations 
on the right-hand side interval of this threshold value,
we conduct the next test, in the interval $[0, \ 75000.3]$,
for the null hypothesis that $s=1$ versus $s=2$.
The corresponding value of the test statistic is 27.34, which again rejects the null.
The second-round estimated threshold value $\hat{\gamma}_2 = 42,600$ (68th percentile).  We now conduct the test for the null hypothesis that
$s=2$ versus $s=3$ in the joint interval of 
$[0, \ 42600]$ and $[42600, \ 75000.3]$.
The value of the joint test statistic is 2.62.
Thus, we reject the null, 
and then estimate the threshold value in this joint interval according to Theorem 12.
We obtain $\hat{\gamma}_3 = 31,836$ (50th percentile).
Since there are insufficient observations in the intervals
$[31836, \ 42600]$ and $[42600, \ 75000.3]$,
we only conduct our next test to detect whether an extra threshold exists
in the interval $[0, \ 31836]$.
Finally, we conduct the test for
$s=3$ versus $s=4$ in the interval of 
$[0, \ 31836]$ 
Here, we do not reject the null because the test statistic with the value 0.85
is less than the critical value.
We also conduct robustness checks by using different bandwidth values with $c=1.05$ and $c=0.95$.
The corresponding three threshold values found are  the same as those found with $c=1$.
In short, our testing procedure allows us to identify four threshold regions and the estimated income threshold values are
$\$ 31,836 \ (50\%)$, 
$\$ 42,600 \ (68\%)$, and
$\$ 75,000.3 \ (92\%)$.
The crucial income threshold values are therefore all above the median income values.

\section{Conclusion}

In this study, we identify the number of thresholds and estimate
the threshold values for  a nonparametric regression with multiple thresholds. 
The significance test of A\"{\i}t-Sahalia  et al.~(2001) is modified to detect the existence of an extra threshold (i.e., $s$ versus $s+1$ thresholds).
The asymptotic properties of the modified tests are then established. 
Based on the modified test, a procedure for determining the number of thresholds is suggested. Accordingly, we then carry out the sequential method to estimate 
the unknown threshold values. We also derive the asymptotic properties of the corresponding threshold value estimator. Our simulation results signify that the proposed estimators perform adequately in  finite samples. To illustrate our testing procedure, we present an empirical analysis of the 401(k) plan with income thresholds.

\pagebreak

\noindent
{\Large\bf Appendix}

\medskip
\noindent
{\bf Proof of Theorem 1.}

\noindent
The kernel density estimator is defined by
\begin{eqnarray*}
\hat{f}_{\gamma_{j}}(\bx) = \frac{1}{n} \sum^{n}_{i=1} \bK_{h}(\bX_{i}-\bx)I_{\gamma_{j}}(Q_{i}).
\end{eqnarray*}
Suppose the kernel satisfies the conditions in Assumption 2 and is a second-order ($r = 2$) kernel function and that
Assumptions 1-1 to 1-4 hold. Then, $\hat{f}_{\gamma_{j}}(\bx)$ has the expectation
\begin{eqnarray}
E[\hat{f}_{\gamma_{j}}(\bx)]
= f_{\gamma_{j}}(\bx)  + \frac{h^{2}}{2} \sum^{p}_{l=1} f^{(2)}_{\gamma_{j},l}(\bx) C_{1} + o(h^{2}) \label{Ker-1}
\end{eqnarray}
and the variance
\begin{eqnarray}
\mbox{V}(\hat{f}_{\gamma_{j}}(\bx))
&=& \frac{1}{n} \mbox{V}( \bK_{h}(\bX_{i}-\bx) I_{\gamma_{j}}(Q_{i}) ) \nonumber \\
& &  + \sum^{M(n)}_{l=1} 2\, \frac{n-l}{n^{2}}
     \COV[\bK_{h}(\bX_{1}-\bx) I_{\gamma_{j}}(Q_{1})), \bK_{h}(\bX_{1+l}-\bx) I_{\gamma_{j}}(Q_{1+l}))] \nonumber \\
& &  + 2\, \sum^{n-1}_{l = M(n)+1} 2 \frac{n-l}{n^{2}}
     \COV[\bK_{h}(\bX_{1}-\bx) I_{\gamma_{j}}(Q_{1})), \bK_{h}(\bX_{1+l}-\bx) I_{\gamma_{j}}(Q_{1+l}))] \nonumber \\
&=& V_{1} + V_{2} + V_{3}.
\end{eqnarray}
Assuming $M(n)$ satisfies
\begin{eqnarray*}
as \ n \to \infty, \ M(n) \to \infty, \ and \ M(n)h^{p} \to 0,
\end{eqnarray*}
we have
\begin{eqnarray}
V_{1}
&=& \frac{1}{n} \left[ \mbox{E}(\bK_{h}(\bX_{i}-\bx)I_{\gamma_{j}}(Q_{i}))^{2}
    - [\mbox{E}(\bK_{h}(\bX_{i}-\bx)I_{\gamma_{j}}(Q_{i}))]^{2} \right]  \nonumber \\
&=& \frac{1}{nh^{p}} C_{2} f_{\gamma_{j}}(\bx) + o(nh^{p}) \label{Ker-2-1}
\end{eqnarray}
By denoting $M_{1} = \max_{l \in [1, \ldots , M(n)]} \COV[\bK_{h}(\bX_{1}-\bx) I_{\gamma_{j}}(Q_{1})), \bK_{h}(\bX_{1+l}-\bx)
I_{\gamma_{j}}(Q_{1+l}))]$, we obtain
\begin{eqnarray}
V_{2}
&=& \sum^{M(n)}_{l=1} 2 \frac{n-l}{n^{2}} \COV[\bK_{h}(\bX_{1}-\bx) I_{\gamma_{j}}(Q_{1})),\bK_{h}(\bX_{1+l}-\bx) I_{\gamma_{j}}(Q_{1+l}))]  \nonumber \\
&\leq&  \frac{1}{n} M(n) M_{1} = o((nh^{p})^{-1}). \label{Ker-2-2}
\end{eqnarray}
Denote $W_{ni}(\bx) = \bK_{h}(\bX_{i}-\bx) I_{\gamma_{j}}(Q_{i})) - \mbox{E}[\bK_{h}(\bX_{i}-\bx) I_{\gamma_{j}}(Q_{i}))]$
and for any $\delta >0$, the upper bound of the covariance terms can be obtained by Lemma A.0 of Fan and Li(1999) as
\begin{eqnarray*}
\COV[\bK_{h}(\bX_{1}-\bx) I_{\gamma_{j}}(Q_{1})), \bK_{h}(\bX_{1+l}-\bx) I_{\gamma_{j}}(Q_{1+l}))]
 \leq 4M_2^{1/(1+\delta)} \beta^{\delta/(1+\delta)}(l)
\end{eqnarray*}
where $M_2$ is defined as
\begin{eqnarray*}
max\left( \mbox{E}|W_{1i}(\bx) W_{(1+l)i}(\bx) |^{1+\delta},
\int \int |W_{1i}(\bx) W_{(1+l)i}(\bx)|^{1+\delta}) dF(\bX_{1},Q_{1}) dF(\bX_{1+l},Q_{1+l} \right).
\end{eqnarray*}
Furthermore, given that Assumption 1-1 holds, 
\begin{eqnarray}
V_{3}
&=&  \sum^{n-1}_{l = M(n)+1} 2 \frac{n-l}{n^{2}}
   \COV[\bK_{h}(\bX_{1}-\bx) I_{\gamma_{j}}(Q_{1})), \bK_{h}(\bX_{1+l}-\bx) I_{\gamma_{j}}(Q_{1+l}))]  \nonumber \\
&\leq& \frac{1}{n} M_{2} \sum^{\infty}_{l = M(n)+1} \beta^{\delta/(1+\delta)}(l)  = o((nh^{p})^{-1}). \label{Ker-2-3}
\end{eqnarray}
By combining ($\ref{Ker-2-1}$), ($\ref{Ker-2-2}$) and ($\ref{Ker-2-3}$), we have the variance of
$\hat{f}_{\gamma_{j}}(\bx)$ as
\begin{eqnarray}
\mbox{V}(\hat{f}_{\gamma_{j}}(\bx)) = \frac{1}{nh^{p}} C_{2} f_{\gamma_{j}}(\bx) + o(nh^{p}). \label{Ker-2}
\end{eqnarray}
In general, if the $r$th-order kernel function is considered,
($\ref{Ker-1}$) becomes
\begin{eqnarray}
\mbox{E}(\hat{f}_{\gamma_{j}}(\bx))
= f_{\gamma_{j}}(\bx)  + O(h^{r}) + o(h^{r}). \label{Ker-r}
\end{eqnarray}
Given the results in ($\ref{Ker-2}$) and ($\ref{Ker-r}$) and that the bandwidth $h$ satisfies Assumption 3-1,
the uniform almost sure convergence rate of a kernel density estimator can be obtained; see Lemma 2
and Lemma 8 in Stone (1983). 
Given the results in ($\ref{Ker-1}$) and 
($\ref{Ker-2}$),
and  that Assumptions 1,  2,  3-1, 
and 5-2 hold, 
the asymptotic sampling distribution
of $\hat{f}_{\gamma_{j}}(\bx)$ is derived by Masry (1996) and Li and Racine~(2007). \quad$\blacksquare$

\bigskip
\noindent
{\bf Proof of Theorem 2.}

\noindent
Given a second-order kernel function as well as
equations  (\ref{Ker-1}) and (\ref{Ker-2}), we  have
\begin{eqnarray}
\hat{f}_{\gamma_{j}}(\bx) = f_{\gamma_{j}}(\bx) + O_{p}( h^{2} + (nh^{p})^{-1/2})
= f_{\gamma_{j}}(\bx) + o_{p}(1). \label{Ker-3}
\end{eqnarray}
Together with (\ref{Ker-3}), 
the local constant estimator  can be rewritten as
\begin{eqnarray*}
\hat{m}_{\gamma_{j}}(\bx) - m_{\gamma_{j}}(\bx)
&=& \frac{\sum^{n}_{i=1} \bK_{h}(\bX_{i}-\bx)I_{\gamma_{j}}(Q_{i})[Y_{i} - m_{\gamma_{j}}(\bx)]}
         {\sum^{n}_{i=1} \bK_{h}(\bX_{i}-\bx)I_{\gamma_{j}}(Q_{i})} \\
&=& \left[\frac{1}{n} \frac{\sum^{n}_{i=1} \bK_{h}(\bX_{i}-\bx)I_{\gamma_{j}}(Q_{i})[Y_{i} - m_{\gamma_{j}}(\bx)]}
         {f_{\gamma_{j}}(\bx)}\right]\, \, (1+o_{p}(1)).
\end{eqnarray*}
Under the correct specification of a nonparametric regression with $s$ thresholds, the first term in the previous
result is
\begin{eqnarray*}
\lefteqn{\frac{1}{n} \frac{\sum^{n}_{i=1} \bK_{h}(\bX_{i}-\bx)I_{\gamma_{j}}(Q_{i})[Y_{i} - m_{\gamma_{j}}(\bx)]}
                     { f_{\gamma_{j}}(\bx) }} \\
&=& \frac{1}{n} \frac{\sum^{n}_{i=1} \bK_{h}(\bX_{i}-\bx)I_{\gamma_{j}}(Q_{i})
   [ \sum^{s+1}_{l=1} m_{\gamma_{l}}(\bX_{i}) I_{\gamma_{l}}(Q_{i}) - m_{\gamma_{j}}(\bx)] }{ f_{\gamma_{j}}(\bx) } \\
& & + \frac{1}{n} \frac{\sum^{n}_{i=1} \bK_{h}(\bX_{i}-\bx)I_{\gamma_{j}}(Q_{i}) e_{i}  }
                       { f_{\gamma_{j}}(\bx) } \\
&=& AB_{\gamma_{j}}(\bx) + AV_{\gamma_{j}}(\bx).
\end{eqnarray*}
From Assumption 1-1, we have
\begin{eqnarray}
E[AB_{\gamma_{j}}(\bx)]
&=& \frac{1}{n} E \left[ \frac{\sum^{n}_{i=1} \bK_{h}(\bX_{i}-\bx)I_{\gamma_{j}}(Q_{i})
    [\sum^{s+1}_{l=1} m_{\gamma_{l}}(\bX_{i}) I_{\gamma_{l}}(Q_{i}) - m_{\gamma_{j}}(\bx)] }
     { f_{\gamma_{j}}(\bx) }\right] \nonumber \\
&=& E \left[ \frac{ \bK_{h}(\bX_{i}-\bx)I_{\gamma_{j}}(Q_{i})[ \sum^{s+1}_{l=1} m_{\gamma_{l}}(\bX_{i})
    I_{\gamma_{l}}(Q_{i}) - m_{\gamma_{j}}(\bx) ]}{ f_{\gamma_{j}}(\bx) } \right] \label{Ker-4}
\end{eqnarray}
where
$P[I_{\gamma_{j}}(Q_{i}) \times I_{\gamma_{l}}(Q_{i})=0]=1, j \neq l$;
$I_{\gamma_{j}}(Q_{i}) \times I_{\gamma_{l}}(Q_{i}) = I_{\gamma_{j}}(Q_{i}), j = l$.
Thus, (\ref{Ker-4}) becomes
\begin{eqnarray}
\lefteqn{E [AB_{\gamma_{j}}(\bx)]} \nonumber \\
&=&  E \left[ \frac{ \bK_{h}(\bX_{i}-\bx)I_{\gamma_{j}}(Q_{i}) [\sum^{s+1}_{l=1} m_{\gamma_{l}}(\bX_{i})
     I_{\gamma_{l}}(Q_{i}) - m_{\gamma_{j}}(\bx)] }{ f_{\gamma_{j}}(\bx) } \right] \nonumber \\
&=&  \frac{h^{2}}{2} C_{1} \sum^{p}_{l=1} [ m^{(2)}_{\gamma_{j},l}(\bx) f_{\gamma_{j}}(\bx)
    + 2 m^{(1)}_{\gamma_{j},l}(\bx) f^{(1)}_{\gamma_{j},l}(\bx) ] /f_{\gamma_{j}}(\bx)  + o(h^{2}). \qquad \label{Ker-5}
\end{eqnarray}
Further, the asymptotic variance term is
\begin{eqnarray*}
\mbox{V}(AV_{\gamma_{j}}(\bx))
&=& \frac{1}{n} \mbox{E} \left(\frac{\bK_{h}(\bX_{i}-\bx)I_{\gamma_{j}}(Q_{i}) e_{i}}{ f_{\gamma_{j}}(\bx) } \right)^{2}  \\
& & + \frac{1}{n} \sum^{n-1}_{l=1} \frac{n-l}{n} \COV \left(\frac{\bK_{h}(\bX_{1}-\bx)I_{\gamma_{j}}(Q_{1}) e_{1}}
                                                           { f_{\gamma_{j}}(\bx) }
                , \frac{\bK_{h}(\bX_{1+l}-\bx)I_{\gamma_{j}}(Q_{1+l}) e_{1+l}}{ f_{\gamma_{j}}(\bx) } \right)  \\
                &=& V_{1} + V_{2}
\end{eqnarray*}
with
\begin{eqnarray}
V_{1}
&=& \frac{1}{n^{2} f^{2}_{\gamma_{j}}(\bx) }
    E\left[ \sum^{n}_{i=1} \bK^{2}_{h}(\bX_{i}-\bx)I^{2}_{\gamma_{j}}(Q_{i}) e^{2}_{i} \right] \nonumber \\
&=& \frac{1}{n^{2}  f^{2}_{\gamma_{j}}(\bx) } \int  \bK^{2}_{h}(\bX_{i}-\bx) \nonumber \\
& & \times  \int ( y_{i} -  m_{\gamma_{j}}(\bX_{i}))^{2}_{i} f_{\gamma_{j}}(y_{i}|\bx_{i})  dy_{i} f_{\gamma_{j}}(\bx_{i}) d\bx_{i} \nonumber \\
&=& \frac{\sigma^{2}_{\gamma_{j}}(\bx)}{nh^{p} f_{\gamma_{j}}(\bx)  }  \int  \bK^{2}(u)   du + o(\frac{1}{nh^{p}}). \label{Ker-6}
\end{eqnarray}
Given that Assumption 1-1 holds, 
and from arguments similar to the proof for Theorem 1, 
the covariance term $V_{2} = o\left((nh^p)^{-1}\right)$. We have
\begin{eqnarray}
\mbox{V}(AV_{\gamma_{j}}(\bx)) = \frac{\sigma^{2}_{\gamma_{j}}(\bx)}{nh^{p} f_{\gamma_{j}}(\bx)}
C_{2} + o(nh^{p}), \label{Ker-6}
\end{eqnarray}
and the covariance terms are
\begin{eqnarray}
\lefteqn{\COV(AV_{\gamma_{j}}(\bx),AV_{\gamma_{k}}(\bx))} \nonumber \\
&=& \frac{1}{n} \frac{1}{ f_{\gamma_{j}}(\bx)f_{\gamma_{k}}(\bx)} \left\{ \mbox{E}[  \bK^{2}_{h}(\bX_{i}-\bx)I_{\gamma_{j}}(Q_{i})
I_{\gamma_{k}}(Q_{i}) e^{2}_{i}] \right. \nonumber \\
& & \left. - \mbox{E}[\bK_{h}(\bX_{i}-\bx)I_{\gamma_{j}}(Q_{i}) e_{i}] \mbox{E}[\bK_{h}(\bX_{i}-\bx)I_{\gamma_{k}}(Q_{i}) e_{i}] \right\} \nonumber \\
&=& 0. \label{Ker-cov}
\end{eqnarray}
In general, when the kernel is an $r$th kernel function,  (\ref{Ker-4}) becomes
\begin{eqnarray}
\mbox{E}(AB_{\gamma_{j}}(\bx)) = O(h^{r}). \label{Ker-8}
\end{eqnarray}

Given that (\ref{Ker-6}), (\ref{Ker-8}), and Assumption 3-1 hold, the result of part a) in Theorem 2 is verified
based on Lemmas 2 and 8 of Stone (1983).
Moreover, given (\ref{Ker-5}), (\ref{Ker-6}),  (\ref{Ker-cov}), and that Assumption 3-1 holds, the result of part b) in Theorem 2
holds according to the central limit theorem; see Masry~(1996) and  Li and Racine~(2007). \quad$\blacksquare$

\bigskip
\noindent
{\bf Proof of Theorem 3.}

\medskip
\noindent
By substituting ($\ref{Ker-6}$) and ($\ref{Ker-8}$) into the mean integrated square error, we have the optimal bandwidth  defined as
\begin{eqnarray}
h_{opt}
&=& \arg \min \int \mbox{E} \left[ 
\sum^{s+1}_{j=1}  \left(\hat{m}_{\gamma_{j}}(\bx) - m_{\gamma_{j}}(\bx)\right) 
\right]^{2} w(\bx) d\bx  \nonumber \\
&=& \arg \min \int   \sum^{s+1}_{j=1} \left[ \mbox{E}(AB_{\gamma_{j}}(\bx)  )^{2} + \mbox{V}(AV_{\gamma_{j}}(\bx)) \right]  w(\bx) d\bx.
  \label{Ker-7}
\end{eqnarray}
Taking the first-order derivative of (\ref{Ker-7})
with respect to $h$,
\begin{eqnarray*}
\frac{ d \int \sum^{s+1}_{j=1} \left[ \mbox{E}(AB_{\gamma_{j}}(\bx))^{2} + \mbox{V}(AV_{\gamma_{j}}(\bx)) \right]
     (\gamma_{j-1} - \gamma_{j}) w(\bx) d\bx}{dh} \stackrel{set}{=} 0
\end{eqnarray*}
we then have $h_{opt} = O(n^{\frac{-1}{2r+p}})$.
It is clear that the convergence rate of $h_{opt}$ depends on the dimension of $\bX$, $p$, and the orders of
the kernel function, $r$. It is worth noting that the convergence rate does not depend on the number of thresholds, $s$.
This result suggests that the bandwidth can be selected without considering
the number of thresholds. \quad$\blacksquare$

\bigskip
\noindent
{\bf Proof of Theorem 4.}

\medskip
\noindent
Since
\begin{eqnarray*}
\Gamma(\tau_{j})
&=& \int \int  \left\{ \int \frac{ y f_{\gamma_{j}}(y,\bx) }{f_{\gamma_{j}}(\bx)} dy I_{\gamma_{j}}(q)
        - \int \frac{ y f_{\gamma_{j-1},\tau_{j}}(y,\bx) }{f_{\gamma_{j-1},\tau_{j}}(\bx)} dy
        I_{\gamma_{j-1},\tau_{j}}(q) \right. \\
& & \hspace{15mm} \left. - \int \frac{ y f_{\tau_{j},\gamma_{j}}(y,\bx) }{f_{\tau_{j},\gamma_{j}}(\bx)} dy
        I_{\tau_{j},\gamma_{j}}(q)  \right\}^{2} a(\bx)dF(\bx,q) \\
&=& \Gamma(f_{\gamma_{j}},f_{\gamma_{j-1},\tau_{j}},f_{\tau_{j},\gamma_{j}},F),
\end{eqnarray*}
we have
\begin{eqnarray*}
\tilde{\Gamma}(\tau_{j})
&=& \frac{1}{n} \sum^{n}_{i=1} \left\{ \hat{m}_{ \gamma_{j}}(\bX) I_{ \gamma_{j}}(Q) -
    \hat{m}_{\gamma_{j-1},\tau_{j}}(\bX)I_{\gamma_{j-1},\tau_{j}}(Q) - \hat{m}_{\tau_{j},\gamma_{j}}(\bX)
      I_{\tau_{j},\gamma_{j}}(Q)  \right\}^{2} a(\bX_{i}) \\
&=& \int \int  \left\{ \int \frac{ y \hat{f}_{\gamma_{j}}(y,\bx) }{\hat{f}_{\gamma_{j}}(\bx)} dy I_{\gamma_{j}}(q)
     - \int \frac{ y \hat{f}_{\gamma_{j-1},\tau_{j}}(y,\bx) }{\hat{f}_{\gamma_{j-1},\tau_{j}}(\bx)} dy
     I_{\gamma_{j-1},\tau_{j}}(q) \right. \\
& & \hspace{15mm} \left. - \int \frac{ y \hat{f}_{\tau_{j},\gamma_{j}}(y,\bx) }{\hat{f}_{\tau_{j},\gamma_{j}}(\bx)} dy
     I_{\tau_{j},\gamma_{j}}(q)  \right\}^{2} a(\bx)d \hat{F}(\bx,q) \\
&=& \Gamma(\hat{f}_{\gamma_{j}},\hat{f}_{\gamma_{j-1},\tau_{j}},\hat{f}_{\tau_{j},\gamma_{j}},\hat{F}).
\end{eqnarray*}
Note that
\begin{eqnarray*}
\lefteqn{\Gamma(\hat{f}_{\gamma_{j}},\hat{f}_{\gamma_{j-1},\tau_{j}},\hat{f}_{\tau_{j},\gamma_{j}},F)} \\
&=& \frac{1}{n} \sum^{n}_{i=1} \left\{ \hat{m}_{ \gamma_{j}}(\bX) I_{ \gamma_{j}}(Q) -
          \hat{m}_{\gamma_{j-1},\tau_{j}}(\bX)I_{\gamma_{j-1},\tau_{j}}(Q) - \hat{m}_{\tau_{j},\gamma_{j}}(\bX)
          I_{\tau_{j},\gamma_{j}}(Q)  \right\}^{2} a(\bX_{i}) \\
&=& \int \int  \left\{ \int \frac{ y \hat{f}_{\gamma_{j}}(y,\bx) }{\hat{f}_{\gamma_{j}}(\bx)} dy I_{\gamma_{j}}(q)
  - \int \frac{ y \hat{f}_{\gamma_{j-1},\tau_{j}}(y,\bx) }{\hat{f}_{\gamma_{j-1},\tau_{j}}( \bx)} dy
   I_{\gamma_{j-1},\tau_{j}}(q) \right. \\
& & \hspace{15mm}\left. - \int \frac{ y \hat{f}_{\tau_{j},\gamma_{j}}(y,\bx) }{\hat{f}_{\tau_{j},\gamma_{j}}(\bx)} dy
   I_{\tau_{j},\gamma_{j}}(q)  \right\}^{2} a(\bx)d \hat{F}(\bx,q).
\end{eqnarray*}

We need the following lemmas to complete the proof.
\begin{lemma}{\label{Sahalia-2}}
(Lemma 2 of A\"{\i}t Sahalia et al.~(2001))

Defining
\begin{eqnarray*}
||g_{\gamma_{j}}|| &\equiv& \max \left( \sup_{\bx}| \int y g_{\gamma_{j}}(y,\bx) dy |,\,
                                        \sup_{\bx}|g_{\gamma_{j}}(\bx)| \right) \\
||g_{\gamma_{j-1},\tau_{j}}|| &\equiv& \max \left( \sup_{\bx}| \int y g_{\gamma_{j-1},\tau_{j}}(y,\bx) dy |,\,
                                                   \sup_{\bx}|g_{\gamma_{j-1},\tau_{j}}(\bx)| \right) \\
||g_{\tau_{j},\gamma_{j}}|| &\equiv& \max \left( \sup_{\bx}| \int y g_{\tau_{j},\gamma_{j}}(y,\bx) dy |,\,
                                                 \sup_{\bx}|g_{\tau_{j},\gamma_{j}}(\bx)| \right)
\end{eqnarray*}
where
\begin{eqnarray*}
g_{\gamma_{j}} &=& \hat{f}_{\gamma_{j}} - f_{\gamma_{j}}   \\
g_{\gamma_{j-1},\tau_{j}} &=& \hat{f}_{\gamma_{j-1},\tau_{j}} - f_{\gamma_{j-1},\tau_{j}}  \\
g_{\tau_{j},\gamma_{j}} &=&  \hat{f}_{\tau_{j},\gamma_{j}} - f_{\tau_{j},\gamma_{j}},
\end{eqnarray*}
we have
\begin{eqnarray*}
||g_{\gamma_{j}}|| &=& O_{p}(h^{r} + \ln(n)/(nh^{p})^{1/2}) \\
||g_{\gamma_{j-1},\tau_{j}}|| &=& O_{p}(h^{r} + \ln(n)/(nh^{p})^{1/2}) \\
 ||g_{\tau_{j},\gamma_{j}}|| &=& O_{p}(h^{r} + \ln(n)/(nh^{p})^{1/2}).
\end{eqnarray*}
\end{lemma}

\begin{lemma}{\label{Sahalia-1}}
(Lemma 7 of A\"{\i}t Sahalia et al.~(2001))
\begin{eqnarray*}
\Gamma(\hat{f}_{\gamma_{j}},\hat{f}_{\gamma_{j-1},\tau},\hat{f}_{\tau,\gamma_{j}},\hat{F})
&=& \Gamma(\hat{f}_{\gamma_{j}},\hat{f}_{\gamma_{j-1},\tau},\hat{f}_{\tau,\gamma_{j}},F) + \Lambda_{1,n} + \Lambda_{2,n}
\end{eqnarray*}
with
\begin{eqnarray*}
\Lambda_{1,n}
       &=& \int \int \left\{ \int \alpha_{\gamma_{j}}(\bx) dy I_{ \gamma_{j}}(q) -
                      \int \alpha_{\gamma_{j-1},\tau_{j}}(\bx) dy I_{\gamma_{i-1},\tau_{j}}(q) \right. \\
       & & \left. \hspace{15mm} - \int \alpha_{\tau_{j},\gamma_{j}}(\bx) dy I_{\tau_{j},\gamma_{j}}(q)
           \right\}^{2} a(\bX) ( d\hat{F}(\bx,q) - dF(\bx,q)) \\
       &=& O_{p}(n^{-3}(h^{-3p}) + n^{-1}(h^{2r}))
       = o_{p}(n^{-2}h^{-p}),    \\
\Lambda_{2,n} &=& O_{p}(||\hat{f}_{\gamma_{j}}-f_{\gamma_{j}}||^{3} + ||\hat{f}_{\gamma_{j-1},\tau_{j}}-f_{\gamma_{j-1},\tau_{j}}||^{3}
+ ||\hat{f}_{\tau_{j},\gamma_{j}}-f_{\tau_{j},\gamma_{j}}||^{3} )
\end{eqnarray*}
where $\alpha_{\gamma_{j}}(\bx) = \frac{y - m_{\gamma_{j}}(\bx)}{f_{\gamma_{j}}(\bx)}$,
$\alpha_{\gamma_{j-1},\tau_{j}}(\bx) = \frac{y - m_{\gamma_{j-1},\tau_{j}}(\bx)}{f_{\gamma_{j-1},\tau_{j}}(\bx)}$,
and $\alpha_{\gamma_{j}}(\bx) = \frac{y - m_{\gamma_{j-1},\tau_{j}}(\bx)}{f_{\gamma_{j-1},\tau_{j}}(\bx)}$. \quad$\square$
\end{lemma}

\begin{lemma}\label{Peter} (Hall, 1984).

 Let $\{Z_{i};i=1,\ldots,n \}$ be an i.i.d sequence. Suppose that the U-statistic
 $U_{n}=\sum_{1 \leq i <j \leq n} \tilde{P}_{n}(Z_{i},Z_{j})$
with the symmetric variable function $\tilde{P}_{n}$ being centered (i.e., $\mbox{E}[\tilde{P}_{n}(Z_{1},Z_{2})]=0$)
and degenerate
(i.e.,$\mbox{E}[\tilde{P}_{n}(Z_{1},Z_{2})|Z_{1}=z_{1}]=0$ almost surely for all $z_{1}$). Let
\begin{eqnarray*}
\sigma^{2}_{n} = \mbox{E}[\tilde{P}_{n}(Z_{1},Z_{2})^{2}] , \ \ \ \ \ \ \tilde{\Pi}_{n}(z_{1},z_{2}) = \mbox{E}[\tilde{P}_{n}(Z_{1},z_{1}) \tilde{P}_{n}(Z_{2},z_{2})].
\end{eqnarray*}
Then, if
\begin{eqnarray*}
\lim_{n \rightarrow \infty} \frac{\mbox{E}[\tilde{\Pi}_{n}(z_{1},z_{2})^{2}] + n^{-1} \mbox{E}[\tilde{P}_{n}(Z_{1},Z_{2})^{4}]}{(\mbox{E}[\tilde{P}_{n}(Z_{1},Z_{2})^{2}])^{2}} \rightarrow 0
\end{eqnarray*}
we have that as $n \rightarrow \infty$
\begin{eqnarray*}
\frac{2^{1/2} U_{n} }{n \sigma_{n} } \rightarrow N(0,1). \quad\square
\end{eqnarray*}
\end{lemma}

\bigskip
From Lemma \ref{Sahalia-1}, we have
\begin{eqnarray*}
\Gamma(\hat{f}_{\gamma_{j}},\hat{f}_{\gamma_{j-1},\tau_{j}},f\hat{f}_{\tau_{j},\gamma_{j}},\hat{F})
&=& \Gamma(\hat{f}_{\gamma_{j}},f\hat{f}_{\gamma_{j-1},\tau_{j}},\hat{f}_{\tau_{j},\gamma_{j}},F)
+ \Lambda_{1,n} + \Lambda_{2,n} \\
&=& \Gamma(\hat{f}_{\gamma_{j}},\hat{f}_{\gamma_{j-1},\tau_{j}},\hat{f}_{\tau_{j},\gamma_{j}},F)
+ o_{p}(n^{-2}h^{-p}).
\end{eqnarray*}

To prove this, denote $\Psi(t)$ as
\begin{eqnarray*}
\Psi(t)
&=& \int \int \left\{ \int y \frac{f_{\gamma_{j}}(\bx,y) + t g_{\gamma_{j}}(\bx,y)}
                                  {f_{\gamma_{j}}(\bx) + tg_{\gamma_{j}}(\bx)} dy I_{ \gamma_{j}}(q)  \right. \\
& & \left. -  \int y \frac{f_{\gamma_{j-1},\tau_{j}}(\bx,y) + t g_{\gamma_{j-1},\tau_{j}}(\bx,y)}
    {f_{\gamma_{j-1},\tau_{j}}(\bx) + t g_{\gamma_{j-1},\tau_{j}}(\bx)} dy I_{\gamma_{i-1},\tau_{j}}(q) \right.  \\
& & \left. - \int y \frac{f_{\tau_{j},\gamma_{j}}(\bx,y) + t g_{\tau_{j},\gamma_{j}}(\bx,y)}
   {f_{\tau_{j},\gamma_{j}}(\bx) + t g_{\tau_{j},\gamma_{j}}(\bx)} dy I_{\tau_{j},\gamma_{j}}(q) \right\}^{2} a(\bX) dF(\bx,q). 
\end{eqnarray*}
It can then be seen that when
\begin{eqnarray*}
g_{\gamma_{j}} = \hat{f}_{\gamma_{j}}  - f_{\gamma_{j}} \ ,
g_{\gamma_{j-1},\tau_{j}} = \hat{f}_{\gamma_{j-1},\tau_{j}}  - f_{\gamma_{j-1},\tau_{j}} \ ,
g_{\tau_{j},\gamma_{j}} = \hat{f}_{\tau_{j},\gamma_{j}} - f_{\tau_{j},\gamma_{j}}
\end{eqnarray*}
are specified, we have $\Psi(1)= \Gamma(\hat{f}_{\gamma_{j}},\hat{f}_{\gamma_{j-1},\tau_{j}},\hat{f}_{\tau_{j},\gamma_{j}},F) $ and
$\Psi(0) = \Gamma(f_{\gamma_{j}},f_{\gamma_{j-1},\tau_{j}},f_{\tau_{j},\gamma_{j}},F)$. Then,  by
the Taylor expansion,
\begin{eqnarray*}
\Psi(1) &=& \Psi(0) + \Psi^{(1)}(0) + 1/2 \Psi^{(2)}(0) + 1/6 \Psi^{(3)}(t^{*}),
\end{eqnarray*}
and thus it is equivalent to have
\begin{eqnarray*}
\Gamma(\hat{f}_{\gamma_{j}},\hat{f}_{\gamma_{j-1},\tau_{j}},\hat{f}_{\tau_{j},\gamma_{j}},F) &=&
\Gamma(f_{\gamma_{j}},f_{\gamma_{j-1},\tau_{j}},f_{\tau_{j},\gamma_{j}},F) + \Psi^{(1)}(0) + 1/2 \Psi^{(2)}(0) + 1/6 \Psi^{(3)}(t^{*})
\end{eqnarray*}
where $t^* \in (0, 1)$.
Denote
\begin{eqnarray*}
\psi(t)
&=&  \int y \frac{f_{\gamma_{j}}(\bx,y) + t g_{\gamma_{j}}(\bx,y)}{f_{\gamma_{j}}(\bx) + tg_{\gamma_{j}}(\bx)} dy I_{ \gamma_{j}}(q) -
      \int y \frac{f_{\gamma_{j-1},\tau_{j}}(\bx,y) + t g_{\gamma_{j-1},\tau_{j}}(\bx,y)}{f_{\gamma_{j-1},\tau_{j}}(\bx) + t g_{\gamma_{j-1},\tau_{j}}(\bx)} dy I_{\gamma_{i-1},\tau_{j}}(q) \\
& & - \int y \frac{f_{\tau_{j},\gamma_{j}}(\bx,y) + t g_{\tau_{j},\gamma_{j}}(\bx,y)}{f_{\tau_{j},\gamma_{j}}(\bx) + t g_{\tau_{j},\gamma_{j}}(\bx)} dy I_{\tau_{j},\gamma_{j}}(q),
\end{eqnarray*}
so that $\Psi(t)$ can be written as
\begin{eqnarray*}
\Psi(t) =  \int \int \psi(t)^{2}\, a(\bx)\, dF(\bx,q)
\end{eqnarray*}
where
\begin{eqnarray*}
\Psi^{(1)}(t)
&=&  \int \int  \frac{ \partial  \psi(t)^{2}}{ \partial t} a(\bx)dF(\bx,q)
=  2\int \int \psi(t) \frac{ \partial  \psi(t)}{\partial t}   a(\bx)dF(\bx,q) \\
\Psi^{(2)}(t)
&=& 2\int \int \left\{ \psi(t) \frac{ \partial^{2}  \psi(t)}{\partial t^{2}}
    + \left[ \frac{ \partial  \psi(t)}{\partial t} \right]^{2}  \right \}  a(\bx)dF(\bx,q) \\
\Psi^{(3)}(t)
&=& 2\int \int \left\{ \psi(t) \frac{ \partial^{3}  \psi(t)}{\partial t^{3}}
    + 3 \frac{ \partial  \psi(t)}{\partial t} \frac{ \partial^{2}  \psi(t)}{\partial t^{2}}  \right \}  a(\bx)dF(\bx,q) 
\end{eqnarray*}
in which the first derivative of $\psi(t)$ is
\begin{eqnarray*}
\frac{\partial \psi(t)}{ \partial t }
&=&  \frac{f_{\gamma_{j}}(\bx) \int y g_{\gamma_{j}} (y,\bx) dy - g_{\gamma_{j}}(\bx) \int y f_{\gamma_{j}} (y,\bx) dy}
                          { [f_{\gamma_{j}}(\bx)  - t g_{\gamma_{j}}(\bx) ]^{2} } I_{\gamma_{j}}(q) \\
& & - \frac{f_{\gamma_{j-1},\tau_{j}}(\bx) \int y g_{\gamma_{j-1},\tau_{j}} (y,\bx) dy - g_{\gamma_{j-1},\tau_{j}}(\bx) \int y f_{\gamma_{j-1},\tau_{j}} (y,\bx) dy}
                            { [f_{\gamma_{j-1},\tau_{j}}(\bx)  - t g_{\gamma_{j-1},\tau_{j}}(\bx) ]^{2} } I_{\gamma_{j-1},\tau_{j}}(q) \\
& & - \frac{f_{\tau_{j},\gamma_{j}}(\bx) \int y g_{\tau_{j},\gamma_{j}} (y,\bx) dy - g_{\tau_{j},\gamma_{j}}(\bx) \int y f_{\tau_{j},\gamma_{j}} (y,\bx) dy}
                            { [f_{\tau_{j},\gamma_{j}}(\bx)  - t g_{\tau_{j},\gamma_{j}}(\bx) ]^{2} } I_{\tau_{j},\gamma_{j}}(q) ,
\end{eqnarray*}
the second derivative of $\psi(t)$ is
\begin{eqnarray*}
\lefteqn{\frac{\partial^{2} \psi(t)}{ \partial t^{2}}} \\
&=& -2\,  \frac{ [f_{\gamma_{j}}(\bx) \int y g_{\gamma_{j}} (y,\bx) dy- g_{\gamma_{j}}(\bx) \int y f_{\gamma_{j}} (y,\bx) dy ]\,
    g_{\gamma_{j}}(\bx)}{ [f_{\gamma_{j}}(\bx)  - t g_{\gamma_{j}}(\bx) ]^{3} } I_{\gamma_{j}}(q) \\
& & +2\,  \frac{ [f_{\gamma_{j-1},\tau_{j}}(\bx) \int y g_{\gamma_{j-1},\tau_{j}} (y,\bx) dy- g_{\gamma_{j-1},\tau_{j}}(\bx)
    \int y f_{\gamma_{j-1},\tau_{j}} (y,\bx) dy ]\,
    g_{\gamma_{j-1},\tau_{j}}(\bx)}{ [f_{\gamma_{j-1},\tau_{j}}(\bx)  - t g_{\gamma_{j-1},\tau_{j}}(\bx) ]^{3} } I_{\gamma_{j-1},\tau_{j}}(q) \\
& & +2\,  \frac{ [f_{\tau_{j},\gamma_{j}}(\bx) \int y g_{\tau_{j},\gamma_{j}} (y,\bx) dy- g_{\tau_{j},\gamma_{j}}(\bx)
        \int y f_{\tau_{j},\gamma_{j}} (y,\bx) dy ]\,
    g_{\tau_{j},\gamma_{j}}(\bx)}{ [f_{\tau_{j},\gamma_{j}}(\bx)  - t g_{\tau_{j},\gamma_{j}}(\bx) ]^{3} } I_{\tau_{j},\gamma_{j}}(q) 
\end{eqnarray*}
and the third derivative of $\psi(t)$ is
\begin{eqnarray*}
\lefteqn{\frac{\partial^{3} \psi(t)}{ \partial t^{3}} } \\
&=& 6\, \frac{ [f_{\gamma_{j}}(\bx) \int y g_{\gamma_{j}} (y,\bx) dy - g_{\gamma_{j}}(\bx) \int y f_{\gamma_{j}} (y,\bx) dy ]\,
    g^{2}_{\gamma_{j}}(\bx)}{ [f_{\gamma_{j}}(\bx)  - t g_{\gamma_{j}}(\bx) ]^{4} } I_{\gamma_{j}}(q) \\
& & - 6\, \frac{ [f_{\gamma_{j-1},\tau_{j}}(\bx) \int y g_{\gamma_{j-1},\tau_{j}} (y,\bx) dy -
         g_{\gamma_{j-1},\tau_{j}}(\bx) \int y f_{\gamma_{j-1},\tau_{j}} (y,\bx) dy ]\,
    g^{2}_{\gamma_{j-1},\tau_{j}}}{ [f_{\gamma_{j-1},\tau_{j}}(\bx)  - t g_{\gamma_{j-1},\tau_{j}}(\bx)]^{4} }
        I_{\gamma_{j-1},\tau_{j}}(q) \\
& & - 6\, \frac{ [f_{\tau_{j},\gamma_{j}}(\bx) \int y g_{\tau_{j},\gamma_{j}} (y,\bx) dy -
    g_{\tau_{j},\gamma_{j}}(\bx) \int y f_{\tau_{j},\gamma_{j}} (y,\bx) dy ]\,
    g^{2}_{\tau_{j},\gamma_{j}}(\bx)}{ [f_{\tau_{j},\gamma_{j}}(\bx)  - t g_{\tau_{j},\gamma_{j}}(\bx) ]^{4} }
    I_{\tau_{j},\gamma_{j}}(q).
\end{eqnarray*}
It is clear that $\psi(0)=0$ under the null hypothesis. Therefore, under the null $\Gamma(f_{\gamma_{j}},f_{\gamma_{j-1},\tau_{j}},f_{\tau_{j},\gamma_{j}},F)$, we have
\begin{eqnarray*}
\Gamma(\hat{f}_{\gamma_{j}},\hat{f}_{\gamma_{j-1},\tau_{j}},\hat{f}_{\tau_{j},\gamma_{j}},F)
&=& \Gamma(f_{\gamma_{j}},f_{\gamma_{j-1},\tau_{j}},f_{\tau_{j},\gamma_{j}},F) + \Psi^{(1)}(0) + 1/2 \Psi^{(2)}(0) + 1/6 \Psi^{(3)}(t^{*}) \\
&=& \int \int \left[ \frac{\partial \psi(t)}{ \partial t } |_{t=0}  \right]^{2}\, a(\bx)\, dF(\bx,q) + 1/6 \Psi^{(3)}(t^{*}) \\
&=& I_{n} + 1/6 \Psi^{(3)}(t^{*}).
\end{eqnarray*}
Given Lemma \ref{Sahalia-2},  $\Psi^{(3)}(t^{*})$ satisfies 
\begin{eqnarray*}
| \Psi^{(3)}(t^{*}) | &=& O(||g_{\gamma_{j}}||^{3} + ||g_{\gamma_{j-1},\tau_{j}}||^{3} + ||g_{\tau_{j},\gamma_{j}}|| ) \\
                    &=& O(||\hat{f}_{\gamma_{j}} - f_{\gamma_{j}}||^{3} + ||\hat{f}_{\gamma_{j-1},\tau_{j}} - f_{\gamma_{j-1},\tau_{j}}||^{3} + ||\hat{f}_{\tau_{j},\gamma_{j}} - f_{\tau_{j},\gamma_{j}}||^{3} ). 
\end{eqnarray*}
Given that Assumption 3-2 holds,
\begin{eqnarray*}
|\Psi^{(3)}(t^{*})| = o_{p}(n^{-1} h^{-p}).
\end{eqnarray*}
For the term $\frac{\partial \psi(t)}{ \partial t }|_{t=0}$ in $I_n$, it is clear that
\begin{eqnarray*}
\frac{\partial \psi(t)}{ \partial t }|_{t=0}
&=& \frac{f_{\gamma_{j}}(\bx) \int y g_{\gamma_{j}} (y,\bx) dy I_{\gamma_{j}}(q) -
          g_{\gamma_{j}}(\bx) \int y f_{\gamma_{j}} (y,\bx) dy}
         { [f_{\gamma_{j}}(\bx) ]^{2} } I_{\gamma_{j}}(q) \\
& & - \frac{f(\bx) \int y g_{\gamma_{j-1},\tau_{j}} (y,\bx) dy I_{\gamma_{j-1},\tau_{j}}(q) -
          g_{\gamma_{j-1},\tau_{j}}(\bx) \int y f_{\gamma_{j-1},\tau_{j}} (y,\bx) dy}
         { [f_{\gamma_{j-1},\tau_{j}}(\bx) ]^{2} } I_{\gamma_{j-1},\tau_{j}}(q)  \\
& & - \frac{f_{\gamma_{j}}(\bx) \int y g_{\tau_{j},\gamma_{j}} (y,\bx) dy I_{\tau_{j},\gamma_{j}}(q) -
          g_{\tau_{j},\gamma_{j}}(\bx) \int y f_{\tau_{j},\gamma_{j}} (y,\bx) dy}
           { [f_{\tau_{j},\gamma_{j}}(\bx) ]^{2} } I_{\tau_{j},\gamma_{j}}(q)  \\
&=& \eta_{\gamma_{j}}(y,\bx) I_{\gamma_{j}}(q) - \eta_{\gamma_{j-1},\tau_{j}}(y,\bx) I_{\gamma_{j-1},\tau_{j}}(q) - \eta_{\tau_{j},\gamma_{j}}(y,\bx) I_{\tau_{j},\gamma_{j}}(q)
\end{eqnarray*}
in which
\begin{eqnarray*}
\eta_{\gamma_{j}}
&=& \frac{f_{\gamma_{j}}(\bx) \int y g_{\gamma_{j}} (y,\bx) dy - g_{\gamma_{j}}(\bx) \int y f_{\gamma_{j}} (y,\bx) dy}
         { [f_{\gamma_{j}}(\bx) ]^{2} } \\
&=& \int \frac{y-m_{\gamma_{j}}(\bx)}{f_{\gamma_{j}}(\bx)} \hat{f}_{\gamma_{j}}(y,\bx)  dy -
    \int \frac{y-m_{\gamma_{j}}(\bx)}{f_{\gamma_{j}}(\bx)} f_{\gamma_{j}} (y,\bx) dy.
\end{eqnarray*}
Since $ \int \frac{y-m_{\gamma_{j}}(\bx)}{f_{\gamma_{j}}(\bx)} f_{\gamma_{j}} (y,\bx) dy =0$,
\begin{eqnarray*}
\eta_{\gamma_{j}} &=& \int \frac{y-m_{\gamma_{j}}(\bx)}{f_{\gamma_{j}}(\bx)} \hat{f}_{\gamma_{j}}(y,\bx)  dy \\
                  &=& \int \alpha_{\gamma_{j}}(y,\bx) \hat{f}_{\gamma_{j}}(y,\bx) dy.
\end{eqnarray*}
Similarly,
\begin{eqnarray*}
\eta_{\gamma_{j-1},\tau_{j}} &=& \int \frac{y-m_{\gamma_{j-1},\tau_{j}}(\bx)}{f_{\gamma_{j-1},\tau_{j}}(\bx)} \hat{f}_{\gamma_{j-1},\tau_{j}}(y,\bx)  dy \\
                         &=& \int \alpha_{\gamma_{j-1},\tau_{j}}(y,\bx) \hat{f}_{\gamma_{j-1},\tau_{j}}(y,\bx) dy,
\end{eqnarray*}
and
\begin{eqnarray*}
\eta_{\tau_{j},\gamma_{i}} &=& \int \frac{y-m_{\tau_{j},\gamma_{i}}(\bx)}{f_{\tau_{j},\gamma_{i}}(\bx)} \hat{f}_{\tau_{j},\gamma_{i}}(y,\bx)  dy \\
                       &=& \int \alpha_{\tau_{j},\gamma_{i}}(y,\bx) \hat{f}_{\tau_{j},\gamma_{i}}(y,\bx) dy.
\end{eqnarray*}
Therefore, we obtain
\begin{eqnarray*}
\frac{\partial \psi(t)}{ \partial t }|_{t=0}
&=& \eta_{\gamma_{j}}(y,\bx) I_{\gamma_{j}}(q) - \eta_{\gamma_{j-1},\tau_{j}}(y,\bx) I_{\gamma_{j-1},\tau_{j}}(q) - \eta_{\tau_{j},\gamma_{j}}(y,\bx) I_{\tau_{j},\gamma_{j}}(q) \\
&=& \int \alpha_{\gamma_{j}}(y,\bx) \hat{f}_{\gamma_{j}}(y,\bx) dy I_{\gamma_{j}}(q) -
    \int \alpha_{\gamma_{j-1},\tau_{j}}(y,\bx) \hat{f}_{\gamma_{j-1},\tau_{j}}(y,\bx) dy I_{\gamma_{j-1},\tau_{j}}(q) \\
& & - \int \alpha_{\tau_{j},\gamma_{j}}(y,\bx) \hat{f}_{\tau_{j},\gamma_{j}}(y,\bx) dy I_{\tau_{j},\gamma_{j}}(q) .
\end{eqnarray*}
Specifically,
\begin{eqnarray*}
\lefteqn{\frac{\partial \psi(t)}{ \partial t }|_{t=0}} \\
&=& \int \alpha_{\gamma_{j}}(y,\bx) \hat{f}_{\gamma_{j}}(y,\bx) dy I_{\gamma_{j}}(q) -
    \int \alpha_{\gamma_{j-1},\tau_{j}}(y,\bx) \hat{f}_{\gamma_{j-1},\tau_{j}}(y,\bx) dy I_{\gamma_{j-1},\tau_{j}}(q) \\
& & - \int \alpha_{\tau_{j},\gamma_{i}}(y,\bx) \hat{f}_{\tau_{j},\gamma_{i}}(y,\bx) dy I_{\tau_{j},\gamma_{j}}(q)  \\
&=& \frac{\sum^{n}_{i=1}}{n} \left\{  \int \alpha_{\gamma_{j}}(y,\bx) \bK_{h}(\bX_{i} - \bx) I_{\gamma_{j}}(Q_{i}) \bK_{h}(Y_{i} - y) dy I_{\gamma_{j}}(q) \right. \\
& & \hspace{15mm} - \int \alpha_{\gamma_{j-1},\tau_{j}}(y,\bx) \bK_{h}(\bX_{i} - \bx) I_{\gamma_{j-1},\tau_{j}}(Q_{i}) \bK_{h}(Y_{i} - y) dy I_{\gamma_{j-1},\tau_{j}}(q) \\
& & \hspace{15mm} - \left. \int \alpha_{\tau_{j},\gamma_{i}}(y,\bx) \bK_{h}(\bX_{i} - \bx) I_{\tau_{j},\gamma_{j}}(Q_{i}) \bK_{h}(Y_{i} - y) dy I_{\tau_{j},\gamma_{j}}(q) \right\}  .
\end{eqnarray*}
To simplify the expression, we denote
\begin{eqnarray*}
\lefteqn{r_{\tau_{j}}(Y_{i} , \bX_{i}, Q_{i} ; y,\bx,q)} \\
&=&  \left\{  \int \alpha_{\gamma_{j}}(y,\bx) \bK_{h}(\bX_{i} - \bx) I_{\gamma_{j}}(Q_{i}) \bK_{h}(Y_{i} - y) dy I_{\gamma_{j}}(q) \right. \\
& & \hspace{5mm} - \int \alpha_{\gamma_{j-1},\tau_{j}}(y,\bx)  \bK_{h}(\bX_{i} - \bx) I_{\gamma_{j-1},\tau_{j}}(Q_{i}) \bK_{h}(Y_{i} - y) dy I_{\gamma_{j-1},\tau_{j}}(q) \\
& & \hspace{5mm} - \left. \int \alpha_{\tau_{j},\gamma_{i}}(y,\bx) \bK_{h}(\bX_{i} - \bx) I_{\tau_{j},\gamma_{j}}(Q_{i}) \bK_{h}(Y_{i} - y) dy I_{\tau_{j},\gamma_{j}}(q) \right\}  .
\end{eqnarray*}
and also denote its de-mean as
\begin{eqnarray*}
\tilde{r}_{\tau_{j}}(Y_{i} , \bX_{i},Q_{i} , y,\bx,q)
= r_{\tau_{j}}(Y_{i} , \bX_{i},Q_{i} ; y,\bx,q) - \mbox{E}(r_{\tau_{j}}(Y_{i} ; \bX_{i},Q_{i} , y,\bx,q)).
\end{eqnarray*}
Finally, the term $I_n$ can be expressed as
\begin{eqnarray*}
I_{n}
&=& \frac{1}{n^{2}} \left\{  2 \sum_{1 \leq i < j \leq n } \int \int \tilde{r}_{\tau_{j}}(Y_{i} ; \bX_{i},Q_{i} , y,\bx,q) \tilde{r}_{\tau_{j}}(Y_{j} , \bX_{j},Q_{j} ; y,\bx,q) a(\bx)dF(\bx,q) \right. \\
& & + \sum^{n}_{i=1} \int \int r^{2}_{\tau_{j}}(Y_{i} , \bX_{i},Q_{i} ; y,\bx,q) a(\bx)dF(\bx,q) \\
& & + 2(n-1)\sum^{n}_{i=1} \int \int (\tilde{r}_{\tau_{j}}(Y_{i} , \bX_{i},Q_{i} ; y,\bx,q)\mbox{E}(r_{\tau_{j}}(Y_{i} , \bX_{i},Q_{i} ; y,\bx,q) ) a(\bx)dF(\bx,q) \\
& & \left. + n(n-1)\sum^{n}_{i=1} \int \int [\mbox{E}(r_{\tau_{j}}(Y_{i} , \bX_{i},Q_{i} ; y,\bx,q) ]^{2} a(\bx)dF(\bx,q) \right\}  \\
&=& I_{n1}(\tau_{j}) +I_{n2}(\tau_{j}) +I_{n3}(\tau_{j}) +I_{n4}(\tau_{j}).
\end{eqnarray*}
In the above equation, the term $I_{n1}(\tau_{j})$ is asymptotically normal, $I_{n2}(\tau_{j})$ is the
asymptotic bias, and $I_{n3}(\tau_{j})$
and $I_{n4}(\tau_{j})$ are asymptotically negligible. 

In addition,
\begin{eqnarray}
\mbox{E}(r(Y_{i} , \bX_{i},Q_{i} ; y,\bx,q)) = O(h^{r}) \label{test-2}
\end{eqnarray}
and uniformly in $\bx$ in $S$ from Assumption 3-2,
\begin{eqnarray}
\mbox{E}[|\bar{r}(Y_{i} , \bX_{i},Q_{i} ; y,\bx,q)|] = O(1). \label{test-3}
\end{eqnarray}
Denote
\begin{eqnarray*}
\zeta_{\gamma_{j}}(i) &=& \int \alpha_{\gamma_{j}}(y,\bx)
 \bK_{h}(\bX_{i}-\bx)I_{\gamma_{j}}(q_{i}) \bK_{h}(y_{i}-y) dy I_{\gamma_{j}}(q) \\
\zeta_{\gamma_{j-1},\tau_{j}}(i) &=& \int \alpha_{\gamma_{j-1},\tau_{j}}(y,\bx)
 \bK_{h}(\bX_{i}-\bx)I_{\gamma_{j-1},\tau_{j}}(q_{i}) \bK_{h}(y_{i}-y) dy I_{\gamma_{j-1},\tau_{j}}(q) \\
\zeta_{\tau_{j},\gamma_{j}}(i) &=& \int \alpha_{\tau_{j},\gamma_{j}}(y,\bx)
 \bK_{h}(\bX_{i}-\bx)I_{\tau_{j},\gamma_{j}}(q_{i}) \bK_{h}(y_{i}-y) dy I_{\tau_{j},\gamma_{j}}(q).
\end{eqnarray*}
$I_{n2}(\tau_{j})$ can then be simplified to
\begin{eqnarray*}
\lefteqn{\mbox{E}( \int r^{2}_{\tau_{j}}(Y_{i} , \bX_{i},Q_{i} ;y,\bx,q) a(\bx)dF(\bx,q))} \\
&=& \int \int \int \int \int \left[   \zeta^{2}_{\gamma_{j}}(i)
   - 2 \zeta_{\gamma_{j}}(i) ( \zeta_{\gamma_{j-1},\tau_{j}}(i) + \zeta_{\tau_{j},\gamma_{j}}(i) ) + ( \zeta^{2}_{\gamma_{j-1},\tau_{j}}(i) + \zeta^{2}_{\tau_{j},\gamma_{j}}(i) )  \right] \\
& & \times a(\bx)dF(\bx,q) f(y_{1},\bx_{1},q_{1}) d\bx_{1}dy_{1}dq_{1} \\
&=& \mbox{E}(I_{n21}(\tau_{j})) +\mbox{E}(I_{n22}(\tau_{j})) +\mbox{E}(I_{n23}(\tau_{j})).
\end{eqnarray*}
We then have
\begin{eqnarray*}
\mbox{E}(I_{n21}(\tau_{j}))
&=& h^{-p} C_{2} \int \sigma^{2}_{\gamma_{j}}(\bx) a(\bx) d\bx \\
\mbox{E}(I_{n22}(\tau_{j}))
&=& -2 h^{-p} C_{2} \left[ \int  \frac{f_{\gamma_{j-1}, \tau_{j}}(\bx_{i}) }{f_{\gamma_{j}}(\bx_{i})}
     \sigma^{2}_{\gamma_{j-1}, \tau_{j}}(\bx)    a(\bx) d\bx
              + \int \frac{f_{\tau_{j}, \gamma_{j}}(\bx_{i}) }{f_{\gamma_{j}}(\bx_{i})}
              \sigma^{2}_{\tau_{j}, \gamma_{j}}(\bx)    a(\bx) d\bx \right] \\
\mbox{E}(I_{n22}(\tau_{j})) &=& h^{-p} C_{2} \left[ \int   \sigma^{2}_{\gamma_{j-1}, \tau_{j}}(\bx)    a(\bx) d\bx
              + \int \sigma^{2}_{\tau_{j}, \gamma_{j}}(\bx)  \right]  a(\bx) d\bx
\end{eqnarray*}
and 
\begin{eqnarray*}
\lefteqn{\mbox{V} \left[ n^{2} \sum^{n}_{i=1} \int \int r^{2}_{\tau_{j}}(Y_{i} , \bX_{i},Q_{i} ; y,\bx,q) a(\bx)dF(\bx,q)  \right]} \\
&=& n^{-3} \mbox{V} \left[ \int r^{2}(Y_{i} , \bX_{i},Q_{i} , y,\bx,q) a(\bx)dF(\bx,q)  \right] \\
&=& n^{-3} O(h^{-2p}).
\end{eqnarray*}
From Chebyshev's inequality it follows that
\begin{eqnarray*}
n h^{p/2} \{ I_{n2}(\tau_{j}) - n^{-1}h^{-p} \xi_{1}(\tau_{j}) \} =o_{p}(1).
\end{eqnarray*}
Let $Z_{i} = (Y_{i},\bX_{i},Q_{i})$ and denote
\begin{eqnarray*}
\tilde{P}_{n}(Z_{i},Z_{l}) = \frac{2}{n^{2}} \int \tilde{r}(Y_{i} , \bX_{i},Q_{i} , y,\bx,q)\,
 \tilde{r}(Y_{l} , \bX_{l},Q_{l} , y,\bx,q)\, a(\bx)\, dF(\bx,q)
\end{eqnarray*}
which verifies the centering and degeneracy conditions by construction. In addition, since
\begin{eqnarray*}
\mbox{E}[\tilde{P}_{n}(Z_{1},Z_{2})^{2}]
&=& \left(\frac{1}{n^{2}h^{2p}} \right)^{2} h^{3p} 2 \sigma^{2}_{1}(\tau_{j}) \\
\mbox{E}[\tilde{\Pi}_{n}(z_{1},z_{2})^{2} ]
&\equiv& O \left( \left[\frac{1}{n^{2}h^{2p}}\right]^{4} h^{7p} \right) \\
\mbox{E}[\tilde{P}_{n}(Z_{1},Z_{2})^{4} ]
&\equiv& O \left( \left[\frac{1}{n^{2}h^{2p}}\right]^{4} h^{5p} \right),
\end{eqnarray*}
we have
\begin{eqnarray*}
\lefteqn{\lim_{n \rightarrow \infty} \frac{\mbox{E}[\tilde{\Pi}_{n}(z_{1},z_{2})^{2}]
 + n^{-1} \mbox{E}[\tilde{P}_{n}(Z_{1},Z_{2})^{4}]}{(\mbox{E}[\tilde{P}_{n}(Z_{1},Z_{2})^{2}])^{2}} } \\
&=& \lim_{n \rightarrow \infty} O(h^{p}) + O((nh^{p})^{-1}) \to 0
\end{eqnarray*}
which is the necessary condition for having Lemma \ref{Peter} applicable.

As for $\sigma^{2}_{n} (\tau_{j})$, we have
{\footnotesize
\begin{eqnarray*}
\sigma^{2}_{n} (\tau_{j})
&=& \mbox{E} \left[ \frac{2}{n^{2}} \int \tilde{r}(Y_{i} , \bX_{i},Q_{i} , y,\bx,q)\,
 \tilde{r}(Y_{l} , \bX_{l},Q_{l} , y,\bx,q)\, a(\bx)\, dF(\bx,q)^{2} \right] \\
&=& \mbox{E} \left[ \left(\frac{2}{n^{2}}
    \int r(Y_{i} , \bX_{i},Q_{i} , y,\bx,q) r(Y_{l} , \bX_{l},Q_{l} , y,\bx,q) a(\bx)dF(\bx,q)\right)^{2} \right]
    + o\left(\frac{1}{n^{4}h^{-p}}\right) \\
&=& \frac{4}{n^{4}} \mbox{E} \left\{ \left[\int  \zeta_{\gamma_{j}}(i) \times \zeta_{\gamma_{j}}(l)
     dF(\bx,q) \right]^{2} + 4 \left[ \int  \zeta_{\gamma_{j}}(i) \times
     (\zeta_{\gamma_{j-1},\tau_{j}}(l) + \zeta_{\tau_{j},\gamma_{j}}(l)  )  dF(\bx,q) \right]^{2} \right. \\
& &    + \left[ \int  \zeta_{\gamma_{j-1},\tau_{j}}(i) \times \zeta_{\gamma_{j-1},\tau_{j}}(l)  dF(\bx,q) \right]^{2}
    + \left[ \int  \zeta_{\tau_{j},\gamma_{j}}(i) \times \zeta_{\tau_{j},\gamma_{j}}(l)  dF(\bx,q) \right]^{2}  \\
& & - 4 \left[ \int  \zeta_{\gamma_{j}}(i) \times \zeta_{\gamma_{j}}(l)  dF(\bx,q)
      \times \int \zeta_{\gamma_{j}}(i) \times (\zeta_{\gamma_{j-1},\tau_{j}}(l) + \zeta_{\tau_{j},\gamma_{j}}(l))
      dF(\bx,q) \right]  \\
& & + 2 \left[ \int  \zeta_{\gamma_{j}}(i) \times \zeta_{\gamma_{j}}(l)  dF(\bx,q)
      \times \int (\zeta_{\gamma_{j-1},\tau_{j}}(i) \times \zeta_{\gamma_{j-1},\tau_{j}}(l) +
      \zeta_{\tau_{j},\gamma_{j}}(i) \zeta_{\tau_{j},\gamma_{j}}(l)) dF(\bx,q)  \right] \\
& & - 4 \left[ \int \zeta_{\gamma_{j}}(i) \times \zeta_{\gamma_{j-1},\tau_{j}}(l)   dF(\bx,q)
      \times \int \zeta_{\gamma_{j-1},\tau_{j}}(i) \times \zeta_{\gamma_{j-1},\tau_{j}}(l)  dF(\bx,q) \right.  \\
& &  \left. \left. +  \int  \zeta_{\gamma_{j}}(i) \times \zeta_{\tau_{j},\gamma_{j}}(l)   dF(\bx,q)
      \times \int \zeta_{\tau_{j},\gamma_{j}}(i) \times \zeta_{\tau_{j},\gamma_{j}}(l)  dF(\bx,q)
      \right] \right\} + o \left(\frac{1}{n^{4}h^{-p}}\right) \\
&=&  \frac{4C_{3}}{n^{4}h^{-p}} \left[ \sigma^{2}_{111}(\tau_{j}) + \sigma^{2}_{112}(\tau_{j}) + \sigma^{2}_{113}(\tau_{j})
 + \sigma^{2}_{114}(\tau_{j}) + \sigma^{2}_{115}(\tau_{j}) + \sigma^{2}_{116}(\tau_{j})\right] +
  o\left(\frac{1}{n^{4}h^{-p}}\right)
\end{eqnarray*}
}
Thus, the following are obtained:
\begin{eqnarray*}
\sigma^{2}_{111}(\tau_{j}) &=&  \int \sigma^{4}_{\gamma_{j}}(\bx) a^{2}(\bx) d\bx \\
\sigma^{2}_{112}(\tau_{j}) &=& 4 \{ \int
                     \sigma^{2}_{\gamma_{j-1},\tau_{j}}(\bx) \sigma^{2}_{\gamma_{j}}(\bx)
                      \frac{f_{\gamma_{j-1},\tau_{j}}(\bx)}{ f_{\gamma_{j}}(\bx) } a^{2}(\bx) d\bx  \\
                 & & + \int \sigma^{2}_{\tau_{j},\gamma_{j}}(\bx) \sigma^{2}_{\gamma_{j}}(\bx)
                     \frac{f_{\tau_{j},\gamma_{j}}(\bx)}{f_{\gamma_{j}}(\bx)} a^{2}(\bx) d\bx \} \\
\sigma^{2}_{113}(\tau_{j}) &=&  \{ \int \sigma^{2}_{\gamma_{j-1},\tau_{j}}(\bx) a^{2}(\bx)  d\bx
                        + \int \sigma^{2}_{\tau_{j},\gamma_{j}}(\bx) a^{2}(\bx)  d\bx \} \\
\sigma^{2}_{114}(\tau_{j}) &=& -4
                    \{ \int \sigma^{2}_{\gamma_{j}}(\bx) \sigma^{2}_{\gamma_{j-1},\tau_{j}}(\bx)
                    \frac{f_{\gamma_{j-1},\tau_{j}}(\bx)}{f_{\gamma_{j}}(\bx)} a^{2}(\bx) d\bx \\
                 & & + \int \sigma^{2}_{\gamma_{j}}(\bx) \sigma^{2}_{\tau_{j},\gamma_{j}}(\bx)
                    \frac{f_{\tau_{j},\gamma_{j}}(\bx)}{f_{\gamma_{j}}(\bx)} a^{2}(\bx) d\bx \} \\
\sigma^{2}_{115}(\tau_{j}) &=& 2 \{
                    \int \sigma^{4}_{\gamma_{j-1},\tau_{j}}(\bx)
                      \frac{f_{\gamma_{j}}(\bx)}{f_{\gamma_{j-1},\tau_{j}}(\bx)} a^{2}(\bx) d\bx \\
                 & & + \int \sigma^{4}_{\gamma_{j-1},\tau_{j}}(\bx)
                     \frac{f_{\gamma_{j-1},\tau_{j}}(\bx)}{f_{\gamma_{j}}(\bx)}  a^{2}(\bx) d\bx \} \\
\sigma^{2}_{116}(\tau_{j}) &=& -4
                     \{ \int   \sigma^{4}_{\gamma_{j-1},\tau_{j}}(\bx) \frac{f_{\gamma_{j-1},\tau_{j}}(\bx)}{f_{\gamma_{j}}(\bx)} a^{2}(\bx) d\bx \\
                 & & + \int   \sigma^{4}_{\tau_{j},\gamma_{j}}(\bx) \frac{f_{\tau_{j},\gamma_{j}}(\bx)}{f_{\gamma_{j}}(\bx)} a^{2}(\bx) d\bx \}
\end{eqnarray*}
Hence,
\begin{eqnarray}
\lefteqn{\sigma^{2}_{n}(\tau_{j})} \nonumber \\
&=& \frac{2}{n^{4}} h^{-p} \sigma^{2}_{111}(\tau_{j}) + \sigma^{2}_{112}(\tau_{j}) + \sigma^{2}_{113}(\tau_{j}) + \sigma^{2}_{114}(\tau_{j}) + \sigma^{2}_{115}(\tau_{j}) + \sigma^{2}_{116}(\tau_{j}) \nonumber \\
&=&  \frac{2}{n^{4}} h^{-p} \sigma^{2}_{1}(\tau_{j}), \label{Test-6}
\end{eqnarray}
where
\begin{eqnarray*}
\sigma^{2}_{1} (\tau_{j}) &=& 2 C_{3} [\sigma^{2}_{11}(\tau_{j}) + \sigma^{2}_{12}(\tau_{j})] \\
\sigma^{2}_{11}(\tau_{j}) &=& \int \sigma^{4}_{\gamma_{j}}(\bx) a^{2}(\bx)  d\bx \\
\sigma^{2}_{12}(\tau_{j}) &=& \int (1 - 2 \frac{f_{\gamma_{j-1}, \tau_{j}}(\bx) }{f_{\gamma_{j}}(\bx)}) \sigma^{4}_{\gamma_{j-1}, \tau_{j}}(\bx)    a^{2}(\bx) d\bx
              + \int (1 - 2 \frac{f_{\tau_{j}, \gamma_{j}}(\bx) }{f_{\gamma_{j}}(\bx)}) \sigma^{4}_{\tau_{j}, \gamma_{j}}(\bx)    a^{2}(\bx) d\bx.
\end{eqnarray*}

According to Lemma \ref{Peter} (Hall, 1984), we have
\begin{eqnarray*}
\sigma^{-1}_{1}(\tau_{j}) nh^{p/2} I_{n1}(\tau_{j})  \sim N(0,1).
\end{eqnarray*}
From (\ref{test-3}), we obtain
\begin{eqnarray*}
\mbox{E} \left[ \left( \int \tilde{r}(Y_{i} , \bX_{i},Q_{i} , y,\bx,q) \mbox{E}(\tilde{r}(Y_{i} , \bX_{i},Q_{i} , y,\bx,q)) \right) \right] = O(h^{2r}),
\end{eqnarray*}
and then
\begin{eqnarray*}
\mbox{E}[(nh^{p/2}I_{n3}(\tau_{j}))^{2}] = O(nh^{p}h^{2r}) = o(1)
\end{eqnarray*}
from Chebyshev's inequality
\begin{eqnarray*}
nh^{p/2} I_{n3}(\tau_{j}) = o_{p}(1)
\end{eqnarray*}
and from (\ref{test-2}), we have the following result for $I_{n4}(\tau_{j})$:
\begin{eqnarray*}
nh^{p/2} I_{n4}(\tau_{j}) = nh^{p/2} O(h^{2r}) = o(1). 
\end{eqnarray*}
Note that this proof is established under  $\{ (Y_i, \bX_i, Q_i), i=1, \ldots, n \}$ are i.i.d.
For mixing data with the $\beta$-coefficient as in Assumption 1-1, A\"{\i}t-Sahalia et al.~(2001),
Fan and Li~(1999), and Dette and Spreckelsen~(2004) point out that this result also holds. \quad$\blacksquare$

\bigskip
\noindent
{\bf Proof of Theorem 5.}

\medskip
\noindent To begin with, we write out the term $\varphi(\tau_{j,l},\tau_{j,k})$ as follows:
 \begin{eqnarray*}
 \varphi(\tau_{j,l},\tau_{j,k}) &=&  \int_{\bx} \sigma^{4}_{\gamma_{j}}(\bx) a^{2}(\bx)\, d\bx  \\
 & & -2 \left\{ \int \sigma^{2}_{\gamma_{j}}(\bx) \sigma^{2}_{\gamma_{j-1},\tau_{j,k}}(\bx)
   \frac{f_{\gamma_{j-1},\tau_{j,k}}(\bx)}{f_{\gamma_{j}}(\bx)}\, a^{2}(\bx)\, d\bx \right.\\
 & & \left. + \int \sigma^{2}_{\gamma_{j}}(\bx) \sigma^{2}_{\tau_{j,k},\gamma_{j}}(\bx)
  \frac{f_{\tau_{j,k},\gamma_{j}}(\bx)}{f_{\gamma_{j}}(\bx)}\, a^{2}(\bx)\, d\bx \right\}  \\
 & & + \int_{\bx} \sigma^{4}_{\gamma_{j-1},\tau_{j,k}}(\bx) \frac{f_{\gamma_{j-1},\tau_{j,k}}(\bx)}
   {f_{\gamma_{j}}(\bx)} a^{2}(\bx)\, d\bx
   + \int_{\bx} \sigma^{4}_{\tau_{j,k},\gamma_{j}}(\bx) \frac{f_{\tau_{j,k},\gamma_{j}}(\bx)}
   {f_{\gamma_{j}}(\bx)} a^{2}(\bx)\, d\bx \\
 & & -2 \left\{ \int_{\bx}
   \sigma^{2}_{\gamma_{j}}(\bx) \sigma^{2}_{\gamma_{j-1},\tau_{j,l}}
   \frac{f_{\gamma_{j-1},\tau_{j,l}}(\bx)}{f_{\gamma_{j}}(\bx)} a^{2}(\bx) d\bx \right. \\
 & & \left. + \int_{\bx} \sigma^{2}_{\gamma_{j}}(\bx) \sigma^{2}_{\tau_{j,l},\gamma_{j}}(\bx)
    \frac{f_{\tau_{j,l},\gamma_{j}}(\bx)}{f_{\gamma_{j}}(\bx)} a^{2}(\bx)\, d\bx \right\}  \\
 & & + 4  \left\{ \int_{\bx} \sigma^{2}_{\gamma_{j}}(\bx) \sigma^{2}_{\gamma_{j-1},\tau_{j,l}}(\bx)
                   \frac{f_{\gamma_{j-1},\tau_{j,l}}(\bx)}{f(\bx)} a^{2}(\bx)\, d\bx \right. \\
 & & \left.   +  \int_{\bx} \sigma^{2}_{\gamma_{j}}(\bx) \sigma^{2}_{\tau_{j,l},\tau_{j,k}}(\bx)
                   \frac{f_{\tau_{j,l},\tau_{j,k}}(\bx)}{f_{\gamma_{j}}(\bx)} a^{2}(\bx)\, d\bx \right. \\
 & &  \left.  +  \int_{\bx} \sigma^{2}_{\gamma_{j}}(\bx) \sigma^{2}_{\tau_{j,k},\gamma_{j}}(\bx)
                   \frac{f_{\tau_{j,k},\gamma_{j}}(\bx)}{f_{\gamma_{j}}(\bx)} a^{2}(\bx)\, d\bx \right\}  \\
 & & -2 \left\{ \int_{\bx} \sigma^{2}_{\gamma_{j-1},\tau_{j,k}}(\bx) \sigma^{2}_{\gamma_{j-1},\tau_{j,l}}(\bx)
                   \frac{f_{\gamma_{j-1},\tau_{j,l}}(\bx)}{f(\bx)} a^{2}(\bx)\,  d\bx \right. \\
 & & \left.   +  \int_{\bx} \sigma^{2}_{\gamma_{j-1},\tau_{j,k}}(\bx) \sigma^{2}_{\tau_{j,l},\tau_{j,k}}(\bx)
                   \frac{f_{\tau_{j,l},\tau_{j,k}}(\bx)}{f_{\gamma_{j}}(\bx)} a^{2}(\bx)\, d\bx \right. \\
 & & \left.   +  \int_{\bx} \sigma^{4}_{\tau_{j,k},\gamma_{j}}(\bx)
                   \frac{f_{\tau_{j,k},\gamma_{j}}(\bx)}{f_{\gamma_{j}}(\bx)} a^{2}(\bx)\, d\bx \right\} \\
 & & + \left\{ \int_{\bx} \sigma^{4}_{\gamma_{j-1},\tau_{j,l}}(\bx)
      \frac{f_{\gamma_{j-1},\tau_{j,l}}(\bx)}{f_{\gamma_{j}}(\bx)} a^{2}(\bx)\,   d\bx
     + \int_{\bx} \sigma^{4}_{\tau_{j,l},\gamma_{j}}(\bx)
     \frac{f_{\tau_{j,l},\gamma_{j}}(\bx)}{f_{\gamma_{j}}(\bx)} a^{2}(\bx)\, d\bx \right\} \\
 & & -2  \left\{ \int_{\bx} \sigma^{4}_{\gamma_{j-1},\tau_{j,l}}(\bx)
                   \frac{f_{\gamma_{j-1},\tau_{j,l}}(\bx)}{f(\bx)} a^{2}(\bx)\, d\bx \right. \\
 & &  \left.  +  \int_{\bx} \sigma^{2}_{\tau_{j,k},\gamma_{j}}(\bx) \sigma^{2}_{\tau_{j,l},\tau_{j,k}}(\bx)
            \frac{f_{\tau_{j,l},\tau_{j,k}}(\bx)}{f_{\gamma_{j}}(\bx)} a^{2}(\bx)\, d\bx \right. \\
 & &  \left.    +  \int_{\bx} \sigma^{2}_{\tau_{j,l},\gamma_{j}}(\bx) \sigma^{2}_{\tau_{j,k},\gamma_{j}}(\bx)
                   \frac{f_{\tau_{j,k},\gamma_{j}}(\bx)}{f_{\gamma_{j}}(\bx)} a^{2}(\bx)\, d\bx \right\} \\
 & &  + \left\{ \int_{\bx} \sigma^{4}_{\gamma_{j-1},\tau_{j,l}}(\bx)
          \frac{f_{\gamma_{j-1},\tau_{j,l}}(\bx)}{f_{\gamma_{j-1},\tau_{j,k}}(\bx)} a^{2}(\bx)\, d\bx \right. \\
 & &  \left.    +  \int_{\bx} \sigma^{4}_{\tau_{j,l},\tau_{j,k}}(\bx)
                   \frac{\{f_{\tau_{j,l},\tau_{j,k}}(x)\}^{2}}
                   {f_{\gamma_{j-1},\tau_{j,k}}(\bx) f_{\tau_{j,l},\gamma_{j}}(\bx)} a^{2}(\bx)\, d\bx \right. \\
& &  \left.  +  \int_{\bx} \sigma^{4}_{\tau_{j,k},\gamma_{j}}(\bx)
                       \frac{f_{\tau_{j,k},\gamma_{j}}(\bx)}{f_{\tau_{j,l},\gamma_{j}}(\bx)} a^{2}(\bx)\, d\bx \right\}. 
 \end{eqnarray*}

Let
\begin{eqnarray*}
\gamma_{j-1} < \tau_{j,l} < \tau_{j,k} < \gamma_{j}.
\end{eqnarray*}
By definition,
\begin{eqnarray*}
I_{n1}(\tau_{j,l}) =  \frac{2}{n^{2}}  \sum_{1 \leq i < j \leq n } \int \int \tilde{r}_{\tau_{j,l}}(Y_{i} , \bX_{i},Q_{i} , y,\bx,q)
                 \tilde{r}_{\tau_{j,l}}(Y_{j} , \bX_{j},Q_{j} , y,\bx,q) a(\bx)dF(\bx,q),
\end{eqnarray*}
where
\begin{eqnarray*}
\lefteqn{r_{\tau_{j,l}}(Y_{i} , \bX_{i},Q_{i} , y,\bx,q)} \\
&=&  \left\{  \int \alpha_{\gamma_{j}}(y,\bx) K_{h}(\bX_{i} - \bx) I_{\gamma_{j}}(Q_{t}) K_{h}(Y_{i} - y) dy I_{\gamma_{j}}(q) \right. \\
& & - \int \alpha_{\gamma_{j-1},\tau_{j,l}}(y,\bx)  K_{h}(\bX_{i} - \bx) I_{\gamma_{j-1},\tau_{j,l}}(Q_{i}) K_{h}(Y_{i} - y) dy I_{\gamma_{j-1},\tau_{j,l}}(q) \\
& & - \left. \int \alpha_{\tau_{j,l},\gamma_{j}}(y,\bx) K_{h}(\bX_{i} - \bx) I_{\tau_{j,l},\gamma_{j}}(Q_{i}) K_{h}(Y_{i} - y) dy I_{\tau_{j,l},\gamma_{j}}(q) \right\} 
\end{eqnarray*}
and
\begin{eqnarray*}
\lefteqn{\tilde{r}_{\tau_{j,l}}(Y_{i} , \bX_{i},Q_{i} , y,\bx,q)} \\
&=& r_{\tau_{j,l}}(Y_{i} , \bX_{i},Q_{i} , y,\bx,q) - \mbox{E}(r_{\tau_{j,l}}(Y_{i} , \bX_{i},Q_{i} , y,\bx,q)).
\end{eqnarray*}
\begin{eqnarray*}
I_{n1}(\tau_{j,k}) = \frac{2}{n^{2}}   \sum_{1 \leq i < j \leq n } \int \int \tilde{r}_{\tau_{j,k}}(Y_{i} , \bX_{i},Q_{i} , y,\bx,q)
                 \tilde{r}_{\tau_{j,k}}(Y_{j} , \bX_{j},Q_{j} , y,\bx,q) a(\bx)dF(\bx,q),
\end{eqnarray*}
where
\begin{eqnarray*}
\lefteqn{r_{\tau_{j,k}}(Y_{i} , \bX_{i},Q_{i} , y,\bx,q)} \\
&=&  \left\{  \int \alpha_{\gamma_{j}}(y,\bx) \bK_{h}(\bX_{i} - \bx) I_{\gamma_{j}}(Q_{t}) \bK_{h}(Y_{i} - y) dy I_{\gamma_{j}}(q) \right. \\
& & - \int \alpha_{\gamma_{j-1},\tau_{j,k}}(y,\bx)  \bK_{h}(\bX_{i} - \bx) I_{\gamma_{j-1},\tau_{j,k}}(Q_{i}) \bK_{h}(Y_{i} - y) dy I_{\gamma_{j-1},\tau_{j,k}}(q) \\
& & - \left. \int \alpha_{\tau_{j,k},\gamma_{j}}(y,\bx) \bK_{h}(\bX_{i} - \bx)
   I_{\tau_{j,k},\gamma_{j}}(Q_{i}) \bK_{h}(Y_{i} - y) dy I_{\tau_{j,k},\gamma_{j}}(q) \right\}
\end{eqnarray*}
and
\begin{eqnarray*}
& & \tilde{r}_{\tau_{j,k}}(Y_{i} , \bX_{i},Q_{i} , y,\bx,q) \\
&=& r_{\tau_{j,k}}(Y_{i} , \bX_{i},Q_{i} , y,\bx,q) - \mbox{E}(r_{\tau_{j,k}}(Y_{i} , \bX_{i},Q_{i} , y,\bx,q)).
\end{eqnarray*}
Denote
\begin{eqnarray*}
a(i) &=& \int \alpha_{\gamma_{j}}(y,\bx) \bK_{h}(\bX_{i} - \bx) I_{\gamma_{j}}(Q_{t}) \bK_{h}(Y_{i} - y) dy I_{\gamma_{j}}(q) \\
b(i) &=&  \int \alpha_{\gamma_{j-1},\tau_{j,l}}(y,\bx)  \bK_{h}(\bX_{i} - \bx) I_{\gamma_{j-1},\tau_{j,l}}(Q_{i}) \bK_{h}(Y_{i} - y) dy I_{\gamma_{j-1},\tau_{j,l}}(q) \\
     & & +  \int \alpha_{\tau_{j,l},\gamma_{j}}(y,\bx) \bK_{h}(\bX_{i} - \bx) I_{\tau_{j,l},\gamma_{j}}(Q_{i}) \bK_{h}(Y_{i} - y) dy I_{\tau_{j,l},\gamma_{j}}(q)  \\
     &=& b_{1}(i) + b_{2}(i) \\
c(i) &=&  \int \alpha_{\gamma_{j-1},\tau_{j,k}}(y,\bx)  \bK_{h}(\bX_{i} - \bx) I_{\gamma_{j-1},\tau_{j,k}}(Q_{i}) \bK_{h}(Y_{i} - y) dy I_{\gamma_{j-1},\tau_{j,k}}(q) \\
     & & +  \int \alpha_{\tau_{j,k},\gamma_{j}}(y,\bx) K_{h}(\bX_{i} - \bx) I_{\tau_{j,k},\gamma_{j}}(Q_{i}) K_{h}(Y_{i} - y) dy I_{\tau_{j,k},\gamma_{j}}(q)  \\
     &=& c_{1}(i) + c_{2}(i).
\end{eqnarray*}
Therefore, the variance-covariance is
\begin{eqnarray}
\lefteqn{\COV(\delta(\tau_{j,l}),\delta(\tau_{j,k}))}  \nonumber \\
&=& \sigma^{-1/2}_{1}(\tau_{j,l}) \sigma^{-1/2}_{1}(\tau_{j,k}) h^{p} \times
     \mbox{E} \left( \int a(i)a(j) dF(\bx,q) \int a(i)a(j) dF(\bx,q) \right. \nonumber \\
& &  \left. -2 \int a(i)a(j) dF(\bx,q) \int  a(i)c(j) dF(\bx,q)
   + \int a(i)a(j) dF(\bx,q) \int c(i) c(j) dF(\bx,q) \right. \nonumber \\
& & \left. -2 \int a(i)b(j)  dF(\bx,q) \int a(i)a(j)  dF(\bx,q)
   +4 \int a(i)b(j) dF(\bx,q) \int a(i)c(j) dF(\bx,q) \right. \nonumber \\
& & \left. -2 \int a(i)b(j) dF(\bx,q) \int c(i) c(j) dF(\bx,q)
    + \int b(i) b(j) dF(\bx,q) \int a(i)a(j) dF(\bx,q) \right. \nonumber \\
& & \left. -2 \int b(i) b(j) dF(\bx,q) \int a(i)c(j) dF(\bx,q)
 + \int b(i) b(j) dF(\bx,q) \int c(i) c(j) dF(\bx,q) \right) \nonumber \\
&=& [\sigma^{2}_{11}(\tau_{j,l}) + \sigma^{2}_{12}(\tau_{j,l})]^{-1/2}
    [\sigma^{2}_{11}(\tau_{j,k}) + \sigma^{2}_{12}(\tau_{j,k})]^{-1/2}  \varphi(\tau_{j,l},\tau_{j,k}) \label{Test-7}
\quad\blacksquare
\end{eqnarray}

\bigskip
\noindent
{\bf Proof of Theorem 6.}

\medskip
\noindent
Let
\begin{eqnarray*}
\delta_{\tau_{j}}(\bx,q)
&=& m_{\gamma_{j}}(\bx) I_{\gamma_{j}}(q) - m_{\gamma_{j-1},\tau}(\bx) I_{\gamma_{j-1},\tau}(q)
                     - m_{\tau,\gamma_{j}}(\bx) I_{\tau,\gamma_{j}}(q)  \\
s_{\tau_{j}}(Y_{i} , \bX_{i}, Q_{i} ; y, \bx,q)
&=& s_{1,\tau_{j}}(Y_{i} , \bX_{i}, Q_{i} ; y, \bx,q) + s_{2,\tau_{j}}(Y_{i} , \bX_{i}, Q_{i} ; y, \bx,q) \\
& &           + s_{3,\tau_{j}}(Y_{i} , \bX_{i}, Q_{i} ; y, \bx,q)
\end{eqnarray*}
with
\begin{eqnarray*}
s_{1,\tau_{j}}(Y_{i} , \bX_{i}, Q_{i} ; y, \bx,q)
&=& \frac{g_{\gamma_{j}}(\bx)}{f_{\gamma_{j}}(\bx)} \int \alpha_{\gamma_{j}}(y, \bx)  \bK_{h}(\bX_{i}- \bx)
    \bK_{h}(Y_{i}-y) dy \\
s_{2,\tau_{j}}(Y_{i} , \bX_{i}, Q_{i} ; y, \bx,q)
&=& \frac{g_{\gamma_{j-1},\tau}(\bx)}{f_{\gamma_{j-1},\tau}(\bx)} \int \alpha_{\gamma_{j-1},\tau}(y, \bx)
   \bK_{h}(\bX_{i}-\bx) \bK_{h}(Y_{i}-y) dy \\
s_{3,\tau_{j}}(Y_{i} , \bX_{i}, Q_{i} ; y, \bx,q)
&=& \frac{g_{\tau,\gamma_{j}}(\bx)}{f_{\tau,\gamma_{j}}(\bx)} \int \alpha_{\tau,\gamma_{j}}(y, \bx)
   \bK_{h}(\bX_{i}-\bx) \bK_{h}(Y_{i}-y) dy.
\end{eqnarray*}
It is clear that $\int \int |\delta_{\tau_{j}}(\bx,q)| d\bx dq \neq 0$ when the alternative hypothesis is true.

As in the proof of Theorem 4, we know that
\begin{eqnarray*}
 \Gamma(\hat{f}_{\gamma_{j}},\hat{f}_{\gamma_{j-1},\tau_{j}},\hat{f}_{\tau_{j},\gamma_{j}},F) & \\
 & \hspace{-2cm}=
\Gamma(f_{\gamma_{j}},f_{\gamma_{j-1},\tau_{j}},f_{\tau_{j},\gamma_{j}},F) + \Psi^{(1)}(0) + 1/2 \Psi^{(2)}(0) + 1/6 \Psi^{(3)}(t^{*})
\end{eqnarray*}
where
\begin{eqnarray*}
\Psi^{(1)}(0)
&=&  2\int \int \psi(t) \frac{ \partial  \psi(t)}{\partial t} a(\bx)dF(\bx,q) \\
\Psi^{(2)}(t)
&=& 2\int \int \left\{ \psi(t) \frac{ \partial^{2}  \psi(t)}{\partial t^{2}}
    + \left[ \frac{ \partial  \psi(t)}{\partial t} \right]^{2}  \right \}  a(\bx)dF(\bx,q).
\end{eqnarray*}
It is clear that  $\psi(t) = 0$ under the null and $\psi(t) \neq 0$ under the alternative. Then, under the alternative,
\begin{eqnarray*}
\Psi^{(1)}(0) &=& \frac{1}{n} \sum^{n}_{i=1} \int  \delta_{\tau_{j}}(\bx,q)
  r_{\tau_{j}}(Y_{i} , \bX_{i}, Q_{i} ; y, \bx,q) a(\bx) dF(\bx,q) \\
              &=& O(h^{r}) + O_{p}((nh)^{-1/2})
\end{eqnarray*}
and
\begin{eqnarray*}
& & \int \int \psi(t) \frac{ \partial^{2}  \psi(t)}{\partial t^{2}} a(\bx) dF(\bx,q) \\
&=& \frac{1}{n} \sum^{n}_{i=1}  \int \delta_{\tau_{j}}(\bx,q)
    s_{\tau_{j}}(Y_{i} , \bX_{i}, Q_{i} ; y, \bx,q) a(\bx) dF(\bx,q) \\
&\leq&  O_{p}(h^{r}+\ln(n)/(nh^{p})^{1/2}) [O(h^{r}) + O_{p}((nh)^{-1/2})]
\end{eqnarray*}
Given the following results in the proof of Theorem 4,
\begin{eqnarray*}
\left[ \frac{ \partial  \psi(t)}{\partial t} \right]^{2} &=& O((nh^{p})^{-1}) + O_{p}(n^{-1}h^{-p/2}) \\
\Psi^{(3)}(t^{*}) &=& O(||\hat{f}_{\gamma_{j}} - f_{\gamma_{j}}||^{3} + ||\hat{f}_{\gamma_{j-1},\tau_{j}} - f_{\gamma_{j-1},\tau_{j}}||^{3}
 + ||\hat{f}_{\tau_{j},\gamma_{j}} - f_{\tau_{j},\gamma_{j}}||^{3} ),
\end{eqnarray*}
we have
\begin{eqnarray*}
\Gamma(\hat{f}_{\gamma_{j}},\hat{f}_{\gamma_{j-1},\tau_{j}},\hat{f}_{\tau_{j},\gamma_{j}},F) & \\
 & \hspace{-3.6cm}=
\Gamma(f_{\gamma_{j}},f_{\gamma_{j-1},\tau_{j}},f_{\tau_{j},\gamma_{j}},F) + \Psi^{(1)}(0) + 1/2 \Psi^{(2)}(0) + 1/6 \Psi^{(3)}(t^{*}) \\
&\hspace{-3cm}= O(1) + [O(h^{r}) + O_{p}((nh^{p})^{-1/2})] + [ O((nh^{p})^{-1}) + O_{p}(n^{-1}h^{-p/2}) ]  \\
& \hspace{-2.3cm} + O_{p}(h^{r}+\ln(n)/(nh^{p})^{1/2}) [O(h^{r}) + O_{p}((nh)^{-1/2})] + 1/6 \Psi^{(3)}(t^{*}).
\end{eqnarray*}
Therefore, under the alternative,
\begin{eqnarray*}
\sigma^{-1}_{1}(\tau_{j}) \{ nh^{p/2}\tilde{\Gamma}(\tau_{j}) - h^{-p/2} \xi_{1}(\tau_{j}) \}
 &\\
 & \hspace{-4cm}= \sigma^{-1}_{1}(\tau_{j}) \{ nh^{p/2}
  [\Gamma(f_{\gamma_{j}},f_{\gamma_{j-1},\tau_{j}},f_{\tau_{j},\gamma_{j}},F) + o(1) ]\} \to \infty.
\end{eqnarray*}
When the alternative converges to the null at speed $n^{-1/2}h^{-p/4}$,
we get
$nh^{p/2} \int \int \psi(t) \frac{ \partial  \psi(t)}{\partial t}   a(x)dF(x,q) = [O((n h^{2/p + 2r})^{1/2} )
+ O_{p}((nh)^{-1/2})] = o_{p}(1)$. Similarly, we have
$nh^{p/2} \int \int \psi(t) \frac{ \partial^{2}  \psi(t)}{\partial t^{2}} a(x) dF(x,q) = o_{p}(1)$.
Hence, from  Proposition 2 of  A\"{\i}t Sahalia et~al.~(2001), we have proved Theorem 6. \quad$\blacksquare$

\bigskip
\noindent
{\bf Proof of Theorem 7.}

\medskip
\noindent
Observe that the indicator functions defined on distinct intervals are mutually exclusive.
Therefore the asymptotic covariance between the statistics
$\delta(\tau_{j,l_{1}})$
and $\delta(\tau_{k,l_{2}})$  ($l \neq k$) is zero.
In what follows, we verify this fact.
Let $\tau_{j,l_{1}}$ and $\tau_{k,l_{2}}$, respectively
be the  $l_{1}$ and $l_{2}$ splitting points in the intervals of 
$[\gamma_{j-1}, \gamma_{j})$ and $[\gamma_{k-1}, \gamma_{k})$;
also let $l \neq k$.
Following the proof of Theorem 4, we have
\begin{eqnarray*}
\delta(\tau_{j,l_{1}}) &=& \sigma^{-1/2}_{1}(\tau_{j,l_{1}}) nh^{p/2} I_{n1}(\tau_{j,l_{1}}) + o \left( (nh^{p/2})^{-1} \right),
\end{eqnarray*}
where
\begin{eqnarray*}
I_{n1}(\tau_{j,l_{1}}) &=& \sum \sum \int \tilde{r}_{\tau_{j,l_{1}}}(Y_{i} , \bX_{i},Q_{i} , y,\bx,q)
                    \times \tilde{r}_{\tau_{j,l_{1}}}(Y_{j} , \bX_{j},Q_{j} , y,\bx,q) dF(\bx,q).
\end{eqnarray*}
As those defined in  Theorem 4,
\begin{eqnarray*}
& & \tilde{r}_{\tau_{j,l_{1}}}(Y_{j} , \bX_{j},Q_{j} , y,\bx,q) \\
 &  & \hspace{2cm}= r_{\tau_{j,l_{1}}}(Y_{i} , \bX_{i},Q_{i} , y,\bx,q)
     - \mbox{E} \left[ r_{\tau_{j,l_{1}}}(Y_{i} , \bX_{i},Q_{i} , y,\bx,q) \right], \\
\mbox{and}\\
& & \lefteqn{r_{\tau_{j,l_{1}}}(Y_{i} , \bX_{i},Q_{i} , y,\bx,q)} \\
&&  \hspace{1cm}=\left\{  \int \alpha_{\gamma_{j}}(y,\bx) K_{h}(\bX_{i} - \bx) I_{\gamma_{j}}(Q_{t}) K_{h}(Y_{i} - y) dy I_{\gamma_{j}}(q) \right. \\
& & \hspace{1cm}- \int \alpha_{\gamma_{j-1},\tau_{j,l_{1}}}(y,\bx)  K_{h}(\bX_{i} - \bx) I_{\gamma_{j-1},\tau_{j,l_{1}}}(Q_{i}) K_{h}(Y_{i} - y) dy I_{\gamma_{j-1},\tau_{j,l_{1}}}(q) \\
& & \hspace{1cm}- \left. \int \alpha_{\tau_{j,l_{1}},\gamma_{j}}(y,\bx) K_{h}(\bX_{i} - \bx)
   I_{\tau_{j,l_{1}},\gamma_{j}}(Q_{i}) K_{h}(Y_{i} - y) dy I_{\tau_{j,l_{1}},\gamma_{j}}(q)\right\}.
\end{eqnarray*}
Following the proof of Theorem 5, we denote
\begin{eqnarray*}
a(i) &=& \int \alpha_{\gamma_{j}}(y,\bx) K_{h}(\bX_{i} - \bx) I_{\gamma_{j}}(Q_{t}) K_{h}(Y_{i} - y) dy I_{\gamma_{j}}(q) \\
b(i) &=& \int \alpha_{\gamma_{k}}(y,\bx) K_{h}(\bX_{i} - \bx) I_{\gamma_{k}}(Q_{t}) K_{h}(Y_{i} - y) dy I_{\gamma_{k}}(q) \\
c(i) &=&  \int \alpha_{\gamma_{j-1},\tau_{j,l_{1}}}(y,\bx)  K_{h}(\bX_{i} - \bx) I_{\gamma_{j-1},\tau_{j,l_{1}}}(Q_{i}) K_{h}(Y_{i} - y) dy I_{\gamma_{j-1},\tau_{j,l_{1}}}(q) \\
     & & +  \int \alpha_{\tau_{j,l_{1}},\gamma_{j}}(y,\bx) K_{h}(\bX_{i} - \bx) I_{\tau_{j,l_{1}},\gamma_{j}}(Q_{i}) K_{h}(Y_{i} - y) dy I_{\tau_{j,l_{1}},\gamma_{j}}(q)  \\
d(i) &=&  \int \alpha_{\gamma_{k-1},\tau_{k,l_{2}}}(y,\bx)  K_{h}(\bX_{i} - \bx) I_{\gamma_{j-1},\tau_{k,l_{2}}}(Q_{i}) K_{h}(Y_{i} - y) dy I_{\gamma_{j-1},\tau_{k,l_{2}}}(q) \\
     & & +  \int \alpha_{\tau_{k,l_{2}},\gamma_{j}}(y,\bx) K_{h}(\bX_{i} - \bx) I_{\tau_{k,l_{2}},\gamma_{j}}(Q_{i}) K_{h}(Y_{i} - y) dy I_{\tau_{k,l_{2}},\gamma_{j}}(q),
\end{eqnarray*}
and  obtain
\begin{eqnarray}
\lefteqn{\COV(\delta(\tau_{j,l_{1}}),\delta(\tau_{k,l_{2}}))}  \nonumber \\
&=& \sigma^{-1/2}_{1}(\tau_{j,l_{1}}) \sigma^{-1/2}_{1}(\tau_{k,l_{2}}) h^{p} \mbox{E}(I_{n1}(\tau_{j,l_{1}})I_{n1}(\tau_{k,l_{2}})) + o(1) \nonumber \\
&=&  \sigma^{-1/2}_{1}(\tau_{j,l_{1}}) \sigma^{-1/2}_{1}(\tau_{k,l_{2}}) h^{p} \times \nonumber  \\
& & \mbox{E} \left( \int [a(i) - c(i)][a(j) - c(j)] dF(\bx,q) \int [b(i) - d(i)][b(j) - d(j)]\, dF(\bx,q) \right) + o(1). \nonumber 
\nonumber
\end{eqnarray}
The equation above signifies that
the indicators
$a(i)$, $a(j)$, $c(i)$, and $c(j)$ are mutually exclusive;
$b(i)$, $b(j)$, $d(i)$, and $d(j)$ are also mutually exclusive.
Hence, 
$\COV(\delta(\tau_{j,l_{1}}),\delta(\tau_{k,l_{2}}))$ is of $o_p(1)$.
Further, 
$\delta(\tau_{j,l_{1}})$ and $\delta(\tau_{k,l_{2}})$
are asymptotically normally distributed,
and they thus can be seen as asymptotically independent.
Accordingly, 
with the same assumptions imposed in Theorem 5,
Theorem 7 holds. \quad$\blacksquare$.

\bigskip
\noindent
{\bf Proof of Theorem 8.}

\medskip
\noindent
With $s$ pseudo threshold values, 
$[\tau_{1},\ldots,\tau_{s}]$ in $[\gamma_0, \gamma_{s+1}]$,
the conditional mean estimator is constructed as 
\begin{eqnarray*}
\hat{m}_{\tau_{j}}(\bx) = \frac{\sum \bK_{h}(\bX_{i}-\bx) I_{\tau_{j}}(Q_{i})Y_{i}}{\sum \bK_{h}(\bX_{i}-\bx) I_{\tau_{j}}(Q_{i})}
\end{eqnarray*}
with
\begin{eqnarray*}
I_{\tau_{j}}(Q_{i}) &=& \left\{ \begin{array}{ll}
                                   1 & Q_{i} \in [ \tau_{j-1} , \tau_{j} ) , \\
                                   0 & \mbox{otherwise},
                                   \end{array} \right.
\end{eqnarray*}
and $\tau_0 = \gamma_0$, $\tau_{s+1} = \gamma_{s+1}$.

\medskip
To proceed, we need the following lemmas.
\begin{lemma}{\label{MultEst}}
For any $[\tau_{1},\ldots,\tau_{s}]$, we have
\begin{eqnarray*}
sup |\hat{m}_{\tau_{j}}(\bx) - m_{\tau_{j}}(\bx)| &=& O_{p}(h^{r} + (\ln(n))^{1/2}/(nh^{p})^{1/2})
\end{eqnarray*}
and
\begin{eqnarray}
m_{\tau_{j}}(\bx) &=& \int \sum^{s+1}_{j=1} m_{\gamma_{j}}(\bx) I_{\gamma_{j}}(q)
I_{\tau_{j}}(q) \frac{f(\bx,q)}{f_{\tau_{j}}(\bx)} dq. \quad\square  \label{Multresult1}
\end{eqnarray}
\end{lemma}

\noindent
\begin{proof}:
Since $\hat{m}_{\tau_{j}}(\bx)$ is a local constant estimator, its almost sure 
convergence rate is
$O_{p}(h^{r} + (\ln(n))^{1/2}/(nh^{p})^{1/2})$ from the result of part a) in Theorem 2.
From the definition of $m_{\tau_{j}}(\bx)$,
\begin{eqnarray*}
m_{\tau_{j}}(\bx) &=& \int y \frac{f_{\tau_{j}}(\bx,y)}{f_{\tau_{j}}(\bx)} dy \\
                &=& \int \int y \frac{f(\bx,y,q)}{f(\bx,q)} dy I_{\tau_{j}}(q) \frac{f(\bx,q)}{f_{\tau_{j}}(\bx)} dq  \\
&=& \int \sum^{s+1}_{j=1} m_{\gamma_{j}}(\bx) I_{\gamma_{j}}(q) I_{\tau_{j}}(q) \frac{f(\bx,q)}{f_{\tau_{j}}(\bx)} dq. \quad\blacksquare
\end{eqnarray*}
\end{proof}

\begin{lemma}{\label{exogenous}}
Under the condition that $\bX$ and $Q$ are exogenous, we have
\begin{eqnarray*}
\frac{1}{n} \sum^{n}_{i=1} g(\bX_{i},Q_{i})e_{i} = o_{p}(1). \quad\square
\end{eqnarray*}
\end{lemma}

\noindent
\begin{proof}:
The second moment of $g(\bX_{i},Q_{i})e_{i}$ exists, that is
\begin{eqnarray*}
\int g^{2}(\bx_{i},q_{i}) \sigma^{2}(\bx,q) dF(\bx,q) < \infty.
\end{eqnarray*}
Since $E[g(\bX_{i},Q_{i})e_{i}] = 0$ for $\bX$ and $Q$ being exogenous, from the law of large number, we have
\begin{eqnarray*}
\frac{1}{n} \sum^{n}_{i=1} g(\bX_{i},Q_{i})e_{i} \to E[g(\bX_{i},Q_{i})e_{i}] = 0. \quad\blacksquare
\end{eqnarray*}
\end{proof}

Let $d_{s,j}(\bX_{i},Q_{i}) = \hat{m}_{\tau_{j}}(\bX_{i}) I_{\tau_{j}}(Q_{i})
- m_{\gamma_{j}}(\bX_{i})  I_{\gamma_{j}}(Q_{i})$ and
$G_{\bX,Q}(\tau_{1},\ldots,\tau_{s}) = E \left[\sum^{s+1}_{j=1} d_{s,j}(\bX_{i},Q_{i}) \right]^{2}$.
The estimated sum of squared residuals at threshold values $[\tau_{1},\ldots,\tau_{s}]$ is
\begin{eqnarray*}
\frac{1}{n} \sum^{n}_{i=1} \hat{e}^{2}_{i} (\tau_{1},\ldots,\tau_{s})
&=& \frac{1}{n} \sum^{n}_{i=1} \left\{ e^{2}_{i} - 2 \sum^{s+1}_{j=1} d_{s,j}(\bX_{i},Q_{i}) e_{i}
   + \left[\sum^{s+1}_{j=1} d_{s,j}(\bX_{i},Q_{i}) \right]^{2}  \right\}  \\
& & \hspace{0.6cm}\to^{p} E(e^{2}_{i}) + G_{\bX,Q}(\tau_{1},\ldots,\tau_{s}) = H(\tau_{1},\ldots,\tau_{s}) 
\end{eqnarray*}
with
\begin{eqnarray*}
\frac{1}{n} \sum^{n}_{i=1} e^{2}_{i}
&\to^p& E(e^{2}_{i}) \\
\frac{1}{n} \sum^{n}_{i=1} \sum^{s+1}_{j=1} d_{s,j}(\bX_{i},Q_{i}) e_{i}
& \to^{p}& 0 \hspace{10mm} \mbox{(by\, Lemma 5)}  \\
\frac{1}{n} \sum^{n}_{i=1} \left[\sum^{s+1}_{j=1} d_{s,j}(\bX_{i},Q_{i}) \right]^{2}
&\to^{p}& E \left[\sum^{s+1}_{j=1} d_{s,j}(\bX_{i},Q_{i}) \right]^{2} = G_{\bX,Q}(\tau_{1},\ldots,\tau_{s}). \\
 \nonumber
\end{eqnarray*}
Moreover,
\begin{eqnarray*}
& & G_{\bX,Q}(\tau_{1},\ldots,\tau_{s}) \\
&=& E \left[\sum^{s+1}_{j=1} d_{s,j}(\bX_{i},Q_{i}) \right]^{2} \\
&=& E \left\{ \sum^{s+1}_{j=1} \left\{ [m_{\tau_{j}}(\bX_{i}) -  m_{\gamma_{j}}(\bX_{i})] I_{\gamma_{j}}(Q_{i})
    + m_{\tau_{j}}(\bX_{i})  [I_{\tau_{j}}(Q_{i}) - I_{\gamma_{j}}(Q_{i})] \right\} \right\}^{2}
    + O\left(\frac{1}{nh^{p}}\right).
\end{eqnarray*}
It is clear that, from Lemma 4,  $G_{\bX,Q}(\gamma_{1},\ldots,\gamma_{s})$ and $H(\tau_{1},\ldots,\tau_{s})$
have their minimum at $\tau_{j} = \gamma_{j} , \forall j \in 1, \ldots ,s$. According to Theorem 2.1 of
Newey and McFadden (1994), we then have
\begin{eqnarray*}
[\hat{\gamma}_{1}, \ldots \hat{\gamma}_{s}]
= \arg \min \frac{1}{n} \sum^{n}_{i=1} \hat{e}^{2}_{i} (\tau_{1},\ldots,\tau_{s})
\to^p \arg \min H(\tau_{1},\ldots,\tau_{s}) = [\gamma_{1}, \ldots, \gamma_{s}].\hspace{5mm}
\end{eqnarray*}
This is the proof of part a) of Theorem 8. $\quad\blacksquare$

\bigskip
For the proof of part b) of Theorem 8, without loss of generality, we provide the proof of
$\hat\gamma_2 \to^p \gamma_2$ in a nonparametric regression with three thresholds. Denote
\begin{eqnarray*}
G_{n,2,3}(\tau_{2},\gamma_{2}) &=& \sum^{n}_{i=1} c^{2}_{2,3}(\bX_{i}) I_{\gamma_{2},\tau_{2}}(Q_{i})  \\
J_{n,2,3}(\tau_{2}) - J_{{n,2,3}}(\gamma_{2})
&=& \frac{1}{\sqrt{n}} c_{2,3}(\bX_{i}) e_{i} I_{\tau_{2},\gamma_{2}} (Q_{i}),
\end{eqnarray*}
where $c_{2,3}(\bX_{i}) = m_{\gamma_{2}}(\bX_{i}) - m_{\gamma_{3}}(\bX_{i})$.

\medskip
The following lemmas are needed for our proof.
\begin{lemma}{\label{Bound-1}}
Set  $\bar{v} = \frac{8K}{\eta^{2}d^{2}_{1}(1 - 1/b) \epsilon}$ and
\begin{eqnarray*}
d_{1} &=& \min_{\tau \in R} \mbox{E}(c^{2}_{2,3}(\bX_{i})|Q_{i} = \tau) f(\tau) > 0 \\
d_{2} &=& \max_{\tau \in R} |\mbox{E}(c_{2,3}(\bX_{i})|Q_{i} = \tau)| f(\tau) >0 \\
d_{3} &=& \max_{\tau \in R} f(\tau) >0 .
\end{eqnarray*}
There exist the constants $B>0$, $0<d_{1},d_{2},d_{3}< \infty$,  and $0<c< \infty$, such that for all $\eta>0$ and $\epsilon > 0$,
there exists a $\bar{v} < \infty $ such that for all $n$,
\begin{eqnarray*}
& & P \left( \inf_{\bar{v}/ n \leq |\tau_{2} - \gamma_{2}| \leq B} \frac{G_{n,2,3}(\tau_{2},\gamma_{2})}{n|\tau_{2} - \gamma_{2}|} <(1 - \eta) d_{1}  \right) \leq \epsilon, \\
& & P \left( \sup_{\bar{v}/ n \leq |\tau_{2} - \gamma_{2}| \leq B} \frac{ \sum^{n}_{i=1} | c_{2,3}(\bX_{i}) | | I_{\gamma_{2},\tau_{2}}(Q_{i}) |}{ n |\tau_{2} - \gamma_{2}|} >(1 + \eta) d_{2}  \right) \leq \epsilon, \\
& & P \left( \sup_{\bar{v}/ n \leq |\tau_{2} - \gamma_{2}| \leq B} \frac{ \sum^{n}_{i=1} | I_{\gamma_{2},\tau_{2}}(Q_{i})|}{ n | \tau_{2} - \gamma_{2}|} >(1 + \eta) d_{3}  \right) \leq \epsilon. \quad\square
\end{eqnarray*}
\end{lemma}

\noindent
\begin{proof}:
See Lemma A.7 of Hansen (2000). \quad$\blacksquare$
\end{proof}

\begin{lemma}{\label{Bound-2}}
For all $\eta >0$ and $\epsilon >0 $, there exists some $\bar{v} < \infty$ such
that for any $B < \infty$,
\begin{eqnarray*}
& & P \left( \sup_{\bar{v}/ n \leq |\tau_{2} - \gamma_{2}| \leq B} \frac{|J_{n,2,3}(\tau_{2}) - J_{n,2,3}(\gamma_{2})|}{ \sqrt{n} |\tau_{2} - \gamma_{2}|} < \eta  \right) \leq \epsilon, \\
& & P \left( \sup_{\bar{v}/ n \leq |\tau_{2} - \gamma_{2}| \leq B} \frac{| \sum^{n}_{i=1} |I_{\gamma_{2},\tau_{2}}(Q_{i})| e_{i} |}{ \sqrt{n} |\tau_{2} - \gamma_{2}|} < \eta  \right) \leq \epsilon. \quad\square
\end{eqnarray*}
\end{lemma}
\noindent
\begin{proof}:
See Lemma A.8 of Hansen (2000). \quad$\blacksquare$
\end{proof}

\medskip
Let $E_{n}$ be the intersection sets of
$\max (|\hat{\gamma}_{1} - \gamma_{1}|, |\hat{\gamma}_{2} - \gamma_{2}|, |\hat{\gamma}_{3} - \gamma_{3}|) \leq B $
and $\sup |\hat{c}_{2,3}(\bX_{i}) - c_{2,3}(\bX_{i})| \leq \kappa$.
From Lemmas $\ref{Bound-1}$ and  $\ref{Bound-2}$, we have
\begin{eqnarray*}
& & \inf_{\bar{v}/ n \leq |\tau_{2} - \gamma_{2}| \leq B} \frac{G_{3,2,n}(\tau_{2})}{|\tau_{2} - \gamma_{2}|} >(1 - \eta) d_{1}  , \\
& & \sup_{\bar{v}/ n \leq |\tau_{2} - \gamma_{2}| \leq B} \frac{ \sum^{n}_{i=1} | c_{2,3}(\bX_{i}) | | I_{\gamma_{2},\tau_{2}}(Q_{i})|}{|\tau_{2} - \gamma_{2}|} <(1 + \eta) d_{2}, \\
& & \sup_{\bar{v}/ n \leq |\tau_{2} - \gamma_{2}| \leq B} \frac{ \sum^{n}_{i=1} | I_{\gamma_{2},\tau_{2}}(Q_{i}) |}{|\tau_{2} - \gamma_{2}|} <(1 + \eta) d_{3}, \\
& & \sup_{\bar{v}/ n \leq |\tau_{2} - \gamma_{2}| \leq B} \frac{|J_{n,2,3}(\tau_{2}) - J_{n,2,3}(\gamma_{2})|}{ \sqrt{n} |\tau_{2} - \gamma_{2}|} < \eta , \\
& & \sup_{\bar{v}/ n \leq |\tau_{2} - \gamma_{2}| \leq B} \frac{| \sum^{n}_{i=1} |I_{\gamma_{2},\tau_{2}}(Q_{i})| e_{i} |}{ \sqrt{n} |\tau_{2} - \gamma_{2}|} < \eta.
\end{eqnarray*}
Take $\eta$ and $\kappa$ to be sufficiently  small  such that
\begin{eqnarray*}
(1 - \eta) d_{1} - 2 \eta - 2 \kappa \eta - 2 \kappa (1 + \eta) d_{2} - 2 \kappa^{2} (1 + \eta) d_{3}
 - 2 \kappa^{2} (1 + \eta) d_{3}
    - 2 \kappa (1 + \eta) d_{2} \geq 0.
\end{eqnarray*}
We thus have
\begin{eqnarray*}
& &  \frac{SSR(\tau_{1},\tau_{2}.\tau_{3}) - SSR(\tau_{1},\gamma_{2},\tau_{3})}{n(\tau_{2} - \gamma_{2})} \\
&=& \frac{1}{n(\tau_{2} - \gamma_{2})} \sum^{n}_{i=1} \left\{ [ m_{\gamma_{3}}(\bX_{i}) - m_{\gamma_{2}}(\bX_{i})]^{2}
    I_{\gamma_{2},\tau_{2}}(Q_{i}) \right. \\
& & + \{ [m_{\gamma_{3}}(\bX_{i}) - m_{\gamma_{2}}(\bX_{i})] - [\hat{m}_{\hat{\gamma}_{3}}(\bX_{i}) - \hat{m}_{\hat{\gamma}_{2}}(\bX_{i})] \} I_{\gamma_{2},\tau_{2}}(Q_{i}) \\
& & \times \{ [m_{\gamma_{3}}(\bX_{i}) - m_{\gamma_{2}}(\bX_{i})] + [\hat{m}_{\hat{\gamma}_{3}}(\bX_{i}) - \hat{m}_{\hat{\gamma}_{2}}(\bX_{i})] \} \\
& & - 2 [\hat{m}_{\hat{\gamma}_{3}}(\bX_{i}) - \hat{m}_{\hat{\gamma}_{2}}(\bX_{i})] I_{\gamma_{2},\tau_{2}}(Q_{i}) e_{i} \\
& & \left. + 2 [\hat{m}_{\hat{\gamma}_{3}}(\bX_{i}) - \hat{m}_{\hat{\gamma}_{2}}(\bX_{i})]
        [\hat{m}_{\hat{\gamma}_{2}}(\bX_{i}) - m_{\gamma_{2}}(\bX_{i})] I_{\gamma_{2},\tau_{2}}(Q_{i}) \right\}.
\end{eqnarray*}
Therefore,
\begin{eqnarray*}
& &  \frac{SSR(\tau_{1},\tau_{2}.\tau_{3}) - SSR(\tau_{1},\gamma_{2},\tau_{3})}{n(\tau_{2} - \gamma_{2})} \\
& \geq & \frac{G_{n,2,3}(\gamma_{2},\tau_{2})}{ n |\tau_{2} - \gamma_{2}|} \\
& & - 2 \frac{|J_{n,2,3}(\tau_{2}) - J_{n,2,3}(\gamma_{2})|}{ \sqrt{n} |\tau_{2} - \gamma_{2}|} \\
& & - 2 \frac{  \sum^{n}_{i=1} |\hat{c}_{3,2}(\bX_{i}) - c_{3,2}(\bX_{i}) |  I_{\gamma_{2},\tau_{2}}(Q_{i})  e_{i}  }{\sqrt{n} | \tau_{2} - \gamma_{2} |} \\
& & - 2 \frac{  \sum^{n}_{i=1} n^{-\alpha} |\hat{m}_{\hat{\gamma_{2}}}(\bX_{i}) - m_{\gamma_{2}}(\bX_{i})|  | c_{3,2}(\bX_{i})| |I_{\gamma_{2},\tau_{2}}(Q_{i})|}{ n | \tau_{2} - \gamma_{2} |} \\
& & - 2 \frac{  \sum^{n}_{i=1} n^{-\alpha} |\hat{m}_{\hat{\gamma}_{2}}(\bX_{i}) - m_{\gamma_{2}}(\bX_{i}) | | \hat{c}_{3,2}(\bX_{i}) - c_{3,2}(\bX_{i})  | | I_{\gamma_{2},\tau_{2}}(Q_{i}) ||}{n | \gamma_{2} - \gamma_{2} |} \\
& & -  \frac{  \sum^{n}_{i=1} [ \hat{c}_{3,2}(\bX_{i}) - c(\bX_{i})  ]^{2}  | I_{\gamma_{2},\tau_{2}}(Q_{i}) |}{n | \tau_{2} - \gamma_{2} |} \\
& & - 2 \frac{  \sum^{n}_{i=1} | \hat{c}_{3,2}(\bX_{i}) - c(\bX_{i})  | |c_{3,2}(\bX_{i}) | | I_{\gamma_{2},\tau_{2}}(Q_{i}) |}{n | \tau_{2} - \gamma_{2} |} \\
& \geq& (1 - \eta) d_{1} - 2 \eta - 2 \kappa \eta - 2 \kappa (1 + \eta) d_{2} 
   \\
 & &   \hspace{2cm} - 2 \kappa^{2} (1 + \eta) d_{3}  - 2 \kappa^{2} (1 + \eta) d_{3}   - 2 \kappa (1 + \eta) d_{2} \geq 0.
\end{eqnarray*}
This result indicates, in the event $E_{n}$, $SSR(\tau_{1},\tau_{2},\tau_{3}) - SSR(\tau_{1},\gamma_{2},\tau_{3}) >0$
when $ \tau_{2} \in [\gamma_{2} + \bar{v}/n , \gamma_{2} + B]$ and when
$ \tau_{2} \in [\gamma_{2} - B,\gamma_{2} - \bar{v}/n ]$.
However, this contradicts the fact that $SSR(\tau_{1},\tau_{2},\tau_{3}) - SSR(\tau_{1},\gamma_{2},\tau_{3}) \leq 0$.
Therefore, the foregoing analysis 
implies $|\hat{\gamma}_{2} - \gamma_{2}| \leq \bar{v}/n$,
and then,  $P(E_{n}) \geq 1 - \epsilon$ for $n \leq \bar{n}$. This is equivalent to
$P(n|\hat{\gamma}_{2} - \gamma_{2}| > \bar{v} ) \leq \epsilon$ for $n \geq \bar{n}$. \quad$\blacksquare$

\bigskip
\noindent
{\bf Proof of Theorem 9.}

\medskip
\noindent
The following lemmas are necessary
for proving Theorem 9.

\begin{lemma}{\label{CovSpeed-2}}
Given the existence of the small effect, $\delta_{n,l,k}(\bX_{i}) = a_{n} c_{l,k}(\bX_{i}) \to 0$,
\begin{eqnarray*}
a_{n} ( \hat{\gamma}_{j} - \gamma_{j} ) = O_{p}(1)
\end{eqnarray*}
where $a_{n} = n^{1-2 \alpha}$. \quad$\square$
\end{lemma}

\noindent
\begin{proof}:
The proof is similar to the one in part b) of Theorem 8. \quad$\blacksquare$
\end{proof}

\medskip
Let us fix some new notations before introducing a new lemma.
\begin{eqnarray*}
\mu_{2} &:=& \mbox{E} [ c^{*2}_{2,3}(\bX_{i}) | Q_{i} = \gamma_{2} ]  \\
\lambda_{2}   &:=& \mbox{E}[ c^{*2}_{2,3}(\bX_{i}) e^{2}_{i} |Q_{i} = \gamma_{2} ].
\end{eqnarray*}
\begin{lemma}{\label{mean}}
Let $ G_{n,2,3}(v) = a_{n} \sum^{n}_{i=1} c^{*2}_{2,3}(\bX_{i}) d_{2,i}(v) $ and $d_{2,i}(v) = I_{\gamma_{2}+v/a_{n},\gamma_{j}}(Q_{i})$. We then have 
\begin{eqnarray*}
& &  G_{n,2,3}(v) \to^{p} \mu_{2} |v|. \quad\square
\end{eqnarray*}
\end{lemma}

\noindent
\begin{proof}: Since
\begin{eqnarray*}
\mbox{E}(G_{n,2,3}(v)) &=& v [ \mbox{E}(c^{*2}_{2,3}(\bX_{i}) d_{2,i}(v))] / (v/a_{n}) \\
                   &=& v f_{q}(\gamma_{2}) \mbox{E}[c^{*2}_{2,3}(\bX_{i})|q_{i} = \gamma_{2}]
\end{eqnarray*}
from Lemma A.2 of Hansen (2000)
\begin{eqnarray*}
\mbox{V}(G_{n,2,3}(v)) &=& \mbox{E}[G_{n,2,3}(v) - \mbox{E}(G_{n,2,3}(v)) ]^{2} \\
                   &\leq& \frac{a^{2}_{n}}{n} D |\frac{v}{a_{n}}| = D|v|n^{-2 \alpha} \to 0.
\end{eqnarray*}
Therefore, $G_{n,2,3}(v)(v) \to^{p} \mu_{2} |v|$ according to Chebyshev's inequality. \quad$\blacksquare$
\end{proof}

\bigskip
Let
$R_{n,2,3}(v) = \frac{\sqrt{a_{n}}}{\sqrt{n}} \sum^{n}_{i=1} c^*_{2,3}(\bX_{i}) e_{i} d_{2,i}(v) $. We have
the following functional central limit theorem:
\begin{lemma}{\label{variance}}
\begin{eqnarray*}
& & R_{n,2,3}(v) \to^{d} \sqrt{\lambda_{2}} B(v)
\end{eqnarray*}
and $B(v)$ is a standard Brownian motion. \quad$\square$
\end{lemma}

\noindent
\begin{proof}:
The variance of $R_{n,2,3}(v)$ is
\begin{eqnarray*}
 V_n [R_{n,2,3}(v)] &=& a_{n} \left\{ E[ c_{2,3}^{2}(\bX_{i}) e^{2}_{i} d_{2,i}(v)] \right.\\
 & & \left.   + \sum^{M(n)-1}_{l=1} 2 \frac{n-l}{n} E[c_{2,3}(\bX_{1}) c_{2,3}(\bX_{1+l}) e_{1} e_{1+l} d_{2,1}(v) d_{2,1+l}(v)] \right. \\
& & \left. + \sum^{n-1}_{l=M(n)} 2 \frac{n-l}{n} E[c_{2,3}(\bX_{1}) c_{2,3}(\bX_{1+l}) e_{1} e_{1+l} d_{2,1}(v) d_{2,1+l}(v)] \right\} \\
&=& V_{1n} + V_{2n} + V_{3n}.
\end{eqnarray*}
For any $M(n) \to \infty$ satisfying $M(n)/a_{n} \to 0$,
$V_{1n}$ is
\begin{eqnarray}
V_{1n} &=& v \frac{\mbox{E}(c_{2,3}^{2}(\bX_{i}) I_{\gamma_{2}+v/a_{n}}(Q_{i}) e^{2}_{i}) - \mbox{E}(c_{2,3}^{2}(\bX_{i})
       I_{\gamma_{2}}(Q_{i}) e^{2}_{i})}{v/a_{n}} \nonumber \\
       &\to& \frac{\partial \mbox{E}(c_{2,3}^{2}(\bX_{i}) I_{\gamma}(Q_{i}) e^{2}_{i}|Q_{i}= \gamma ) f(\gamma)}{\partial \gamma}
       |_{Q_{i}=\gamma_{2}} \nonumber \\
       &=& v \mbox{E}(c_{2,3}^{2}(\bX_{i}) e^{2}(\bX_{i})|q_{i} = \gamma_{2}) f_{q}(\gamma_{2})
       = v \lambda_2. \label{gammavar1}
\end{eqnarray}
Furthermore, let
\begin{eqnarray*}
{\bar D}_1 &=& \max_{l \in [1, \ldots , M(n)-1]}   E[c_{2,3}(\bX_{1}) c_{2,3}(\bX_{1+l}) e_{1} e_{1+l} | Q_{1} = \gamma_{2} , Q_{1+l} = \gamma_{2}] < \infty.
\end{eqnarray*}
We then have
\begin{eqnarray}
V_{2n}
&=& \frac{v^{2}}{a_{n}} \sum^{M(n)-1}_{l=1} 2 \frac{n-l}{n} E[c_{2,3}(\bX_{1}) c_{2,3}(\bX_{1+l}) e_{1} e_{1+l} | Q_{1} = \gamma_{2} , Q_{1+l} = \gamma_{2}]
 f_{Q_{1},Q_{1+l}}(\gamma_{2},\gamma_{2})  \nonumber \\
&\leq& 2 \frac{v^{2}}{a_{n}} M(n)\, {\bar D}_1\, f^{2}(\gamma_{2}) = o(1). \label{gammavar2}
\end{eqnarray}
From Lemma A.0 of Fan and Li~(1999), it can then be seen that
\begin{eqnarray*}
& & \hspace{-2cm} E[c_{2,3}(\bX_{1}) c_{2,3}(\bX_{1+l}) e_{1} e_{1+l} | Q_{1} = \gamma_{2} , Q_{1+l} = \gamma_{2}] \\
&=& E[c_{2,3}(\bX_{1}) e_{1} | Q_{1} = \gamma_{2} ]E[c_{2,3}(\bX_{1+l}) e_{1+l} | Q_{1+l} = \gamma_{2}] \\
& &  \quad+ \{ E[c_{2,3}(\bX_{1}) c_{2,3}(\bX_{1+l}) e_{1} e_{1+l} | Q_{1} = \gamma_{2} , Q_{1+l} = \gamma_{2}] \\
& & \quad - E[c_{2,3}(\bX_{1}) e_{1} | Q_{1} = \gamma_{2} ]E[c_{2,3}(\bX_{1+l}) e_{1+l} | Q_{1+l} = \gamma_{2}]\} \\
&\leq& 0 + 4 {\bar D}_2^{1/(1+\delta)} \beta^{\delta/(1+\delta)}(l),
\end{eqnarray*}
where
\begin{eqnarray*}
{\bar D}_2 = sup_{l \in [M(n), \ldots , \infty]} \{ E[| c_{2,3}(\bX_{1}) e_{1} c_{2,3}(\bX_{1+l}) e_{1+l} | Q_{1} = \gamma_{2} , Q_{1+l} = \gamma_{2} |^{1+\delta}], \\
\int \int | c_{2,3}(\bX_{1}) e_{1}\, c_{2,3}(\bX_{1+l}) e_{1+l} |^{1+\delta} Q(\bx_{1},e_{1}|q_{1}=\gamma_{2}) R(\bx_{1+l},e_{1+l}|q_{1+l}=\gamma_{2}) \}.
\end{eqnarray*}
In addition,
\begin{eqnarray}
V_{3n}
&=& a_{n} \sum^{n-1}_{l=M(n)} 2 \frac{n-l}{n} E[c_{2,3}(\bX_{1}) c_{2,3}(\bX_{1+l}) e_{1} e_{1+l} d_{2,1}(v) d_{2,1+l}(v)] \nonumber \\
 f_{Q_{1},Q_{1+l}}(\gamma_{2},\gamma_{2}) \nonumber \\
      &\leq& 8 \frac{v^{2}}{a_{n}} {\bar D}_2^{1/(1+\delta)} \sum^{\infty}_{l=1} l^{2}\beta^{\delta/(1+\delta)}(l) = o(1).  \label{gammavar3}
\end{eqnarray}
By combining ($\ref{gammavar1}$), ($\ref{gammavar2}$), and ($\ref{gammavar3}$), we have
\begin{eqnarray}
V_{n}[R_{n,2,3}(v)] = \lambda_{2} v + o(1) \label{gammavar}.
\end{eqnarray}
Next, the big block and small block method is used to derive the asymptotic normality of $R_{n,2,3}(v)$.
Let $s_{n}$ and $l_{n}$ satisfy
\begin{eqnarray*}
\frac{s_{n}}{l_{n}} \to 0, \ \  \frac{l_{n}}{n} \to 0 , \ \  \frac{l_{n}}{(nh)^{1/2}} \to 0 , \ \ \frac{n}{l_{n}} \alpha(s_{n}) \to 0,
\end{eqnarray*}
where $\alpha$ is the mixing coefficient of $(Y_i, \bX_i, Q_i)$.
Denote
\begin{eqnarray*}
\zeta_{j} &=& \sum^{j(s_{n}+l_{n})+r-1}_{i = j(s_{n}+l_{n})}  \frac{\sqrt{a_{n}}}{\sqrt{n}} \sum^{n}_{i=1}  c_{2,3}(\bX_{i}) e_{i} d_{2,i}(v), \\
\eta_{j} &=& \sum^{(j+1)(s_{n}+l_{n})}_{i = j(s_{n}+l_{n})+r}  \frac{\sqrt{a_{n}}}{\sqrt{n}} \sum^{n}_{i=1} m(\bX_{i}) e_{i} d_{2,i}(v), \\
\xi &=& \sum^{n-1}_{k_{n}(s_{n}+l_{n})} \frac{\sqrt{a_{n}}}{\sqrt{n}} \sum^{n}_{i=1}  c_{2,3}(\bX_{i}) e_{i} d_{2,i}(v),
\end{eqnarray*}
where $k_n = [\frac{n}{s_{n} + l_{n}}]$, $[\cdot]$ is a Gaussian function. Then $R_{n,2,3}(v)$ can be rewritten as
\begin{eqnarray*}
 R_{n,2,3}(v) &=& \sum^{k_n}_{j=0} \zeta_{j} + \sum^{k_n}_{j=0} \eta_{j} + \xi \\
         &=& R^{'}_{n,2,3}(v) + R^{''}_{n,2,3}(v) +  R^{'''}_{n,2,3}(v).
\end{eqnarray*}
The necessary conditions for applying a functional central limit theorem in a big and small block method include
\begin{eqnarray}
& & R^{''}_{n,2,3}(v) \to 0, \ \ R^{'''}_{n,2,3}(v) \to 0 \label{gammaNor1} \\
& & \left| E(e^{R^{'}_{n,2,3}(v)t}) - \Pi^{k(n)}_{i=0} E(e^{\zeta_{j}t}) \right| \to 0 \label{gammaNor2} \\
& & R^{'}_{n,2,3}(v) \to \lambda_{2} v \label{gammaNor3} \\
& & \frac{1}{n} \sum^{k}_{j=0} E(\zeta^{2}_{j} I[|\zeta_{j}| \leq \epsilon \theta \sqrt{n}]) \to 0 \label{gammaNor4} \\
& & P\left( \sup_{v_{1} \leq v \leq \tau v_{1}+ \nu} |R_{n,2,3}(v) - R_{n,2,3}(v_{1})| > \zeta\right) \to 0. \label{gammaNor5}
\end{eqnarray}

From ($\ref{gammavar}$), we have the variance $V(\eta_{j}) = s_{n} v \theta$ and then the variance
$V(R^{''}_{n,2,3}(v)) = n^{-1} k_{n} s_{n} v \lambda_{2}= \frac{s_{n}}{l_{n} + s_{n}}  v \theta = o(1)$.
Similarly, we have $V(R^{'''}_{n,2,3}(v)) = o(1)$. Therefore, it is clear that $(\ref{gammaNor1})$ holds.
In addition, as $V(R^{'}_{n,2,3}(v)) = n^{-1} k_{n} l_{n} v \lambda_{2} = \frac{l_{n}}{l_{n} + s_{n}}
\lambda_{2} v  =  v \lambda_{2}$, it can be seen that ($\ref{gammaNor3}$) also holds.

From Proposition 2.6 of Fan and Yao (2003), we have
\begin{eqnarray*}
\left| E(e^{R^{'}_{n,2,3}(v)t}) - \Pi^{k_n}_{i=0} E(e^{\zeta_{j}t}) \right| \leq 16 k_n \alpha(s_{n}) \to 0,
\end{eqnarray*}
and then $(\ref{gammaNor2})$ also holds. Furthermore,
from Lemma 1 of Hansen~(2000), and by letting $D_{1} = max_{q \in R} E[m(\bX_{i}e_{i}|Q_{i} = q)]$, we obtain
\begin{eqnarray*}
\mbox{E} \left( \left|n^{-1/2} \max_{1 \leq i \leq n} | u_{i,n}(v) | \right|^{4} \right)
&\leq& \frac{1}{n} \mbox{E}|u_{i,n}(v)|^{4} \\
&=& \frac{a^{2}_{n}}{n} \mbox{E}\left(|c_{2,3}^* (\bX_{i}) e_{i}|^{4} |d_{i}(v)|\right) \\
&\leq& \frac{a^{2}_{n}}{n} D_{1} \frac{|v|}{a_{n}} \\
&=& n^{-2 \alpha} D_{1} |v| \to 0
\end{eqnarray*}
and then (\ref{gammaNor4}) holds.
From Lemma 3 of Hansen (1999), we have
\begin{eqnarray*}
& & P\left( \sup_{v_{1} \leq v \leq \tau v_{1}+ \nu} |R_{n}(v) - R_{n}(v_{1})| > \zeta\right) \\
&=&  P\left( \sup_{\tau_{1} \leq \tau \leq \tau_{1}+ \nu/a_{n}} |R_{n}(\tau) - R_{n}(\tau_{1})| > \frac{\zeta}{a^{1/2}_{n}}\right) \\
&\leq& \frac{K_{1} (\frac{\nu}{a_{n}})^{2}}{a^{-2}_{n} \zeta^{4}} \leq v \epsilon,
\end{eqnarray*}
and then ($\ref{gammaNor5}$) also holds.
Finally, combining equations ($\ref{gammaNor1}$) through ($\ref{gammaNor5}$), we have proved
$(\lambda_{2})^{-1/2} R_{n,2,3}(v) \to^d B(v)$.

Given Lemma $\ref{CovSpeed-2}$, the probability of having $\hat{\gamma}_{2}$ in
$(\gamma_{2}-\bar{v}/n, \gamma_{2} + \bar{v}/n)$ is $ 1 - \epsilon$.
Denote $Q_{n}(v) = SSR(\tau_{1},\gamma_{2},\tau_{3}) - SSR(\tau_{1},\gamma_{2}+v/a_{n},\tau_{3})$.
We consequently have
\begin{eqnarray*}
 Q_{n}(v)  &=& \sum^{n}_{i=1} \left\{ -n^{-2\alpha} c_{2,3}^{*2}(\bX_{i}) d_{2,i} (v) \right.\\
 & &\left.        - [ \delta_{n,2,3} (\bX_i) + \hat\delta_{n,2,3} (\bX_i) ]
    d_{2,i} (v) \{ \delta_{n,2,3} (\bX_i) -\hat\delta_{n,2,3} (\bX_i) \} \right. \\
& & \left.+ 2 [\hat{m}_{\hat{\gamma}_{3}}(\bX_{i}) + \hat{m}_{\hat{\gamma}_{2}}(\bX_{i})] d_{2,i} (v) e_{i}\right. \\
& & \left.   + 2 \hat\delta_{n,2,3} (\bX_i)
        [\hat{m}_{\hat{\gamma}_{2}}(\bX_{i}) - m_{\gamma_{2}}(\bX_{i})] d_{2,i} (v) \right\} \\
&=& - G_{n,2,3}(v) + 2 R_{n,2,3}(v) + L_{n,2,3}(v)
\end{eqnarray*}
and
\begin{eqnarray*}
L_{n,2,3}(v)
&\leq& 2 \sqrt{n} \sup \left\{| \hat{\delta}_{n,2,3}(\bX_{i}) - \delta_{n,2,3}(\bX_{i}) |
                              \times |R_{n,2,3}(v)| \right. \\
& & \left. + [2 n^{\alpha} |\hat{m}_{\hat{\gamma}_{2}}(\bX_{i})  - m_{\gamma_{2}}(\bX_{i}) | \right. \\
& & \left.  
  + |c_{2,3}^{*2}(\bX_{i})-{\hat c}_{2,3}^{*2}(\bX_{i})|\times|c_{2,3}^{*2}(\bX_{i})+{\hat c}_{2,3}^{*2}(\bX_{i})|]  |d_{2,i} (v)| \right\} \to 0.
\end{eqnarray*}
Given Lemmas $\ref{mean}$ and  $\ref{variance}$, we have
\begin{eqnarray*}
Q_{n}(v)
 \to^d  - v \mu_{2} + 2 \sqrt{\lambda_{2}} B(v) =  Q(v)
\end{eqnarray*}
and then from Theorem 2.7 of Kim and Pollard~(1990), we obtain
Theorem 1 of Hansen~(2000),
\begin{eqnarray*}
a_{n}(\hat{\gamma}_{2} - \gamma_{2} ) \to^{d} \arg \max_{v \in R} Q_{2}(v). \quad\blacksquare
\end{eqnarray*}
\end{proof}

\bigskip
\noindent {\bf Note for Theorem 10.}

\begin{eqnarray*}
SSR(\gamma) \stackrel{p}{\rightarrow}  S(\gamma) = \sum^{4}_{j=1} b_{j}(\gamma) I_{\gamma_{j}}(\gamma),
\end{eqnarray*}
where
\begin{eqnarray*}
& &b_{1}(\gamma) = E(e_i^{2}) \\
& &+ E \left\{ \left[ c_{1,2}(\bX_{i}) \frac{f_{\gamma_{2}}(\bX_{i})}{f_{\gamma,2}(\bX_{i})}
                      + c_{1,3}(\bX_{i}) \frac{f_{\gamma_{3}}(\bX_{i})}{f_{\gamma,2}(\bX_{i})}
                      + c_{1,4}(\bX_{i}) \frac{f_{\gamma_{4}}(\bX_{i})}{f_{\gamma,2}(\bX_{i})} \right]^{2}
                      I_{\gamma,\gamma_{1}}(Q_{i}) \right\} \\
& & + E \left\{ \left[ - c_{1,2}(\bX_{i}) \frac{f_{\gamma,\gamma_{1}}(\bX_{i})}{f_{\gamma,2}(\bX_{i})}
                      + c_{2,3}(\bX_{i}) \frac{f_{\gamma_{3}}(\bX_{i})}{f_{\gamma,2}(\bX_{i})}
                      + c_{2,4}(\bX_{i}) \frac{f_{\gamma_{4}}(\bX_{i})}{f_{\gamma,2}(\bX_{i})} \right]^{2}
                      I_{\gamma_{2}}(Q_{i}) \right\} \\
& & + E \left\{ \left[ - c_{1,3}(\bX_{i}) \frac{f_{\gamma,\gamma_{1}}(\bX_{i})}{f_{\gamma,2}(\bX_{i})}
                             -  c_{2,3}(\bX_{i}) \frac{f_{\gamma_{2}}(\bX_{i})}{f_{\gamma,2}(\bX_{i})}
                      + c_{3,4}(\bX_{i}) \frac{f_{\gamma_{4}}(\bX_{i})}{f_{\gamma,2}(\bX_{i})} \right]^{2}
                      I_{\gamma_{3}}(Q_{i}) \right\} \\
& & + E \left\{ \left[c_{1,4}(\bX_{i}) \frac{f_{\gamma,\gamma_{1}}(\bX_{i})}{f_{\gamma,2}(\bX_{i})}
                             + c_{2,4}(\bX_{i}) \frac{f_{\gamma_{2}}(\bX_{i})}{f_{\gamma,2}(\bX_{i})}
                      + c_{3,4}(\bX_{i}) \frac{f_{\gamma_{3}}(\bX_{i})}{f_{\gamma,2}(\bX_{i})}\right]^{2}
                      I_{\gamma_{4}}(Q_{i}) \right\},
\end{eqnarray*}
with $I_{\gamma,\gamma_{1}}(Q_{i}) = 1$ for $Q_i \in [\gamma, \gamma_1)$ and 0 otherwise,  and
\begin{eqnarray*}
& &b_{2}(\gamma)
= E  (e^{2}_{i}) \\
& & + E \left\{ c^{2}_{1,2}(\bX_{i}) \frac{f^{2}_{\gamma_{1},\gamma}(\bX_{i})}{f^{2}_{\gamma,1}(\bX_{i})}
       I_{\gamma_{1}}(Q_{i}) \right\}
     + E \left\{ c^{2}_{1,2}(\bX_{i}) \frac{f^{2}_{\gamma_{1}}(\bX_{i})}{f^{2}_{\gamma,1}(\bX_{i})} I_{\gamma_{1},\gamma}(Q_{i}) \right\}  \\
& & +E \left\{ \left[ c_{2,3}(\bX_{i}) \frac{f_{\gamma_{3}}(\bX_i)}{f_{\gamma,2}(\bX_{i})}
    + c_{2,4}(\bX_{i}) \frac{f_{\gamma_{4}}(\bX_i)}{f_{\gamma,2}(\bX_{i})}\right]^{2} I_{\gamma,\gamma_{2}}(Q_{i}) \right\}   \\
& & +  E \left\{ \left[ - c_{2,3}(\bX_{i}) \frac{f_{\gamma,\gamma_{2}}(\bx)}{f_{\gamma,2}(\bX_{i})}
    + c_{3,4} \frac{f_{\gamma_{4}}(\bX_i)}{f_{\gamma,2}(\bX_{i})}\right]^{2} I_{\gamma_{3}}(Q_{i}) \right\}  \\
& & + E \left\{ \left[  c_{2,4}(\bX_{i}) \frac{f_{\gamma,\gamma_{2}}(\bX_i)}{f_{\gamma,2}(\bX_{i})}
    + c_{3,4}(\bX_{i}) \frac{f_{\gamma_{3}}(\bX_i)}{f_{\gamma,2}(\bX_{i})}\right]^{2} I_{\gamma_{4}}(Q_{i}) \right\},
\end{eqnarray*}
with $I_{\gamma_1,\gamma}(Q_{i}) = 1$ for $Q_i \in (\gamma_1, \gamma]$ and 0 otherwise,
$I_{\gamma,\gamma_2}(Q_{i}) = 1$ for $Q_i \in [\gamma, \gamma_2)$ and 0 otherwise, and
\begin{eqnarray*}
& &b_{3}(\gamma)
= E(e^{2}_{i}) \\
& & + E \left\{ \left[ c_{1,2}(\bX_{i}) \frac{f_{\gamma_{2}}(\bX_{i})}{f_{\gamma,1}(\bX_{i})}
     + c_{1,3}(\bX_{i}) \frac{f_{\gamma_{2},\gamma}(\bX_{i})}{f_{\gamma,1}(\bX_{i})} \right]^{2} I_{\gamma_{1}}(Q_{i}) \right\}  \\
& & + E \left\{ \left[ - c_{1,2}(\bX_{i}) \frac{f_{\gamma_{1}}(\bX_{i})}{f_{\gamma,1}(\bX_{i})}
    + c_{2,3}(\bX_{i}) \frac{f_{\gamma_{2},\gamma}(\bX_{i})}{f_{\gamma,1}(\bX_{i})} \right]^{2} I_{\gamma_{2}}(Q_{i}) \right\}   \\
& & + E \left\{ \left[ c_{1,3}(\bX_{i}) \frac{f_{\gamma_{1}}(\bX_{i})}{f_{\gamma,1}(\bX_{i})}
    + c_{2,3}(\bX_{i}) \right] \frac{f_{\gamma_{2}}(\bX_{i})}{f_{\gamma,1}(\bX_{i})} ]^{2} I_{\gamma_{2},\gamma}(Q_{i}) \right\}  \\
& & + E \left\{ \left[ c_{3,4}(\bX_{i}) \frac{f_{\gamma_{4}}(\bX_{i})}{f_{\gamma,2}(\bX_{i})} \right]^{2}
    I_{\gamma,\gamma_{3}}(Q_{i}) \right\}  \\
& & + E \left\{ \left[ c_{3,4}(\bX_{i})
    \frac{f_{\gamma,\gamma_{3}}(\bX_{i})}{f_{\gamma,2}(\bX_{i})} \right]^{2} I_{\gamma_{4}}(Q_{i}) \right\},
\end{eqnarray*}
with $I_{\gamma_2,\gamma}(Q_{i}) = 1$ for $Q_i \in (\gamma_2, \gamma]$ and 0 otherwise,
$I_{\gamma,\gamma_3}(Q_{i}) = 1$ for $Q_i \in [\gamma, \gamma_3)$ and 0 otherwise, and
\begin{eqnarray*}
& &b_{4}(\gamma)
= E( e^{2}_{i}) \\
& & + E \left\{ \left[  c_{1,2} \frac{f_{\gamma_{2}}(\bX_{i})}{f_{\gamma,1}(\bX_{i})}
     + c_{1,3} \frac{f_{\gamma_{3}}(\bX_{i})}{f_{\gamma,1}(\bX_{i})}
     + c_{1,4} \frac{f_{\gamma_{3},\gamma}(\bX_{i})}{f_{\gamma,1}(\bX_{i})} \right]^{2} I_{\gamma_{1}}(Q_{i}) \right\}  \\
& & + E \left\{ \left[ - c_{1,2} \frac{f_{\gamma_{1}}(\bX_{i})}{f_{\gamma,1}(\bX_{i})}
    + c_{2,3} \frac{f_{\gamma_{3}}(\bX_{i})}{f_{\gamma,1}(\bX_{i})}
    + c_{2,4} \frac{f_{\gamma_{3},\gamma}(\bX_{i})}{f_{\gamma,1}(\bX_{i})} \right]^{2} I_{\gamma_{2}}(Q_{i}) \right\}   \\
& & + E \left\{ \left[ - c_{1,3} \frac{f_{\gamma_{1}}(\bX_{i})}{f_{\gamma,1}(\bX_{i})}
    - c_{2,3} \frac{f_{\gamma_{2}}(\bX_{i})}{f_{\gamma,1}(\bX_{i})}
    + c_{3,4} \frac{f_{\gamma_{3},\gamma}(\bX_{i})}{f_{\gamma,1}(\bX_{i})} \right]^{2} I_{\gamma_{3}}(Q_{i}) \right\}  \\
& & + E \left\{ \left[  c_{1,4} \frac{f_{\gamma_{1}}(\bX_{i})}{f_{\gamma,1}(\bX_{i})}
    + c_{2,4} \frac{f_{\gamma_{2}}(\bX_{i})}{f_{\gamma,1}(\bX_{i})}
    + c_{3,4} \frac{f_{\gamma_{3}}(\bX_{i})}{f_{\gamma,1}(\bX_{i})} \right]^{2} I_{\gamma_{3},\gamma}(Q_{i}) \right\},
\end{eqnarray*}
with $I_{\gamma_3,\gamma}(Q_{i}) = 1$ for $Q_i \in (\gamma_3, \gamma)$ and 0 otherwise. \quad$\square$

\newpage
\noindent
{\bf Graphic Description of Theorem 10.}


\begin{picture}(300,400)(40,-200)
\put (40,175){Case of $b(1)$}
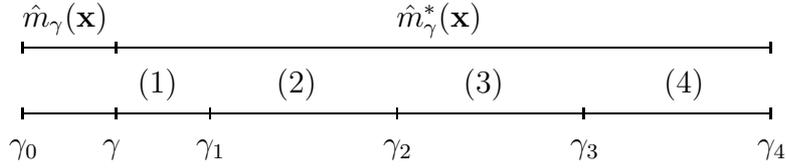
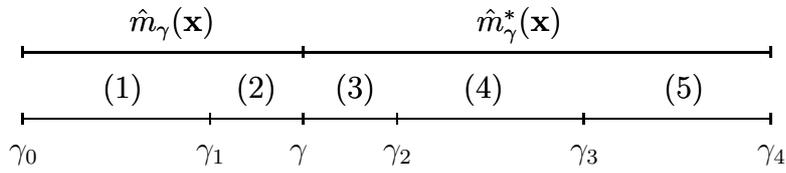
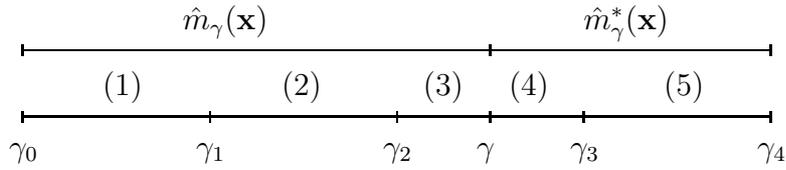
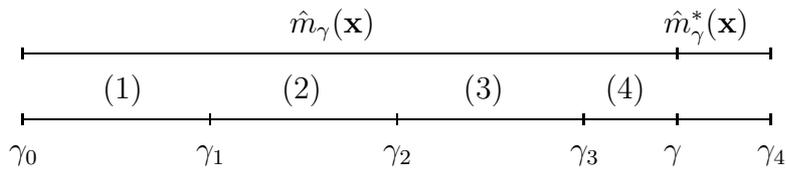
\put(40,120){\line(1,0){280}}
\put (40,118){\line(0,1){4}}
\put (110,118){\line(0,1){4}}
\put (75,118){\line(0,1){4}}
\put (180,118){\line(0,1){4}}
\put (250,118){\line(0,1){4}}
\put (320,118){\line(0,1){4}}
\put (35,105){$\gamma_0$}
\put (105,105){$\gamma_1$}
\put (70,105){$\gamma$}
\put (175,105){$\gamma_2$}
\put (245,105){$\gamma_3$}
\put (315,105){$\gamma_4$}
\put(40,145){\line(1,0){280}}
\put (40,153){$\hat{m}_{\gamma}(\bx)$}
\put (180,153){$\hat{m}_{\gamma}^* (\bx)$}
\put (83,128){$(1)$}
\put (135,128){$(2)$}
\put (205,128){$(3)$}
\put (280,128){$(4)$}
\put (40,143){\line(0,1){4}}
\put (75,143){\line(0,1){4}}
\put (320,143){\line(0,1){4}}
\put (80,53){$\hat{m}_{\gamma}(\bx)$}
\put (70,28){$(1)$}
\put (120,28){$(2)$}
\put (157,28){$(3)$}
\put (205,28){$(4)$}
\put (280,28){$(5)$}
\put (210,53){$\hat{m}_{\gamma}^*(\bx)$}

\put (40,75){Case of $b(2)$}
\put(40,20){\line(1,0){280}}
\put (40,18){\line(0,1){4}}
\put (110,18){\line(0,1){4}}
\put (145,18){\line(0,1){4}}
\put (180,18){\line(0,1){4}}
\put (250,18){\line(0,1){4}}
\put (320,18){\line(0,1){4}}
\put (35,5){$\gamma_0$}
\put (105,5){$\gamma_1$}
\put (140,5){$\gamma$}
\put (175,5){$\gamma_2$}
\put (245,5){$\gamma_3$}
\put (315,5){$\gamma_4$}

\put(40,45){\line(1,0){280}}
\put (40,43){\line(0,1){4}}
\put (145,43){\line(0,1){4}}
\put (320,43){\line(0,1){4}}
\put (80,53){$\hat{m}_{\gamma}(\bx)$}
\put (70,28){$(1)$}
\put (120,28){$(2)$}
\put (157,28){$(3)$}
\put (205,28){$(4)$}
\put (280,28){$(5)$}
\put (210,53){$\hat{m}_{\gamma}^* (\bx)$}

\put (40,-25){Case of $b(3)$}
\put (40,-80){\line(1,0){280}}
\put (40,-82){\line(0,1){4}}
\put (110,-82){\line(0,1){4}}
\put (215,-82){\line(0,1){4}}
\put (180,-82){\line(0,1){4}}
\put (250,-82){\line(0,1){4}}
\put (320,-82){\line(0,1){4}}
\put (35,-95){$\gamma_0$}
\put (105,-95){$\gamma_1$}
\put (210,-95){$\gamma$}
\put (175,-95){$\gamma_2$}
\put (245,-95){$\gamma_3$}
\put (315,-95){$\gamma_4$}

\put (40,-55){\line(1,0){280}}
\put (40,-57){\line(0,1){4}}
\put (215,-57){\line(0,1){4}}
\put (320,-57){\line(0,1){4}}
\put (100,-47){$\hat{m}_{\gamma}(\bx)$}
\put (70,-72){$(1)$}
\put (137,-72){$(2)$}
\put (190,-72){$(3)$}
\put (222,-72){$(4)$}
\put (280,-72){$(5)$}
\put (250,-47){$\hat{m}_{\gamma}^* (\bx)$}

\put (40,-125){Case of $b(4)$}
\put (40,-180){\line(1,0){280}}
\put (40,-182){\line(0,1){4}}
\put (110,-182){\line(0,1){4}}
\put (285,-182){\line(0,1){4}}
\put (180,-182){\line(0,1){4}}
\put (250,-182){\line(0,1){4}}
\put (320,-182){\line(0,1){4}}
\put (35,-195){$\gamma_0$}
\put (105,-195){$\gamma_1$}
\put (280,-195){$\gamma$}
\put (175,-195){$\gamma_2$}
\put (245,-195){$\gamma_3$}
\put (315,-195){$\gamma_4$}

\put (40,-155){\line(1,0){280}}
\put (40,-157){\line(0,1){4}}
\put (285,-157){\line(0,1){4}}
\put (320,-157){\line(0,1){4}}
\put (140,-147){$\hat{m}_{\gamma}(\bx)$}
\put (70,-172){$(1)$}
\put (137,-172){$(2)$}
\put (205,-172){$(3)$}
\put (258,-172){$(4)$}
\put (280,-147){$\hat{m}_{\gamma}^* (\bx)$}
\end{picture}

Given the three true threshold values $\gamma_1$, $\gamma_2$, and $\gamma_3$, the threshold value $\gamma$ of a
mis-specified nonparametric regression with one threshold may be  in $[\gamma_0, \gamma_1)$, or in
$(\gamma_1, \gamma_2)$, or in $(\gamma_2, \gamma_3)$, or in $(\gamma_3, \gamma_4]$. For $\gamma \in [\gamma_0, \gamma_1)$,
there is no model miss-specified error for $Q_i \in [\gamma_0, \gamma]$ but the miss-specified errors are
\begin{enumerate}
\item $Q_i \in [\gamma, \gamma_1]$ is $\hat{m}_{\gamma}^* (\bx) - m_{\gamma_1} (\bx)$,
\item $Q_i \in [\gamma_1, \gamma_2)$ is $\hat{m}_{\gamma}^*(\bx) - m_{\gamma_2} (\bx)$,
\item $Q_i \in [\gamma_2, \gamma_3)$  is $\hat{m}_{\gamma}^*(\bx) - m_{\gamma_3} (\bx)$, and
\item $Q_i \in [\gamma_3, \gamma_4]$  is $\hat{m}_{\gamma}^*(\bx) - m_{\gamma_4} (\bx)$,
\end{enumerate}
as shown in the first graph of Case  $b(1)$.
For $\gamma \in (\gamma_1, \gamma_2)$, the mis-specified errors are
\begin{enumerate}
\item for $Q_i \in [\gamma_0, \gamma_1)$ is $\hat{m}_{\gamma}(\bx) - m_{\gamma_1} (\bx)$ and is denoted as (1),
\item for $Q_i \in [\gamma_1, \gamma)$ is $\hat{m}_{\gamma}(\bx) - m_{\gamma_2} (\bx)$ and is denoted as (2),
\item for $Q_i \in [\gamma, \gamma_2)$ is $\hat{m}_{\gamma}^*(\bx) - m_{\gamma_2} (\bx)$ and is denoted as (3),
\item for $Q_i \in [\gamma_2, \gamma_3)$  is $\hat{m}_{\gamma}^*(\bx) - m_{\gamma_3} (\bx)$ and is denoted as (4),
\item for $Q_i \in [\gamma_3, \gamma_4]$  is $\hat{m}_{\gamma}^*(\bx) - m_{\gamma_4} (\bx)$ and is denoted as (5),
\end{enumerate}
as shown in the second graph of Case  $b(2)$.
For $\gamma \in (\gamma_2, \gamma_3)$, the mis-specified errors are
\begin{enumerate}
\item for $Q_i \in [\gamma_0, \gamma_1)$ is $\hat{m}_{\gamma}(\bx) - m_{\gamma_1} (\bx)$ and is denoted as (1),
\item for $Q_i \in [\gamma_1, \gamma_2)$ is $\hat{m}_{\gamma}(\bx) - m_{\gamma_2} (\bx)$ and is denoted as (2),
\item for $Q_i \in [\gamma_2, \gamma)$ is $\hat{m}_{\gamma}(\bx) - m_{\gamma_3} (\bx)$ and is denoted as (3),
\item for $Q_i \in [\gamma, \gamma_3)$  is $\hat{m}_{\gamma}^*(\bx) - m_{\gamma_3} (\bx)$ and is denoted as (4),
\item for $Q_i \in [\gamma_3, \gamma_4]$  is $\hat{m}_{\gamma}^*(\bx) - m_{\gamma_4} (\bx)$ and is denoted as (5),
\end{enumerate}
as shown in the third graph of Case  $b(3)$.
For $\gamma \in (\gamma_3, \gamma_4)$, the mis-specified errors are
\begin{enumerate}
\item for $Q_i \in [\gamma_0, \gamma_1)$ is $\hat{m}_{\gamma}(\bx) - m_{\gamma_1} (\bx)$ and is denoted as (1),
\item for $Q_i \in [\gamma_1, \gamma_2)$ is $\hat{m}_{\gamma}(\bx) - m_{\gamma_2} (\bx)$ and is denoted as (2),
\item for $Q_i \in [\gamma_2, \gamma_3)$ is $\hat{m}_{\gamma}(\bx) - m_{\gamma_2} (\bx)$ and is denoted as (3),
\item for $Q_i \in [\gamma_3, \gamma)$  is $\hat{m}_{\gamma}(\bx) - m_{\gamma_3} (\bx)$ and is denoted as (4),
\end{enumerate}
as shown in the last graph of Case  $b(3)$. Note that there is no model mis-specified error for
$Q_i \in [\gamma, \gamma_4]$ in this case.

\bigskip
As to the cases of $\gamma = \gamma_1$,  $\gamma_2$, or $\gamma_3$, the model mis-specification errors are
{\small
\begin{eqnarray*}
& &S(\gamma_{1})
= E(e_i^{2}) + E \left\{ \left[ c_{2,3}(\bX_{i}) \frac{f_{\gamma_{3}}(\bX_{i})}{f(\bX_{i}) - f_{\gamma_{1}}(\bX_{i})}
+ c_{2,4}(\bX_{i}) \frac{f_{\gamma_{4}}(\bX_{i})}{f(\bX_{i}) - f_{\gamma_{1}}(\bX_{i})} \right]^{2} I_{\gamma_{2}}(Q_{i}) \right\} \\
& & + E \left\{ \left[ -  c_{2,3}(\bX_{i}) \frac{f_{\gamma_{2}}(\bX_{i})}{f(\bX_{i}) - f_{\gamma_{1}}(\bX_{i})}
+ c_{3,4}(\bX_{i}) \frac{f_{\gamma_{4}}(\bX_{i})}{f(\bX_{i}) - f_{\gamma_{1}}(\bX_{i})} \right]^{2} I_{\gamma_{3}}(Q_{i}) \right\} \\
& & + E \left\{ \left[  c_{2,4}(\bX_{i}) \frac{f_{\gamma_{2}}(\bX_{i})}{f(\bX_{i}) - f_{\gamma_{1}}(\bX_{i})}
+ c_{3,4}(\bX_{i}) \frac{f_{\gamma_{3}}(\bX_{i})}{f(\bX_{i}) - f_{\gamma_{1}}(\bX_{i})} \right]^{2}
  I_{\gamma_{4}}(Q_{i}) \right\}, \\
& &S(\gamma_{2})
= E(e^{2}_{i})
+ E \left\{ \left[ c_{1,2}(\bX_{i}) \frac{f_{\gamma_{2}}(\bX_{i})}{f_{\gamma_{1}}(\bX_{i}) + f_{\gamma_{2}}(\bX_{i})} \right]^{2}
  I_{\gamma_{1}}(Q_{i}) \right\}  \\
& & + E \left\{ \left[ - c_{1,2}(\bX_{i}) \frac{f_{\gamma_{1}}(\bX_{i})}{f_{\gamma_{1}}(\bX_{i}) + f_{\gamma_{2}}(\bX_{i})} \right]^{2}
   I_{\gamma_{2}}(Q_{i}) \right\}   \\
& & + E \left\{ \left[ c_{3,4}(\bX_{i}) \frac{f_{\gamma_{4}}(\bx)}{f_{\gamma_{3}}(\bX_{i}) + f_{\gamma_{4}}(\bX_{i})} \right]^{2}
   I_{\gamma,\gamma_{3}}(Q_{i}) \right\}  \\
& & + E \left\{ \left[ c_{3,4}(\bX_{i}) \frac{f_{\gamma_{3}}(\bX_{i})}{f_{\gamma_{3}}(\bX_{i}) + f_{\gamma_{4}}(\bX_{i})} \right]^{2}
   I_{\gamma_{4}}(Q_{i}) \right\},  \\
& &S(\gamma_{3}) = E(e^{2}_{i})
 + E \left\{ \left[ c_{1,2}(\bX_{i}) \frac{f_{\gamma_{2}}(\bX_{i})}{f(\bX_{i}) - f_{\gamma_{4}}(\bX_{i})}
  + c_{1,3}(\bX_{i}) \frac{f_{\gamma_{3}}(\bX_{i})}{f(\bX_{i}) - f_{\gamma_{1}}(\bX_{i})} \right]^{2}
  I_{\gamma_{1}}(Q_{i}) \right\}  \\
& & + E \left\{ \left[ - c_{1,2}(\bX_{i}) \frac{f_{\gamma_{1}}(\bX_{i})}{f(\bX_{i}) - f_{\gamma_{1}}(\bX_{i})}
 + c_{2,3}(\bX_{i}) \frac{f_{\gamma_{3}}(\bX_{i})}{f(\bX_{i}) - f_{\gamma_{1}}(\bX_{i})} \right]^{2}
  I_{\gamma_{2}}(Q_{i}) \right\}   \\
& & + E \left\{ \left[ c_{1,3}(\bX_{i}) \frac{f_{\gamma_{1}}(\bX_{i})}{f(\bX_{i}) - f_{\gamma_{1}}(\bX_{i})}
 + c_{2,3}(\bX_{i}) \frac{f_{\gamma_{2}}(\bX_{i})}{f(\bX_{i}) - f_{\gamma_{1}}(\bX_{i})} \right]^{2}
 I_{\gamma_{3}}(Q_{i}) \right\}. \quad\square
\end{eqnarray*}
}

\bigskip
\noindent
{\bf Proof of Theorems 10 and 11.}

\medskip
\noindent
From Lemma $\ref{MultEst}$,
\begin{eqnarray*}
sup | \hat{m}_{\gamma}(\bx)  - m_{\gamma}(\bx)| &=&  O_{p}(h^{r} + (\ln(n))^{1/2}/(nh^{p})^{1/2}) \\
sup | \hat{m}^{*}_{\gamma}(\bx) - m^{*}_{\gamma}(\bx)| &=&  O_{p}(h^{r} + (\ln(n))^{1/2}/(nh^{p})^{1/2}),
\end{eqnarray*}
where
\begin{eqnarray}
 & & \hspace{-1cm} m_{\gamma}(\bx)\notag \\
&=& m_{\gamma_{1}} I_{\gamma_{1}}(\gamma)  + \left[m_{\gamma_{1}}(\bx) \frac{f_{\gamma_{1}}(\bx)}{f_{\gamma,1}(\bx)}
   + m_{\gamma_{2}}(\bx) \frac{f_{\gamma_{1},\gamma}(\bx)}{f_{\gamma,1}(\bx)} \right] I_{\gamma_{2}}(\gamma)  \nonumber  \\
& & + \left[m_{\gamma_{1}}(\bx) \frac{f_{\gamma_{1}}(\bx)}{f_{\gamma,1}(\bx)}
    + m_{\gamma_{2}}(\bx) \frac{f_{\gamma_{2}}(\bx)}{f_{\gamma,1}(\bx)}
    + m_{\gamma_{3}}(\bx) \frac{f_{\gamma_{2},\gamma}(\bx)}{f_{\gamma,1}(\bx)} \right] I_{\gamma_{3}}(\gamma) \nonumber \\
& & + \left[m_{\gamma_{1}}(\bx) \frac{f_{\gamma_{1}}(\bx)}{f_{\gamma,1}(\bx)}
    + m_{\gamma_{2}}(\bx) \frac{f_{\gamma_{2}}(\bx)}{f_{\gamma,1}(\bx)}
    + m_{\gamma_{3}}(\bx) \frac{f_{\gamma_{3}}(\bx)}{f_{\gamma,1}(\bx)}
    + m_{\gamma_{4}}(\bx) \frac{f_{\gamma_{3},\gamma}(\bx)}{f_{\gamma,1}(\bx)}\right] I_{\gamma_{4}}(\gamma) \nonumber 
                                                                                                            \label{Multest-1}
\end{eqnarray}
and
\begin{eqnarray}
& &  \hspace{-1cm} m^{*}_{\gamma}(\bx)\notag  \\
 &=&  [m_{\gamma_{1}}(\bx) \frac{f_{\gamma,\gamma_{1}}(\bx)}{f_{\gamma,2}(\bx)}
                             + m_{\gamma_{2}}(\bx) \frac{f_{\gamma_{2}}(\bx)}{f_{\gamma,2}(\bx)}
                      + m_{\gamma_{3}}(\bx) \frac{f_{\gamma_{3}}(\bx)}{f_{\gamma,2}(\bx)}
                      + m_{\gamma_{4}}(\bx) \frac{f_{\gamma_{4}}(\bx)}{f_{\gamma,2}(\bx)}]I_{\gamma_{1}}(\gamma) \nonumber \\
                & & + [m_{\gamma_{2}}(\bx) \frac{f_{\gamma,\gamma_{2}}(\bx)}{f_{\gamma,2}(\bx)}
                             + m_{\gamma_{3}}(\bx) \frac{f_{\gamma_{3}}(\bx)}{f_{\gamma,2}(\bx)}
                             + m_{\gamma_{4}}(\bx) \frac{f_{\gamma_{4}}(\bx)}{f_{\gamma,2}(\bx)}] I_{\gamma_{2}}(\gamma) \nonumber \\
                & & + [ \frac{f_{\gamma,\gamma_{3}}(\bx)}{f_{\gamma,2}(\bx)} m_{\gamma_{3}}  + m_{\gamma_{4}}(\bx)
                     \frac{f_{\gamma_{4}}(\bx)}{f_{\gamma,2}(\bx)}]I_{\gamma_{3}}(\gamma)   + m_{\gamma_{4}}(\bx) I_{\gamma_{4}}(\gamma). \label{Multest-2}
\end{eqnarray}
Denote $\gamma$ as a pseudo threshold value considered in a mis-specified nonparametric regression with one threshold and
assume $\gamma \in [\gamma_{0},\gamma_{1})$.
From 
($\ref{Multest-2}$), we have
\begin{eqnarray}
& & \frac{1}{n} \sum^{n}_{i=1} \{ Y_{i} - \hat{m}_{\gamma}(\bX_{i})I_{\gamma}(Q_{i}) - \hat{m}^{*}_{\gamma}(\bX_{i}) (1-I_{\gamma}(Q_{i}) )  \}^{2} \nonumber \\
&=& \frac{1}{n} \sum^{n}_{i=1} \{ e_{i} + \left\{ c_{1,2}(\bX_{i}) \frac{f_{\gamma_{2}}(\bX_{i})}{f_{\gamma,2}(\bX_{i})}
                      + c_{1,3}(\bX_{i}) \frac{f_{\gamma_{3}}(\bX_{i})}{f_{\gamma,2}(\bX_{i})}
                      + c_{1,4}(\bX_{i}) \frac{f_{\gamma_{4}}(\bX_{i})}{f_{\gamma,2}(\bX_{i})}  \right\} I_{\gamma,\gamma_{1}}(Q_{i}) \nonumber \\
& & + \left\{ - c_{1,2}(\bX_{i}) \frac{f_{\gamma,\gamma_{1}}(\bX_{i})}{f_{\gamma,2}(\bX_{i})}
                      + c_{2,3}(\bX_{i}) \frac{f_{\gamma_{3}}(\bX_{i})}{f_{\gamma,2}(\bX_{i})}
                      + c_{2,4}(\bX_{i}) \frac{f_{\gamma_{4}}(\bX_{i})}{f_{\gamma,2}(\bX_{i})} \right\} I_{\gamma_{2}}(Q_{i}) \nonumber \\
& & + \left\{ - c_{1,3}(\bX_{i}) \frac{f_{\gamma,\gamma_{1}}(\bX_{i})}{f_{\gamma,2}(\bX_{i})}
                             -  c_{2,3}(\bX_{i}) \frac{f_{\gamma_{2}}(\bX_{i})}{f_{\gamma,2}(\bX_{i})}
                      + c_{3,4}(\bX_{i}) \frac{f_{\gamma_{4}}(\bX_{i})}{f_{\gamma,2}(\bX_{i})} \right\} I_{\gamma_{3}}(Q_{i}) \nonumber  \\
& & + \left\{ c_{1,4}(\bX_{i}) \frac{f_{\gamma,\gamma_{1}}(\bX_{i})}{f_{\gamma,2}(\bX_{i})}
                             + c_{2,4}(\bX_{i}) \frac{f_{\gamma_{2}}(\bX_{i})}{f_{\gamma,2}(\bX_{i})}
                      + c_{3,4}(\bX_{i}) \frac{f_{\gamma_{3}}(\bX_{i})}{f_{\gamma,2}(\bX_{i})}] \right\} I_{\gamma_{4}}(Q_{i})  \}^{2}. \label{multcal-1}
\end{eqnarray}
Based on Lemma $\ref{exogenous}$, the limit of the cross products of $e_i$ with the other terms in the above  equation 
will be $o_{p}(1)$. Note that the cross products among these terms  converge to zero.
Therefore, the limit of ($\ref{multcal-1}$) is
\begin{eqnarray*}
& & \frac{1}{n} \sum^{n}_{i=1} \{ Y_{i} - \hat{m}_{\gamma}(\bX_{i})I_{\gamma}(Q_{i}) - \hat{m}^{*}_{\gamma}(\bX_{i}) (1-I_{\gamma}(Q_{i}) )  \}^{2} \\
&\to^{P}& \mbox{E}(e^{2}) + \mbox{E} \left\{ [ c_{1,2}(\bX_{i}) \frac{f_{\gamma_{2}}(\bX_{i})}{f_{\gamma,2}(\bX_{i})}
                      + c_{1,3}(\bX_{i}) \frac{f_{\gamma_{3}}(\bX_{i})}{f_{\gamma,2}(\bX_{i})}
                      + c_{1,4}(\bX_{i}) \frac{f_{\gamma_{4}}(\bX_{i})}{f_{\gamma,2}(\bX_{i})} ]^{2} I_{\gamma,\gamma_{1}}(Q_{i}) \right\} \\
& & + \mbox{E} \left\{ [ - c_{1,2}(\bX_{i}) \frac{f_{\gamma,\gamma_{1}}(\bX_{i})}{f_{\gamma,2}(\bX_{i})}
                      + c_{2,3}(\bX_{i}) \frac{f_{\gamma_{3}}(\bX_{i})}{f_{\gamma,2}(\bX_{i})}
                      + c_{2,4}(\bX_{i}) \frac{f_{\gamma_{4}}(\bX_{i})}{f_{\gamma,2}(\bX_{i})} ]^{2} I_{\gamma_{2}}(Q_{i}) \right\} \\
& & + \mbox{E} \left\{ [ - c_{1,3}(\bX_{i}) \frac{f_{\gamma,\gamma_{1}}(\bX_{i})}{f_{\gamma,2}(\bX_{i})}
                             -  c_{2,3}(\bX_{i}) \frac{f_{\gamma_{2}}(\bX_{i})}{f_{\gamma,2}(\bX_{i})}
                      + c_{3,4}(\bX_{i}) \frac{f_{\gamma_{4}}(\bX_{i})}{f_{\gamma,2}(\bX_{i})} ]^{2} I_{\gamma_{3}}(Q_{i}) \right\} \\
& & + \mbox{E} \left\{ [c_{1,4}(\bX_{i}) \frac{f_{\gamma,\gamma_{1}}(\bX_{i})}{f_{\gamma,2}(\bX_{i})}
                             + c_{2,4}(\bX_{i}) \frac{f_{\gamma_{2}}(\bX_{i})}{f_{\gamma,2}(\bX_{i})}
                      + c_{3,4}(\bX_{i}) \frac{f_{\gamma_{3}}(\bX_{i})}{f_{\gamma,2}(\bX_{i})}]^{2} I_{\gamma_{4}}(Q_{i}) \right\} \\
&=& b_{1}(\gamma).
\end{eqnarray*}
The limiting properties of $b_{2}(\gamma)$, $b_{3}(\gamma)$, and $b_{3}(\gamma)$ can be derived in the same manner. \quad$\blacksquare$

\bigskip
\noindent
{\bf Proof of Theorem 12.}

\medskip
\noindent
The slope of $b_1(\gamma)$ for $\gamma \in [\gamma_{0},\gamma_{1})$ is
\begin{eqnarray}
& & \hspace{-1cm} \frac{d b_{1}(\gamma)}{d \gamma} \nonumber \\
&=& - \int [ c_{1,2}(\bx_{i}) f_{\gamma_{2}}(\bx_{i}) + c_{1,3}(\bx_{i}) f_{\gamma_{3}}(\bx_{i})
                 + c_{1,4}(\bx_{i}) f_{\gamma_{4}}(\bx_{i})  ]^{2} \frac{f(\bx_{i},q_{i} = \gamma)}{f^{2}_{\gamma,2}(\bx_{i})} d\bx_{i} \nonumber \\
&=& - \mbox{E} \left\{ [ c_{1,2}(\bX_{i}) \frac{f_{\gamma_{2}}(\bX_{i})}{f_{\gamma,2}(\bX_{i})}
 + c_{1,3}(\bX_{i}) \frac{f_{\gamma_{3}}(\bX_{i})}{f_{\gamma,2}(\bX_{i})}
  + c_{1,4}(\bX_{i}) \frac{f_{\gamma_{4}}(\bX_{i})}{f_{\gamma,2}(\bX_{i})}  ]^{2} |q_{i} = \gamma \right\} f(\gamma). \qquad \label{mult-1}
\end{eqnarray}
The slope of $b_2 (\gamma)$ for $\gamma \in [\gamma_{1},\gamma_{2})$ is
\begin{eqnarray}
& & \hspace{-1.4cm} \frac{d b_{2}(\gamma)}{d \gamma}\notag \\
&=& \int c^{2}_{1,2}(\bx_{i})  f^{2}_{\gamma_{1}}(\bx_{i}) \frac{1}{f^{2}_{\gamma,1}(\bx_{i})} \nonumber \\
& & - ( c_{2,3}(\bx_{i}) f_{\gamma_{3}}(\bx_{i}) + c_{2,4}(\bx_{i}) f_{\gamma_{4}}(\bx_{i}))^{2}
 \frac{1}{f^{2}_{\gamma,2}(\bx_{i})} f(\bx_{i},q_{i}=\gamma) d\bx_{i} \nonumber \\
&=& \mbox{E} [c^{2}_{1,2}(\bX_{i})   \frac{f^{2}_{\gamma_{1}}(\bX_{i})}{f^{2}_{\gamma,1}(\bX_{i})} | Q_{i} = \gamma] f(\gamma) \nonumber \\
& & - \mbox{E} \left\{ [ c_{2,3}(\bX_{i}) \frac{f_{\gamma_{3}}(\bX_{i})}{f_{\gamma,2}(\bX_{i})} +
       c_{2,4}(\bx_{i})  \frac{f_{\gamma_{4}}(\bX_{i})}{f_{\gamma,2}(\bX_{i})}]^{2} | Q_{i} = \gamma \right\}f(\gamma). \quad \label{mult-2}
\end{eqnarray}
The slope of $b_3 (\gamma)$ for $\gamma \in [\gamma_{2},\gamma_{3})$ is
\begin{eqnarray}
& &  \hspace{-2.2cm}  \frac{d b_{3}(\gamma) }{d \gamma}\notag\\
&=& \int \{  c_{1,3}f_{\gamma_{1}}(\bx_{i}) + c_{2,3} f_{\gamma_{2}}(\bx_{i})  \}^{2} \frac{1}{f^{2}_{\gamma,1}(\bx_{i})} f(\bx_{i},q_{i}=\gamma)d\bx_{i} \nonumber \\
& & - \int c^{2}_{3,4} \frac{f^{2}_{\gamma_{4}}(\bx)}{f^{2}_{\gamma,2}(\bx)} f(\bx_{i},q_{i}=\gamma)d\bx_{i} \nonumber \\
&=&  \mbox{E} \left\{ [ c_{1,3} \frac{f_{\gamma_{1}}(\bx_{i})}{f_{\gamma,1}(\bx_{i})}
           + c_{2,3} \frac{f_{\gamma_{2}}(\bx_{i})}{f_{\gamma,1}(\bx_{i})} ]^{2} |Q_{i} = \gamma \right\} f(\gamma) \nonumber  \\
& & - \mbox{E} [ c^{2}_{3,4} \frac{f^{2}_{\gamma_{4}}(\bx)}{f^{2}_{\gamma,2}(\bx)} |Q_{i} = \gamma ] f(\gamma).
\label{mult-3}
\end{eqnarray}
Finally, the slope of $b_4 (\gamma)$ for $\gamma \in [\gamma_{3},\gamma_{4})$ is
\begin{eqnarray}
& &  \hspace{-1cm} \frac{d b_{4}(\gamma)}{d \gamma } \nonumber \\
&=& \int \left\{  c_{1,4}(\bx_{i}) f_{\gamma_{1}}(\bx_{i})
    + c_{2,4}(\bx_{i}) f_{\gamma_{2}}(\bx_{i})
    + c_{3,4}(\bx_{i}) f_{\gamma_{3}}(\bx_{i}) \right\}^{2} \frac{1}{f^{2}_{\gamma,1}(\bx_{i})}
    f(\bx_{i},q_{i}=\gamma) d\bx_{i} \nonumber \\
&=& \mbox{E} \left\{  c_{1,4}(\bX_{i}) \frac{f_{\gamma_{1}}(\bX_{i})}{f^{2}_{\gamma,1}(\bX_{i})}
    + c_{2,4}(\bX_{i}) \frac{f_{\gamma_{2}}(\bX_{i})}{f^{2}_{\gamma,1}(\bX_{i})}
    + c_{3,4}(\bX_{i}) \frac{f_{\gamma_{3}}(\bX_{i})}{f^{2}_{\gamma,1}(\bX_{i})} |Q_{i} = \gamma ]^{2}   \right\}
    f(\gamma).  \  \label{mult-4}
\end{eqnarray}

From  ($\ref{mult-1}$), the slope is a strictly decreasing function in $\gamma$ for
$\gamma \in [\gamma_{0},\gamma_{1})$. Thus, $S(\gamma_{1})$ is the smallest value of the model mis-specification error
for $\gamma \in [\gamma_{0},\gamma_{1})$. For $\gamma \in [\gamma_{1},\gamma_{2})$, we denote
\begin{eqnarray*}
\pi_{2}(\bx_{i},\gamma) = c^{2}_{1,2}(\bx_{i})  f^{2}_{\gamma_{1}}(\bx_{i}) \frac{1}{f^{2}_{\gamma,1}(\bx_{i})}  -
 [c_{2,3}(\bx_{i}) f_{\gamma_{3}}(\bx_{i}) + c_{2,4}(\bx_{i}) f_{\gamma_{4}}(\bx_{i})]^{2}
 \frac{1}{f^{2}_{\gamma,2}(\bx_{i})}
\end{eqnarray*}
$\forall \bx_{i} \in R^p$. The partial effect of $\gamma$ on $\pi(\bx_{i})$ is
\begin{eqnarray*}
\frac{\partial \pi_{2}(\bx_{i},\gamma)}{\partial \gamma}  =
	 - 2 c^{2}_{1,2}(\bx_{i})  f^{2}_{\gamma_{1}}(\bx_{i}) \frac{f(\bx,\gamma)}{f^{3}_{\gamma,1}(\bx_{i})} \\
 &     \hspace{-2.8cm} - 2 [ c_{2,3}(\bx_{i}) f_{\gamma_{3}}(\bx_{i}) + c_{2,4}(\bx_{i}) f_{\gamma_{4}}(\bx_{i})]^{2}
 \frac{f(\bx,\gamma)}{f^{3}_{\gamma,2}(\bx_{i})} \leq 0.
\end{eqnarray*}
We have $\int \frac{\partial \pi_{2}(\bx_{i},\gamma)}{\partial \gamma} f(\bx_{i},\gamma) d\bx_{i} < 0$.
This result indicates that the minimum of $b_2 (\gamma)$ is either at $\gamma_1$ or at $\gamma$ in spite of
the initial value of $b_1 (\gamma)$ being positive or negative. In other words, either $S(\gamma_1)$ or
$S(\gamma_2)$ must be the minimal value of the model mis-specification error for
$\gamma \in [\gamma_{1},\gamma_{2})$. In the same manner,
either $S(\gamma_2)$ or $S(\gamma_3)$ must be the minimal value of the model mis-specification error for
$\gamma \in [\gamma_{2},\gamma_{3})$.
Finally, from ($\ref{mult-4}$), the slope is a strictly increasing function in $\gamma$ for
$\gamma \in [\gamma_{3},\gamma_{4})$. This fact implies that the minimal value of the model mis-specification error takes place at $\gamma_3$, which is equal
to $S(\gamma_3)$. Therefore, the minimal value among $S(\gamma_1)$, $S(\gamma_2)$, and $S(\gamma_3)$ is the global
minimum of the model mis-specification error for $\gamma \in [\gamma_0, \gamma_4]$. This is the proof of part a) in Theorem 12.

Since $\min(S(\gamma_{1}),S(\gamma_{2}),S(\gamma_{3})) = S(\gamma_{1})$ is assumed, $S(\gamma_1)$ is the global
minimum of the model mis-specification error for $\gamma \in [\gamma_0, \gamma_4]$.
Therefore, from Theorem 2.1 of Newey and McFadden (1994), we have
\begin{eqnarray*}
\hat\gamma_1 = \arg \min \frac{1}{n} SSR (\gamma) \to^p \gamma_1 = \arg \min \frac{1}{n} S(\gamma).
\end{eqnarray*}
This completes the proof of parts b) and c) in Theorem 12. \quad$\blacksquare$

\newpage
\begin{apabib}
A\"{\i}t-Sahalia, Y., Bickel, P.J., Stoker, T.M., 2001. Goodness-of-fit tests for kernel regression with
    an application to option implied volatilities, Journal of Econometrics 105, 363--412.
    
Angrist, J.D., Pischke, J., 2009. Mostly Harmless Econometrics, New Jersey: Princeton University Press.

Bai, J., 1997. Estimating multiple breaks one at a time, Econometric Theory 13, 315--352.

Bai, J., Perron, P., 1998. Estimating and testing linear models with multiple structural changes,
     Econometrica 66, 47--78.

Bai, J., Perron, P., 2003. Computation and analysis of multiple structural change
        models, Journal of Applied Econometrics 18, 1--22.

Bhattacharya, P.K., Brockwell, P.J., 1976. The minimum of an additive process with applications to signal
  estimation and storage theory, Z. Wahrschein. Verw. Gebiete  37, 51--75.

Chan, K.S., 1993. Consistency and limiting distribution of the least squares estimator of a threshold
      autoregressive model, The Annals of Statistics 21, 520--533.

Chen, B., Hong, Y., 2012. Testing for smooth structural changes in time series models via nonparametric regression, Econometrica 80, 1157--1183.

Chen, B., Hong, Y., 2013. Nonparametric testing for smooth structural change in panel data models,
   Working Paper, Department of Economics, University of Rochester.
   
Chen, J.-E., 2008. Estimating and testing quantile regression with structural changes, Working Paper, Department of Economics, NYU.

Chernozhukov, V., Chetverikov, D., Demirer, M., Duflo, E., Hansen, C., Newey, W., 2016. Double machine learning for treatment and causal parameters,
Cemmap Working Paper CWP49/16.

Chernozhukov, V., Hansen, C., 2004. The impact of 401(k) participation on the wealth distribution: An instrumental quantile regression analysis, Review of Economics and Statistics 86, 735--751.

Chernozhukov, V., Hansen, C., 2013. High-Dimensional Methods: Examples for Inference on Structural Effects, NBER Summer Institute.

Dette, H., Spreckelsen, I., 2004. Some comments on specification tests in nonparametric
   absolutely regular processes, Journal of Time Series Analysis 25, 159--172.

Fan, Y., Li, Q., 1999.  Central limit theorem for degenerate U-statistics of absolutely regular
 processes with applications to model specification testing, Journal of Nonparametric Statistics 10, 245--271.

Fan, J., Yao, Q., 2003. Nonlinear Time Series: Nonparametric and Parametric Methods, New York: Springer-Verlag.

Hall, P., 1984. Central limit theorem for integrated squared error of multivariate nonparametric
   density estimators, Journal of Multivariate Analysis 14, 1--16.

Hansen, B.E., 1999. Threshold effects in non-dynamic panels:Estimation, testing, and inference,
      Journal of Econometrics 93, 345--368.

Hansen, B.E., 2000. Sample splitting and threshold estimation,  Econometrica 68, 575--603.

Henderson, D.J., Parmeter, C.F., Su, L., 2014. Nonparametric threshold regression: Estimation
   and inference, Working Paper,  Department of Economics, University of Miami.

Li, Q., Racine, J.S., 2007. Nonparametric Econometrics: Theory and Practice,
   Princeton, NJ: Princeton University Press.

Masry, E., 1996. Multivariate regression estimation local polynomial fitting for time series,
   Stochastic Processes and their Applications 65, 81--101.

Masry, E., Fan, J., 1997. Local polynomial estimation of regression functions for mixing processes,
   Scandinavian Journal of Statistics 24, 165--179.

Newey, W. K., McFadden, D.L., 1994. Large sample estimation and hypothesis testing,
     \underline{Handbook of Econometrics: Vol. IV}, ed. by R. F. Engle and D. L. McFadden,
     New York: Elsevier,  2113 -- 2245.

Oka, T., Qu, Z., 2011. Estimating structural changes in regression quantiles,
    Journal of Econometrics 162, 248--267.
    
Poterba, J.M., Venti, S.F., Wise, D.A., 1994a. 401(k) plans and tax-deferred savings, Studies in the Economics of Aging, Chicago: University of Chicago Press, 105--142.

Poterba, J.M., Venti, S.F., Wise, D.A., 1994b. Do 401(k) contributions crowd out other personal saving?, Journal of Public Economics 58, 1--32.

Qu, Z., 2008. Testing for structural change in regression quantiles, Journal of
   Econometrics 146, 170--184.

Qu, Z., Perron P., 2007.  Estimating and testing structural changes in multivariate regressions,
   Econometrica  75, 459--502.

Stone, C.J., 1983. Optimal uniform rate of convergence for nonparametric estimators of a
   density function or its derivatives, Recent Advances in Statistics, 393--406. Academic Press,
     New York.

Su, L.,  Xiao, Z., 2008. Testing structural change in time-series nonparametric regression models,
    Statistics and Its Interface 1, 347--366.
    
Yu, P.,  Philips, P.C.B., 2015. Threshold regression with endogeneity,
    Cowles Foundation Discussion Paper no.\ 1966.

\end{apabib}

\pagebreak
\end{document}